%% file: MUO-19-001_temp.tex
\pdfoutput=1
\documentclass[11pt,twoside,a4paper,cmspaper,final,collab]{cms-tdr}
\def\svnVersion{dd99c62-D}\def\svnDate{2021/06/28}\def\cmsCernNoTag{CERN-EP-2021-013}\def\cmsCernDate{\today}\def\cmsMessage{"Published in the Journal of Instrumentation as \href{http://dx.doi.org/10.1088/1748-0221/16/07/P07001}{\doi{10.1088/1748-0221/16/07/P07001}.}"}
\begin{document}\cmsNoteHeader{MUO-19-001}

\cmsNoteHeader{MUO-19-001}

\title{Performance of the CMS muon trigger system in proton-proton collisions at \texorpdfstring{$\sqrt{s} = 13\TeV$}{sqrt(s) = 13 TeV}}

\date{\today}

\abstract{The muon trigger system of the CMS experiment uses a combination of hardware and software to identify events containing a muon.
   During Run 2 (covering 2015--2018) the LHC achieved  instantaneous luminosities as high as $2\times 10^{34}\percms$  while delivering proton-proton collisions at $\sqrt{s} = 13\TeV$.
   The challenge for the trigger system of the CMS experiment is to reduce the registered event rate from about 40\unit{MHz} to about 1\unit{kHz}. Significant improvements important for the success of the CMS physics program have been made to the muon trigger system via improved muon
   reconstruction and identification algorithms since the end of Run 1 and throughout the Run 2 data-taking period. The new algorithms maintain
   the acceptance of the muon triggers at the same or even lower rate throughout the data-taking period despite the increasing number of
   additional proton-proton interactions in each LHC bunch crossing. In this paper, the algorithms used in 2015 and 2016 and their improvements throughout 2017 and 2018 are described. Measurements of the CMS muon trigger performance for this data-taking period are presented, including efficiencies, transverse momentum resolution, trigger rates, and the purity of the selected muon sample. This paper focuses on the single- and double-muon triggers with the lowest sustainable transverse momentum thresholds used by CMS. The efficiency is measured in a transverse momentum range from 8 to several hundred\GeV.
}

\hypersetup{ 
pdfauthor={CMS Collaboration},
pdftitle={Performance of the CMS muon trigger system in proton-proton collisions at sqrt{s} = 13 TeV},
pdfsubject={CMS},
pdfkeywords={CMS, muon, trigger}}

\maketitle 

\section{Introduction}
Muons feature prominently in many of the most important signatures studied at the CERN LHC,
including final states from {\cPqb} hadron, vector boson, and Higgs boson decays, as well as in searches for
new physics beyond the standard model. Since they are the only detectable particles to traverse
the whole detector without significant loss of energy, muons are ideal probes to identify
interesting interactions among the large number of proton-proton ($\Pp\Pp$) collisions provided by
the LHC. The ability to reconstruct and identify muons with high efficiency and precision while
maintaining low misidentification rates is therefore crucial to the success of the CMS physics program.

In the $\Pp\Pp$ collisions, muons originate from a variety of processes. The dominant source is decay in flight of pions and kaons, followed by semileptonic decays of b and c hadrons. Muons from the decay of vector or Higgs bosons contribute less than 1\% to the overall muon sample~\cite{Chatrchyan:2012xi}.

In Run 2 (years 2015--2018), the center-of-mass energy ($\sqrt{s}$) of the $\Pp\Pp$ collisions delivered by the LHC increased from 8 to 13\TeV. The small amount of data recorded in 2015 is disregarded in the following. The peak instantaneous luminosity achieved during the data
taking in Run 2 increased from $1.5\times
10^{34}\percms$  in 2016 to $2.0\text{--}2.1\times
10^{34}\percms$ in 2017 and 2018. The average
number of simultaneous 
$\Pp\Pp$ interactions per bunch crossing (pileup) increased from 23 in 2016 to 32 in 2017 and 2018. However, because of two distinct beam configurations employed during 2017, the pileup distribution exhibits a double peak structure with a secondary peak at pileup $\approx$55. This posed a
significant challenge to the CMS trigger system, which was tasked to maintain acceptable event rates for storage
and analysis in a pileup environment  far in excess of the design value of 20--25.

The trigger system comprises two
levels~\cite{Khachatryan:2016bia}. The Level-1 (L1) trigger system consists
of custom hardware processors that receive data from calorimeter and muon systems and reduce the event
rate from 40\unit{MHz} to about 100\unit{kHz}. The software-based high-level trigger (HLT) system, which operates on a computing farm with up to 30\,000 CPU cores (at the end of Run 2), further reduces the rate
to approximately 1\unit{kHz}. The algorithms and techniques used in these triggers, both at L1 and HLT, evolved
during the data taking.

The muon reconstruction in the HLT is split into two steps: the Level-2 (L2) reconstruction that uses
information from the muon detectors, and the Level-3 (L3) reconstruction that incorporates additional information
from the inner tracking detectors. These reconstruction steps are applied in sequence, followed by muon identification and isolation requirements, if applicable, defining individual ``triggers''. If all the requirements in a given trigger are fulfilled, the event will be accepted by the trigger system.  In this paper, we will focus on triggers requiring the presence of either at least one (``single-muon trigger'') or two (``double-muon trigger'') muons fulfilling certain transverse momentum (\pt) requirements. These thresholds are chosen to control the rate of accepted events. Dedicated triggers targeting muons from low mass resonances or requiring the presence of at least one muon in conjunction with other objects, such as electrons or tau leptons, which also use the same muon reconstruction, are not discussed in this paper. At both L1 and HLT, triggers can be prescaled, so that only a predefined fraction of events fulfilling the trigger condition are accepted. 

During Run 2, the L3 reconstruction algorithms were
significantly improved with two objectives in mind:
maintaining a single-muon trigger with a \pt threshold low enough to retain a large
fraction of $\PW\to \mu\nu$ decays, and extending the muon reconstruction to use the additional
pixel layers that were installed during the pixel detector upgrade~\cite{Dominguez:1481838} between the 2016 and 2017 data taking. Another consideration was to bring the trigger-level reconstruction closer to the offline one~\cite{Sirunyan:2018fpa} by reducing the reliance on specialized code. With the improved
algorithms,
a significant increase in the purity of the muon sample selected by the single-muon triggers was achieved without a significant reduction in trigger efficiency.

In this paper, we summarize the reconstruction algorithms deployed in the HLT throughout the data-taking period,
focusing on the improvements made since the start of Run 2. The selection and reconstruction efficiencies of
the different levels of the trigger systems (L1, L2, and L3), as well as the \pt resolution for L2 and L3 muons, are presented. For the unprescaled single- and double-muon triggers with
the lowest \pt thresholds sustainable at the full instantaneous luminosity during Run 2, the overall performance of the muon trigger system is measured
in terms of the efficiency, covering a \pt range from 8 to several hundred~GeV.
Additionally, the impact of the algorithmic improvements on the purity of the selected muon sample, the event processing time, and the trigger rate
is discussed. 

The description and performance of the muon trigger system used during Run 1 (years 2010--2012), for both the L1 and the HLT, are presented
in Ref.~\cite{Khachatryan:2016bia}.  For Run 2, a detailed description of the L1 trigger system and its performance is presented
in Ref.~\cite{TRG-17-001}. The muon HLT performance for muons with very high \pt is included in Ref.~\cite{MUO-17-001}.
This paper completes the documentation of the CMS muon trigger system, as used during Run 2, with detailed studies
of the HLT performance for a more general selection of triggers. 

This paper is organized as follows. Section~\ref{sec:cms} describes the CMS detector with a focus
on the subdetectors relevant for muon reconstruction. Section~\ref{sec:DataAndTnP} describes the data used
together with a brief explanation of the efficiency measurement
techniques. The hardware-based L1 trigger system and its performance are
discussed in Section~\ref{sec:L1}. The L2 and the various L3 muon reconstruction algorithms used in the
HLT throughout the data
taking are described in Section~\ref{sec:HLT}. Section~\ref{sec:performance} presents the overall muon
trigger system performance combining the full L1--L3 chain, followed by a summary in Section~\ref{sec:summary}.

\section{The CMS detector}
\label{sec:cms}
The central feature of the CMS apparatus is a superconducting solenoid of 6\unit{m} internal diameter,
providing a magnetic field of 3.8\unit{T}. Within the solenoid volume are a silicon pixel and strip tracker,
a lead tungstate crystal electromagnetic calorimeter (ECAL), and a brass and scintillator hadron calorimeter (HCAL), each
composed of a barrel and two endcap sections. Forward calorimeters extend the pseudorapidity ($\eta$) coverage provided
by the barrel and endcap detectors. Muons are detected in gas-ionization chambers embedded in the steel flux-return
yoke outside the solenoid. A more detailed description of the CMS detector, together with a definition of the
coordinate system used and the relevant kinematic variables, can be found in Ref.~\cite{Chatrchyan:2008zzk}. 

In the region $\abs{\eta} < 1.74$, the HCAL cells have widths of 0.087 in both $\eta$ and azimuth ($\phi$). In the $\eta$-$\phi$ plane, and for $\abs{\eta} < 1.48$, the HCAL cells map onto $5{\times}5$ arrays of ECAL crystals to form calorimeter towers projecting radially outwards from close to the nominal interaction point. For $\abs{ \eta } > 1.74$  , the granularity of the towers increases progressively to a maximum of 0.174 in $\Delta \eta$ and $\Delta \phi$. 

The silicon tracker measures the trajectories of charged particles within $\abs{\eta} < 2.5$. During the data-taking period in 2016, the silicon tracker consisted of 1440 silicon pixel and 15\,148 silicon strip detector modules. They were arranged in concentric layers, three pixel and 10 strips, around the beam axis in the central region of the detector, and in disks, two pixel and 12 strips, perpendicular to it in the forward directions. For nonisolated particles (i.e., with no isolation requirement applied) with $1 < \pt < 10\GeV$ and $\abs{\eta} < 1.4$, reconstructed using the full offline event reconstruction, the track resolutions are typically 1.5\% in \pt and 25--90 (45--150)\mum in the transverse (longitudinal) impact parameter \cite{Chatrchyan:2014fea}. Before the data-taking period in 2017, an upgraded pixel detector consisting of 1856 modules was installed, adding an additional barrel layer closer to the interaction point and additional disks in the two forward parts of the detector~\cite{Dominguez:1481838}. With the upgraded detector, the transverse impact parameter resolution for nonisolated particles of $1 < \pt < 10\GeV$ and $\abs{\eta} < 2.5$ improved to 20--75\mum~\cite{CMS-DP-2020-032}.

\begin{figure} \centering
\includegraphics[width=1.0\textwidth]{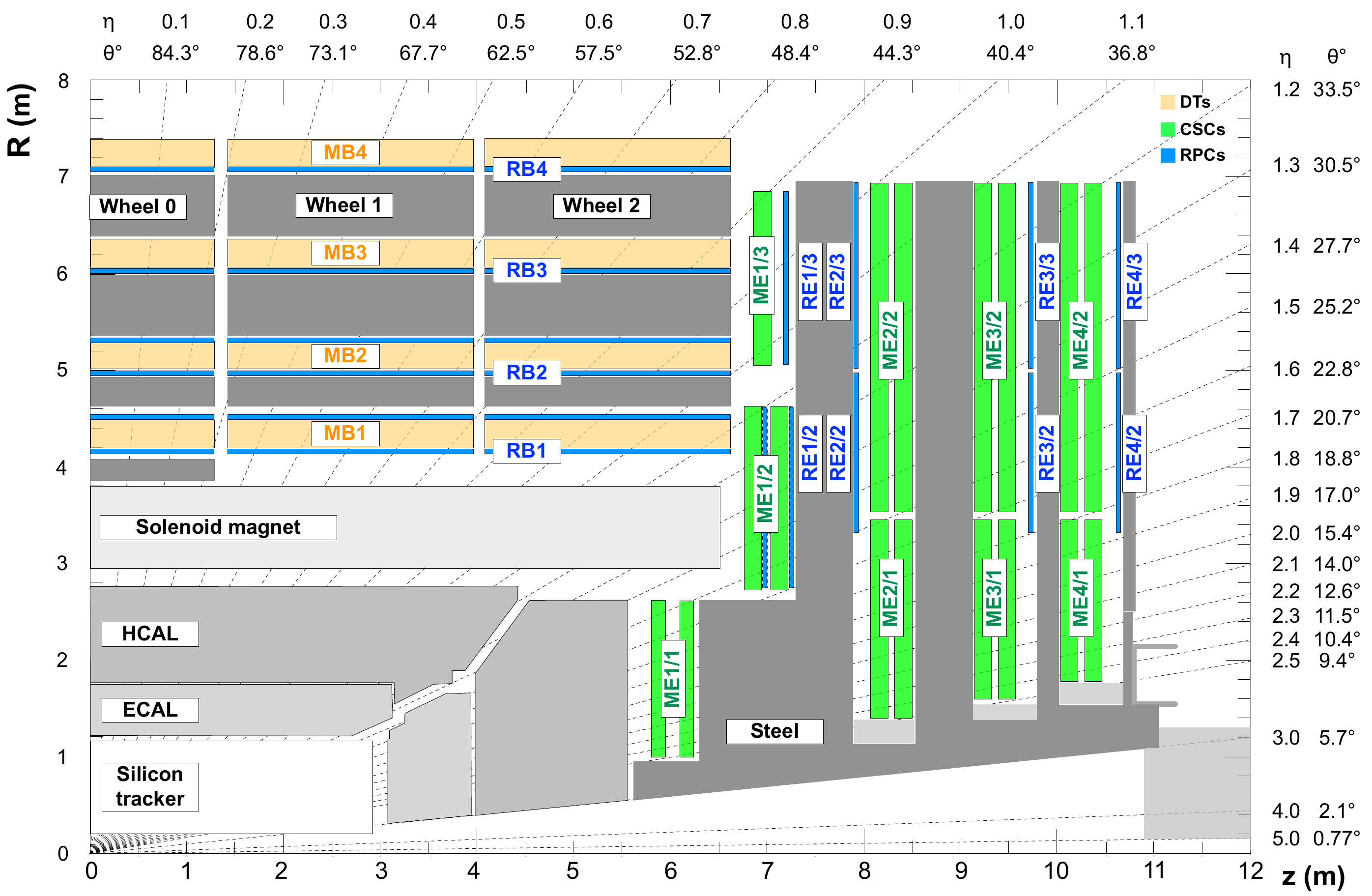}
\caption{\textit{R-z} quadrant of the CMS detector during Run 2. The proton beams travel along the $z$ axis and cross at the interaction point (0,0). The three CMS muon subdetectors typically have four stations:
  DT in yellow, labeled MB; CSC in green, labeled ME; and RPC in blue, labeled RB or RE.}
\label{fig:CMS}
\end{figure}

Three types of gas-ionization chambers make up the CMS muon system: drift tube chambers (DTs),
cathode strip chambers (CSCs), and resistive-plate chambers (RPCs). The geometrical arrangement of these three muon
subdetectors in a quadrant of the CMS detector is shown in Fig.~\ref{fig:CMS}. A detailed description of these chambers,
including the gas composition and operating voltage, is reported in Ref.~\cite{DPGPerformance}. The DTs are segmented
into drift cells; the position of the muon is determined by measuring the drift time to an anode wire of a cell
with a shaped electric field. The CSCs operate as standard multi-wire proportional counters but add a finely
segmented cathode strip readout, which yields a precise measurement of the position of the bending plane
($R$-$\phi$) coordinate at which the muon crosses the gas volume. The RPCs are double-gap chambers operated in
avalanche mode and are primarily designed to provide timing information for the muon trigger. The DT and CSC
chambers are located in the regions  $\abs{\eta} < 1.2$ and $0.9 < \abs{\eta} < 2.4$, respectively, and are complemented
by RPCs in the range $\abs{\eta} < 1.9$.  The chambers are arranged to maximize the coverage and to provide some overlap
where possible. In both the barrel and endcap regions the chambers are grouped into four ``muon stations'', separated by the steel absorber of the magnet return yoke. 
In the barrel region, the DT and RPC stations are arranged in five wheels along the $z$ direction. 

The thickness of the detector in radiation lengths is greater than 25\,$X_0$ for the ECAL, and the thickness in interaction lengths varies from 7-11\,$\lambda_0$ for HCAL depending on $\eta$. The material depth, in radiation lengths, at the muon stations varies from 100 for the first muon station up to 280 for the outermost muon station~\cite{Bayatian:2006nff}.

Muon trajectories are measured in the range $\abs{\eta} < 2.4$, using detection planes comprised of 
the three muon subdetectors in up to four muon stations. Using the full offline event reconstruction, the efficiency to reconstruct and identify muons is greater 
than 96\%. Matching muons to tracks measured in the silicon tracker results in a relative \pt resolution, for muons
with \pt up to 100\GeV, of 1\% in the barrel and 3\% in the endcaps. The \pt resolution in the barrel is
better than 7\% for muons with \pt up to 1\TeV~\cite{Sirunyan:2018fpa,MUO-17-001}.

\section{Data samples and trigger efficiencies}\label{sec:DataAndTnP}
The results described in this paper are based on the $\Pp\Pp$ collision data collected during the Run 2 of the LHC by the CMS experiment, corresponding to a total integrated luminosity of 137\fbinv distributed as follows: 36\fbinv (2016), 41\fbinv (2017), and 60\fbinv (2018). Because the instantaneous luminosity of the LHC increased during the data taking, the average pileup in the 2017 and 2018 data samples is about 40\% higher than in the 2016 sample. For performance measurements of the full trigger, all available data for each year are used. However, a fair comparison of the algorithmic performance of the different steps of the reconstruction between different years is difficult because of the continually evolving nature of the detector and the trigger algorithms and configurations. For this reason, smaller data samples representing the best detector performance of each year are chosen for these measurements. For 2016 and 2018, the last 5\fbinv of the data of the relevant year were used since they provide data sets with stable and well understood detector conditions. For 2017, two different data sets are used. For L1 and L2 algorithms that rely solely on the muon detectors, the last 5\fbinv collected in that year show the best performance. The later part of the 2017 data is affected by inactive pixel detector modules caused by a powering issue  related to radiation-induced damage in the FEAST DC-DC converter ASICs~\cite{DCDC}, which affected about 7.5\% of modules at the end of the year. Therefore, a data set of 4\fbinv from early in the data taking is used for L3 and isolation performance measurements. 

The selected data samples for most of the studies shown in this paper were recorded using a single-muon trigger; they include events with a pair of reconstructed muons with a dimuon invariant mass consistent with the \PZ boson resonance. Throughout this paper, efficiencies with respect to offline muons are measured with $\PZ \rightarrow \PGm\PGm$ data events using a ``tag-and-probe'' (T\&P) technique~\cite{CMS_WZ}. Pairs of muons reconstructed by the CMS offline event reconstruction are selected, where at least one of the muons has to pass the strict ``tag'' criteria of  $\pt > 26\GeV, \abs{\eta} < 2.4$, and tight identification and isolation requirements. The identification requirements are based on the number of hits in the tracker and muon systems associated with the muon, the quality of the muon track fit and its impact parameters; the isolation requirements are based on tracks and energy deposits in the calorimeters in the vicinity of the muon~\cite{Sirunyan:2018fpa}. The tag is also required to have passed an isolated single-muon trigger requirement to ensure that the efficiency measurement remains unbiased. The other muon is considered the ``probe'' muon, and it is used to measure the efficiency of a given reconstruction step relative to a probe selection by counting the number of probe muons that pass the condition under study and dividing by the overall number of probes. Probe muons are required to fulfill $\pt > 24\GeV$ for most measurements. The threshold is lowered to 10\GeV for the measurement of the trigger efficiency at low \pt and to $12\GeV$ for some of the L1 efficiency measurements. The probe muon must fulfill $\abs{\eta} < 2.4$ and the same tight identification and isolation criteria as the tag muon. For L2 and L3 reconstruction, the probe muon must be matched to an L1 muon with $\pt > 22\GeV$ that passes tight L1 quality requirements. For the isolation efficiency measurement, the probe has to satisfy the single-muon trigger requirements without an isolation requirement. The tag and probe muons are required to be separated by $\Delta R = \sqrt{\smash[b]{(\Delta\eta)^2+(\Delta\phi)^2}} > 0.3$, which is sufficient to remove biases due to nearby muons. To pass the probe condition, the probe muon is required to geometrically match the trigger object that is being studied by requiring $\Delta R < 0.1$. This criterion is relaxed to 0.5 (0.3) for the  L1 (L2)  muon efficiency measurement, since these objects have a larger directional resolution. Table \ref{tab:TnPDefs} summarizes the different probe objects and the matching criteria for the probe. Finally, the tag and probe muon pair is required to have an invariant mass in the range  $81 < m_{\mu\mu} < 101 \GeV$. This mass window size is chosen to efficiently reject non-\PZ boson backgrounds while retaining large enough data sets for the measurement. After the full selection, the contribution to the data samples from non-DY background is negligible.

\begin{table}
\centering
\topcaption{Parameters of the tag-and-probe method used in this paper. $\Delta R$ is the maximum allowed angular separation between the tag muon and the probe muon.}
\label{tab:TnPDefs}
\begin{tabular}{lcc}
  Efficiency         & Probe object & $\Delta R$ trigger-probe \\
  \hline
  L1 reconstruction  & Probe muon   &  0.5\\
  L2 reconstruction  & Probe muon matched to L1 muon&  0.3 \\
  L3 reconstruction  & Probe muon matched to L1 muon&  0.1 \\
  L3 isolation       &  Probe muon matched to L3 muon  &  0.1 \\
  HLT (single-muon)   & Probe muon &  0.1 \\
  HLT (double-muon)  & Probe muon with relaxed \pt requirement  &  0.1 \\
\end{tabular}
\end{table}

\section{The Level-1 trigger}
\label{sec:L1}

During Run~1, the L1 muon trigger system included three separate hardware muon track finders, each of which
reconstructed muons using input from a single muon subdetector: DT, CSC, or RPC~\cite{Khachatryan:2016bia}.
The upgraded L1 trigger system in Run~2~\cite{L1Phase1UpgradeTDR} combines
inputs from all geometrically available subdetectors and is segmented
into three regional track finders: the barrel muon track finder (BMTF) 
covering $0 < \abs{\eta} < 0.83$, the overlap muon track finder
(OMTF) covering $0.83 < \abs{\eta} < 1.24$, and the endcap muon
track finder (EMTF) covering 
$1.24 < \abs{\eta} < 2.40$. The track finders receive hits or short track segments
called ``trigger primitives'' (TPs) from each station for every subdetector in that region.
The TPs carry position ($\theta$ and $\phi$) coordinates, direction, and timing information (correlated
with a collision bunch crossing). 
In an improvement over Run 1, adjacent RPC hits are clustered into TPs before being used in track building and RPC TPs are combined with nearby DT segments in the barrel to improve the overall TP efficiency and timing for the BMTF.

Each of the three track finders reconstructs muons on processor boards with field-programmable
gate arrays (FPGAs),
following a similar sequence. Trigger primitives aligned in $\theta$ and $\phi$ are grouped to form tracks,
which traverse the four stations in each subdetector. The angular deflection between stations
is used to assign a \pt value, primarily using the $\phi$ coordinate, since the magnetic
field causes charged-particle tracks to curve in the transverse plane with a radius proportional to \pt. However,
because of significant differences between the subdetectors, and large variations in the magnetic
field strength and background muon rate, the exact track building and \pt assignment logic
for each track finder is unique.  

The Global Muon Trigger collects reconstructed muons from the three track finders and removes geometrically overlapping tracks using information based on \pt and quality,
sending the remainder to the L1 Global Trigger, which decides whether to keep the event for processing by the HLT. The quality is assigned based on the number and location of TPs on a given track~\cite{TRG-17-001}. Tracks passing the tight L1 quality requirement  have the best \pt resolution and are therefore used for single-muon L1 seeds. For multi-muon L1 seeds, the criteria are relaxed to improve the efficiency, allowing for tracks with a lower number of TPs to pass. This only affects the overlap and endcap regions since all L1 tracks found by the BMTF pass the tight L1 quality selection. The L1 muon reconstruction is described in greater detail in dedicated publications~\cite{Khachatryan:2016bia,TRG-17-001}. 

Some features of the upgraded system were implemented or improved progressively throughout Run 2.  The BMTF switched from DT-only segments to combined DT+RPC TPs at the end of 2016.  The OMTF developed a track building algorithm that was less reliant on RPC TPs in 2018.  In the EMTF, track timing was improved during 2016; RPC TPs were added to the CSC-dominated track reconstruction before 2017 collisions, along with a more accurate \pt assignment algorithm; and in 2018 modifications were made to reduce the impact of extra TPs from pileup.

The efficiency of  the L1 muon trigger is measured using the T\&P technique. The numerator of the efficiency includes all probe muons matched  within $\Delta R < 0.5$ to an L1 trigger with $\pt>22$ (7)\GeV and passes the tight (medium) L1 quality criteria~\cite{TRG-17-001}. These requirements are
applied to select L1 muons, called L1 seeds, to initiate single (double) muon reconstruction at HLT. In particular, these \pt thresholds correspond to the lowest threshold used in an unprescaled single-muon L1 seed, and the lowest threshold used in an unprescaled double-muon L1 seed during Run 2, respectively.

\begin{figure} \centering
\includegraphics[width=0.45\textwidth]{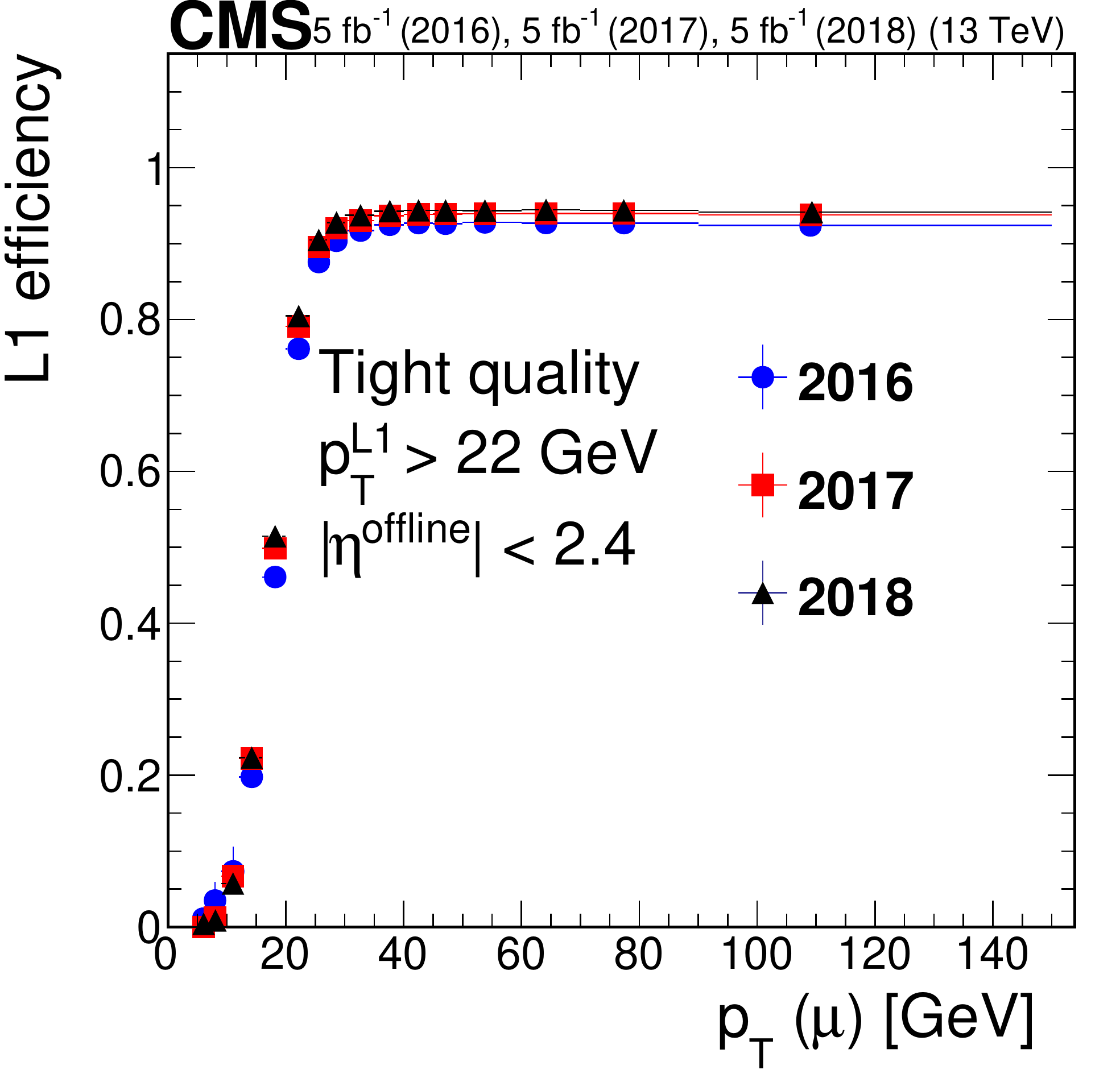}
\includegraphics[width=0.45\textwidth]{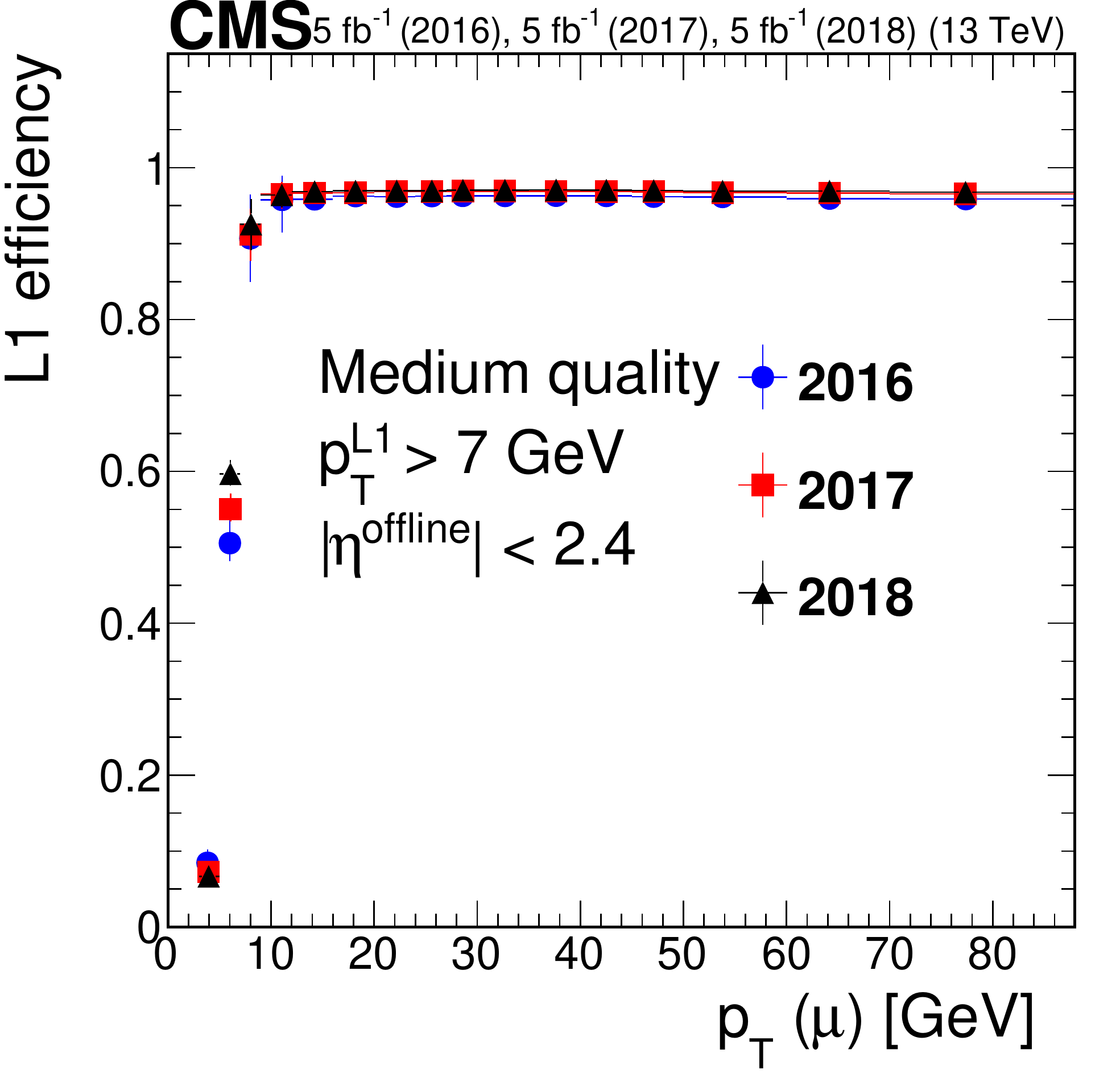}
\includegraphics[width=0.45\textwidth]{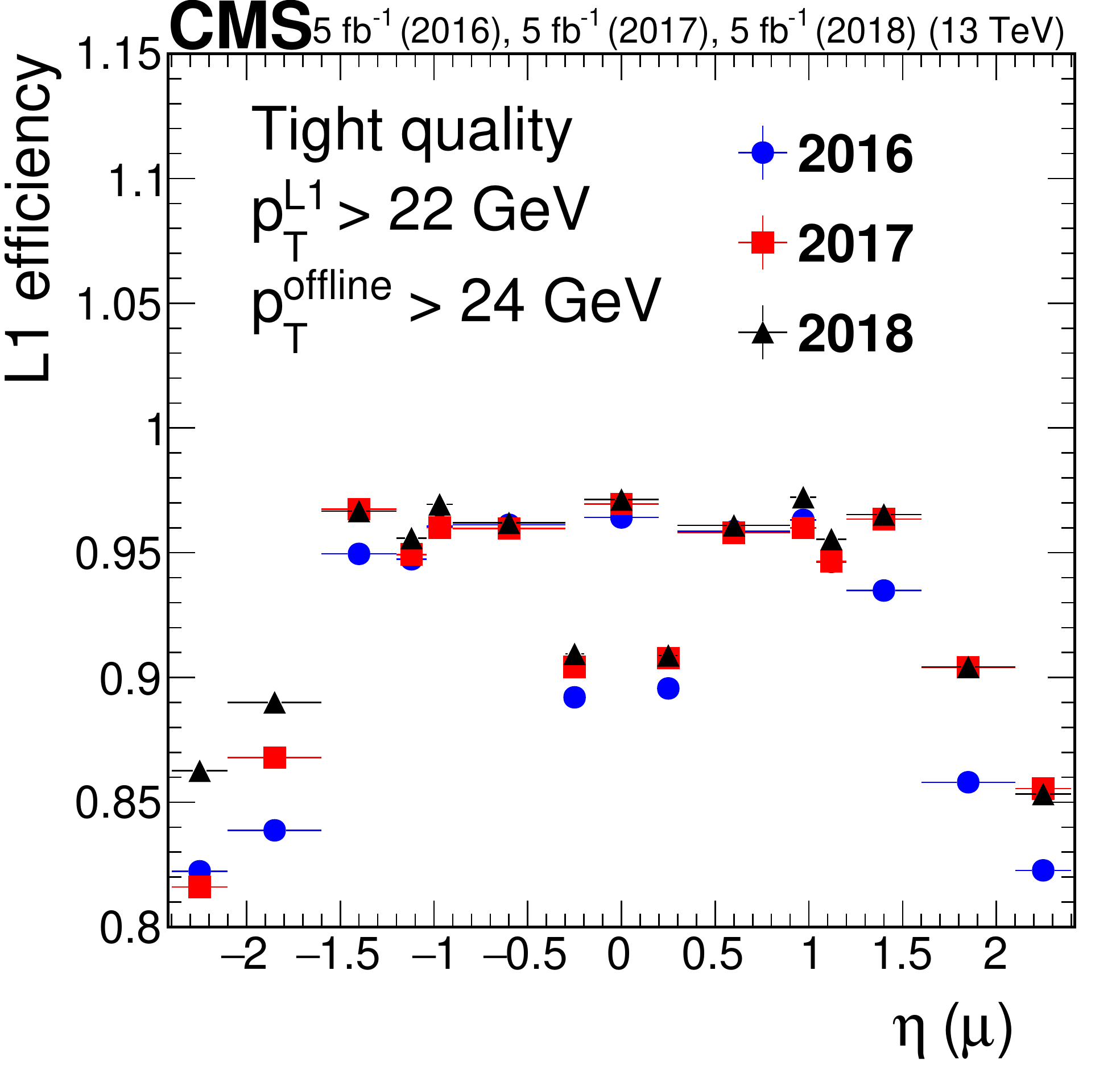}
\includegraphics[width=0.45\textwidth]{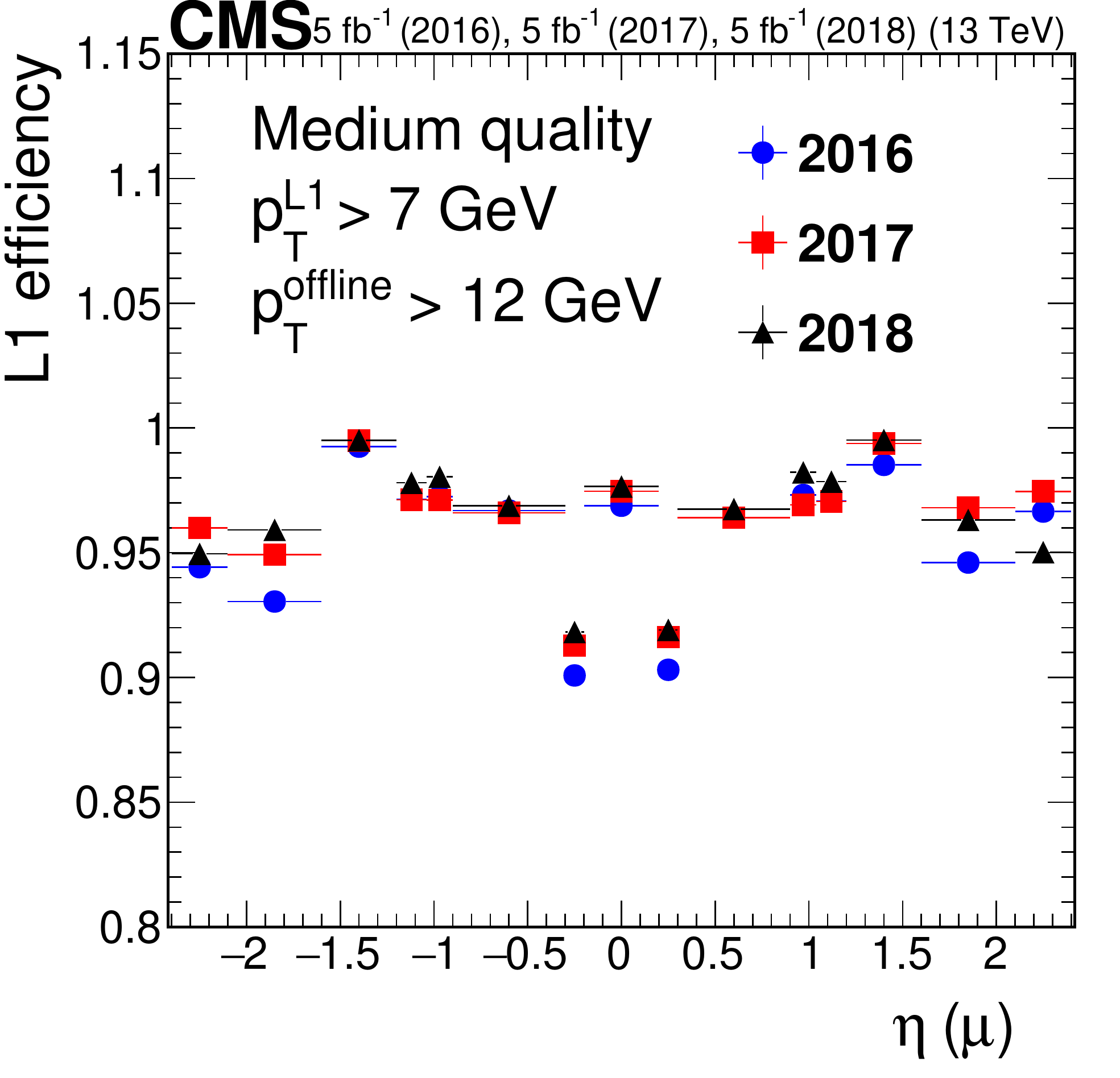}
\includegraphics[width=0.45\textwidth]{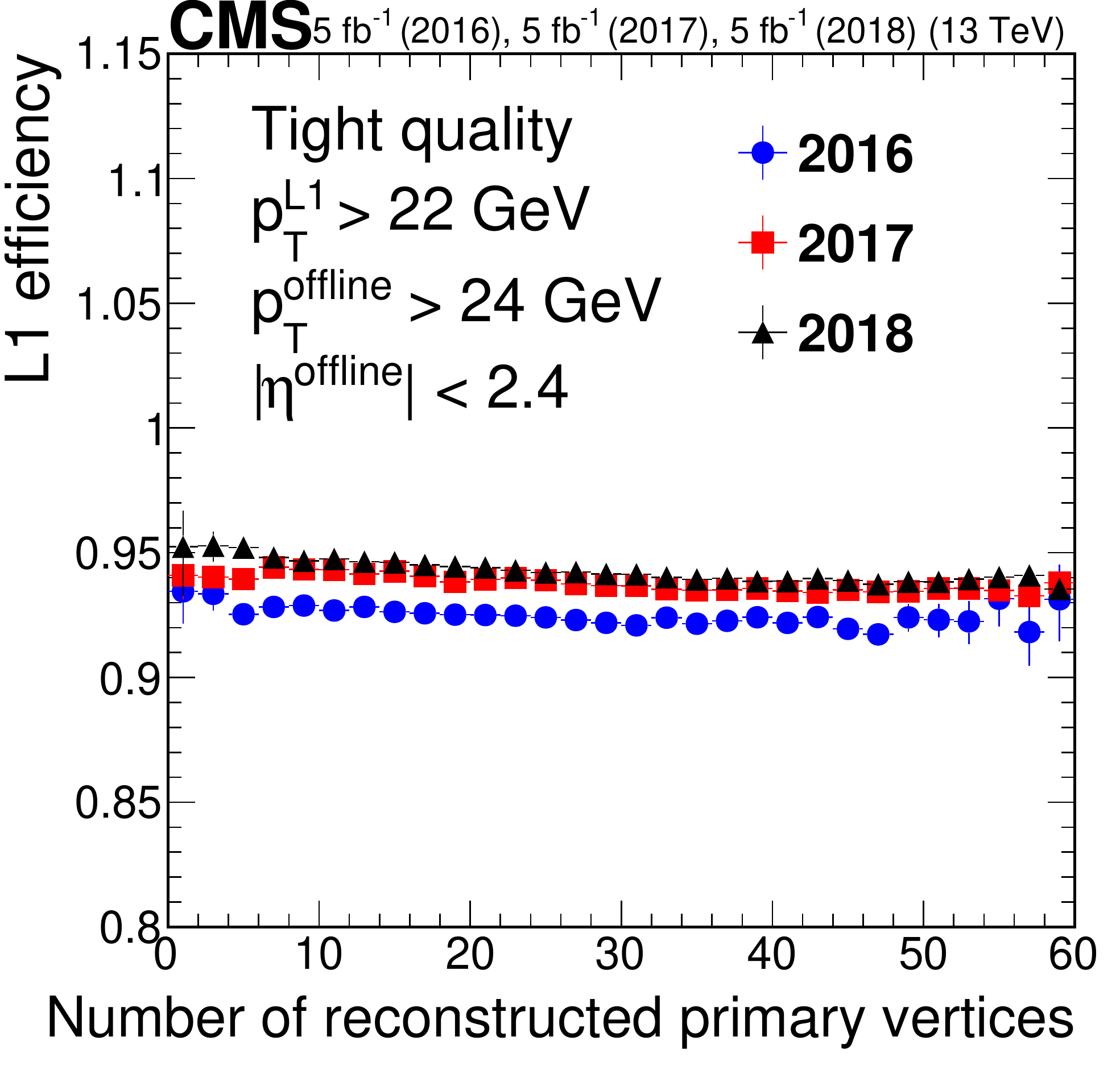} 
\includegraphics[width=0.45\textwidth]{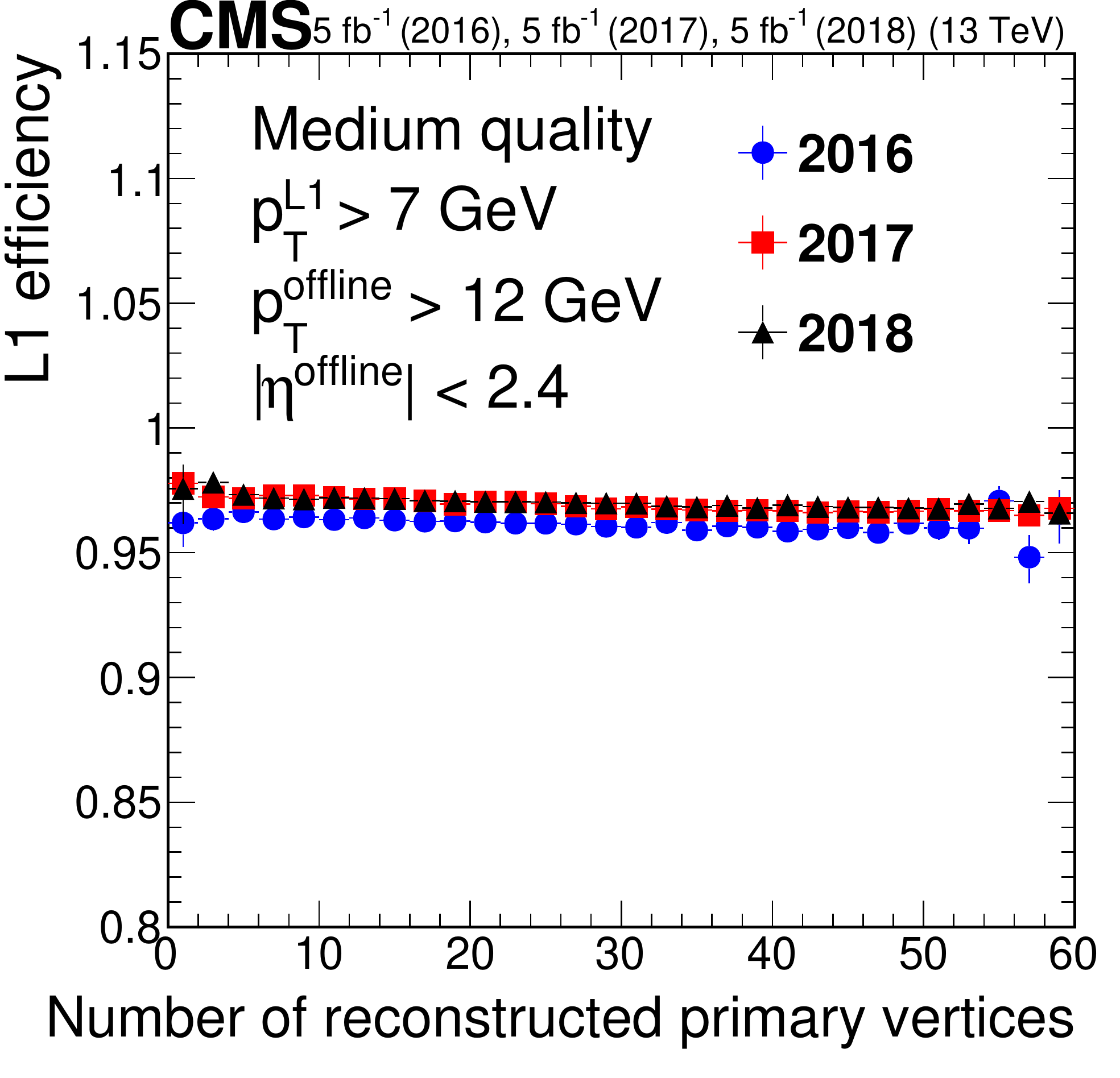}
\caption{Efficiency of the L1 muon trigger requiring the tight (left) and medium (right)  quality criteria as required in single-muon and double-muon seeds, respectively, for $\PZ\to \mu\mu$ events, as a function of \pt (upper row), muon $\eta$ (middle row), and number of primary vertices (lower row). The efficiency for 2016 data is shown in blue circles, for 2017 data in red squares, and for 2018 data in black triangles. Only statistical uncertainties are shown.}
\label{fig:perf:L1Eff}
\end{figure}

The L1 efficiency in data is shown in Fig.~\ref{fig:perf:L1Eff}, as a
function of the offline \pt and $\eta$ of the muon, as well as the number of primary vertices, the latter being all vertices reconstructed in the immediate $\Pp\Pp$ interaction region, including pileup interactions. This excludes secondary vertices from meson decays in flight. Since very low \pt muons are absorbed in the calorimeters and the magnet cryostat before they reach the muon system, no measurement is performed for offline $\pt < 4\GeV$ for the medium quality. The
efficiency reaches about 93--95 (96--97)\%
for tight (medium) L1 quality and, above 35 (12)\GeV, is virtually independent of muon \pt (up to 150\GeV). The performance of the L1 muon trigger for muons with a \pt of several hundred~GeV is reduced by the high probability for radiative energy losses in
the yoke of the CMS magnet. The impact on the trigger efficiency for
muons with momenta up to 1.5\TeV was studied in detail in Ref.~\cite{MUO-17-001}.
The conclusion is that the efficiency has been generally stable during the entire data-taking period; it is higher for central muons and
drops for $\abs{\eta} > 1.6$, especially for tighter quality requirements.
The small variations in the efficiency as a function of $\eta$ reflect the geometric 
structure of the CMS muon system, in particular the cracks in
between the wheels at $\abs{\eta} \approx 0.2$. In general, for $\abs{\eta} > 1.6$ the efficiency is larger in 2017 and 2018 compared to 2016 because of the addition of RPC TPs in the EMTF. However, for 2017 the efficiency is slightly lower in
the negative endcap because of two disabled muon CSC	 chambers in the first station, each
covering $10^{\circ}$ in azimuth. Both 2017 and 2018 show slightly higher performance in
the 
endcaps, which arises from the series of improvements to the L1
reconstruction algorithms during the data-taking period. 
Because the requirements on the number of TPs are more stringent in the
case of the tight L1 quality criteria, the gain in efficiency is larger
in this case.
Since the particles from pileup interactions are mostly absorbed before they reach the muon system, the L1 trigger efficiency is roughly independent of pileup and decreases 
by only a small amount as a function of the number of reconstructed
vertices. This small loss of performance as a function of the pileup is limited to the endcap
regions, where particles from pileup are concentrated. 

Because of the limited time resolution of the muon detectors, L1
muons might mistakenly be assigned to an earlier or later LHC bunch
crossing (BX). This will lead to the wrong BX being accepted by the L1
trigger system, an effect known as ``trigger prefiring'' if an earlier BX is accepted. When such an event
is accepted, the CMS L1 trigger
rules~\cite{Khachatryan:2016bia,TRG-17-001} prevent the following two BXs
from generating a valid trigger accept signal, resulting in the correct event
being lost. Furthermore, while L2 muons might be reconstructed from the muon system information now assigned to the wrong BX, the reconstruction of L3 muons using information from the tracker will fail. Thus the BX mistakenly accepted by the L1 will be rejected by the HLT. 
The associated inefficiency cannot be measured using the standard T\&P 
technique since events containing a prefiring L1 muon are usually not 
recorded at all.
There exist a few rare situations where such events are recorded~\cite{TRG-17-001}, allowing the assessment of the size of the effect. The overall prefire probability of 1.5\% per muon at the beginning of 2016 was reduced to 0.5\% after the L1 timing measurements were improved by the use of RPC information in the BMTF and EMTF, as mentioned above. Since in the tag and probe measurement the tag muon is required to be matched to a single-muon trigger, these events cannot be lost as a result of the muon being wrongly assigned to a later BX. Trigger inefficiency due to these prefired muons is therefore not included in the trigger efficiencies presented above.

\section{Muon reconstruction and selection in the high-level trigger}
\label{sec:HLT}
Depending on the year, 20--25\% of all objects accepted by the single-muon L1 trigger do not correspond to offline reconstructed muons (as detailed in Section~\ref{sec:rate}). More sophisticated muon reconstruction and identification algorithms are employed in the HLT to achieve the necessary precision in the reconstructed muon candidates to reject these objects and to reduce the trigger rate.

The reconstruction of the muon trajectory is performed in two steps: a reconstruction within the muon system only (L2 muons) and a reconstruction in the inner tracking detector, which is then combined with the information from the muon spectrometer to reconstruct the full trajectory of the muon through the detector (L3 muons).
The goal of the L2 reconstruction is to refine the initial estimate of the muon trajectory from the L1 reconstruction by using more precise algorithms that cannot be implemented in the hardware-based L1 trigger system. This is especially important, since the L2 muons are used to seed the track reconstruction in the inner tracker in most L3 reconstruction algorithms. The limitations on the computing time per event in the HLT do not allow for track reconstruction to be performed in the whole volume of the inner tracking detector. Therefore, the L3 reconstruction algorithms have to be limited to small regions in the detector based on the presence of L1 or L2 muons to achieve high reconstruction efficiency while keeping the use of computing resources at acceptable levels. 

The single-muon trigger is allocated the largest share of the total available trigger bandwidth (roughly 15\%), so it is crucial to ensure that only well-measured muons are accepted. To this end, identification and isolation criteria can be applied to the L3 muons. This helps to reject trajectories that do not correspond to genuine muons, as well as to reject muons inside jets or from heavy flavor decays, thus reducing the trigger rate while keeping a high acceptance for the muons of interest.

\subsection{Level-2 muon reconstruction}
\label{sec:l2reco}	
The reconstruction of L2 muons is identical to that used for
offline ``standalone muons''~\cite{Bayatian:922757,Chatrchyan:2009ae,Chatrchyan:2012xi,Sirunyan:2018fpa}. It has been stable throughout Run 2 and showed excellent performance despite the harsher pileup conditions of Run 2. This is because the muon detectors are not significantly affected by the harsher pileup conditions of Run 2, since the particles from pileup collisions are absorbed in the calorimeters, the magnet cryostat, and the steel return yoke.

The DT and CSC segments are taken as inputs and combined
to produce a set of initial track states, referred to as ``L2 seeds'', which are the starting point
for the reconstruction of muon tracks. Here a ``state'' refers to the five-dimensional parameterization of a trajectory on a specific detector surface, such as one of the muon chambers. The five parameters are the ratio of charge over momentum, the transverse and longitudinal angles with respect to the beam direction, and the transverse and longitudinal impact parameters with respect to the beam spot, defined as the center of the $\Pp\Pp$ interaction region. 

A track seed is built using a pattern of segments across the muon stations.
L2 seeds are geometrically matched to an L1 muon trigger candidate within $\Delta R < 0.3$ and only the closest seed used. 
For a given pattern, consisting of at least one segment, the \pt of the seed is
estimated using a parameterization linear in $\Delta\phi$, which is either the angle of the segment with respect to a straight line from the interaction point to the position of the segment, defined as the center of the CMS coordinate system, or the difference in bending angle
between two segments in different stations.

Standalone tracks are built using the Kalman filter
technique~\cite{Fruhwirth:1987fm}, a recursive algorithm that performs pattern
recognition detector layer by detector layer and, at the same time,
updates the trajectory parameters, where for the muon system this corresponds to the four stations of muon chambers.

The track-building algorithm starts from the seed state (position,
direction, \pt).
The parameters of the seed are extrapolated to the innermost
compatible muon chamber and are used to identify a measurement
(track segment in the DTs and CSCs, hit in the RPCs) consistent with
the muon trajectory.
The information from the selected measurement (position, direction) is
used to update the state parameters.
If multiple compatible segments or hits are found in the same muon
chamber, the measurement that minimizes the $\chi^2$ of the track 
is selected. The updated track is then extrapolated to the next compatible chamber and the same procedure is repeated until the outermost chamber of the detector is reached.
The track building is performed twice:
first by proceeding from the innermost layer of the muon detector
toward the outermost, and second in the opposite direction.
This allows the algorithm to remove possible biases from the initial
seed.

If the track segments for
one muon are not merged into a single seed, multiple L2 seeds and subsequently duplicate tracks can be built from a single muon.
Since they typically share a fraction of hits or segments, all the track candidates that
share at least one hit are compared with each other and
only the L2 candidate of best quality is kept,
on the basis of its $\chi^2$ and hit multiplicity.

As a last step, 
the beam spot position is used to constrain the track parameters to improve the momentum resolution of tracks.
However, this constraint is not applied in dedicated triggers targeting
cosmic muons or muons not originating from the $\Pp\Pp$ interaction region, because it would reduce efficiency for these types of muons. These dedicated triggers rely purely on the unconstrained L2 information and do not use the L3 track reconstruction steps described in the upcoming sections.

\begin{figure} \centering
  \includegraphics[width=0.45\textwidth]{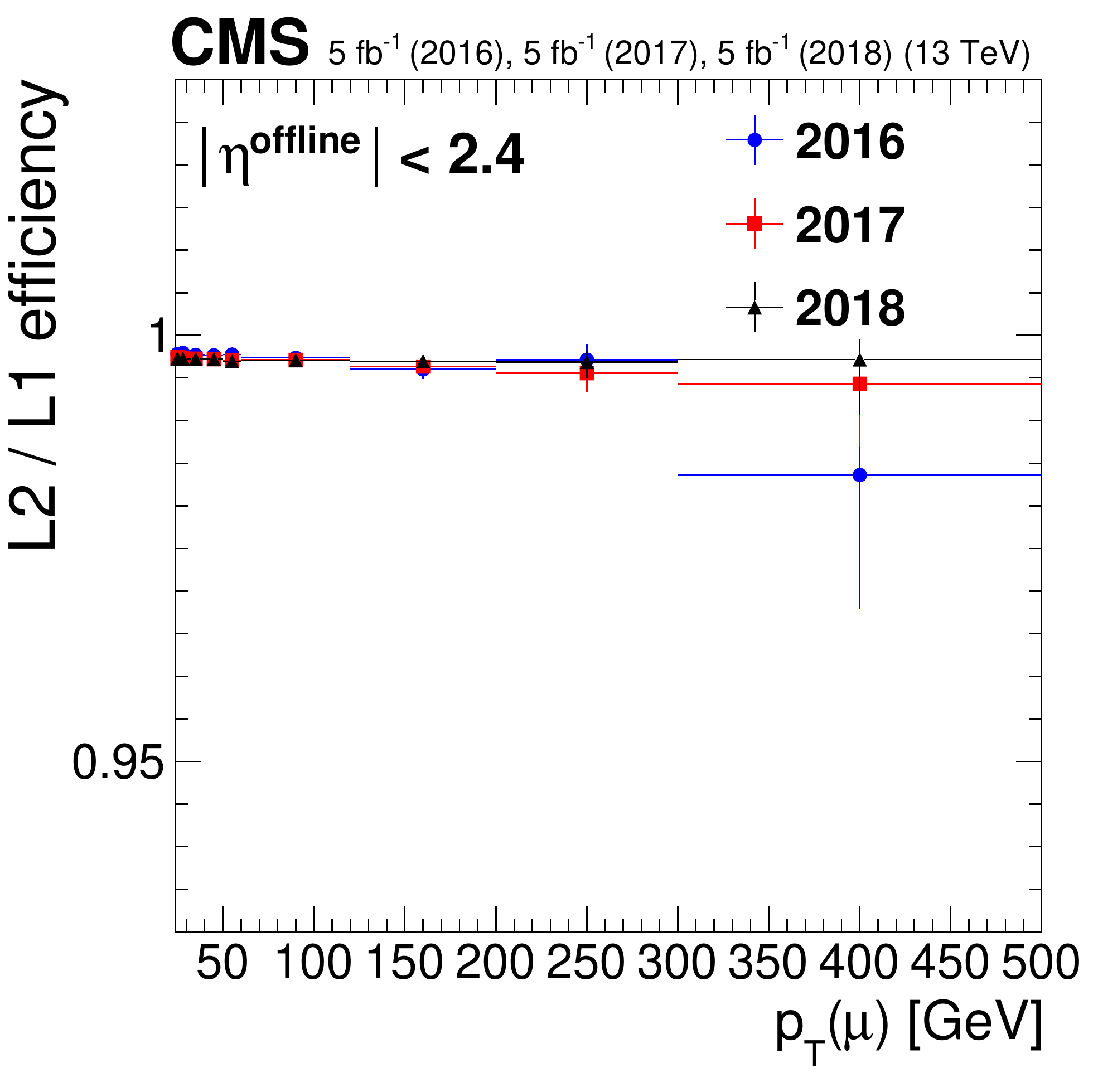}
  \includegraphics[width=0.45\textwidth]{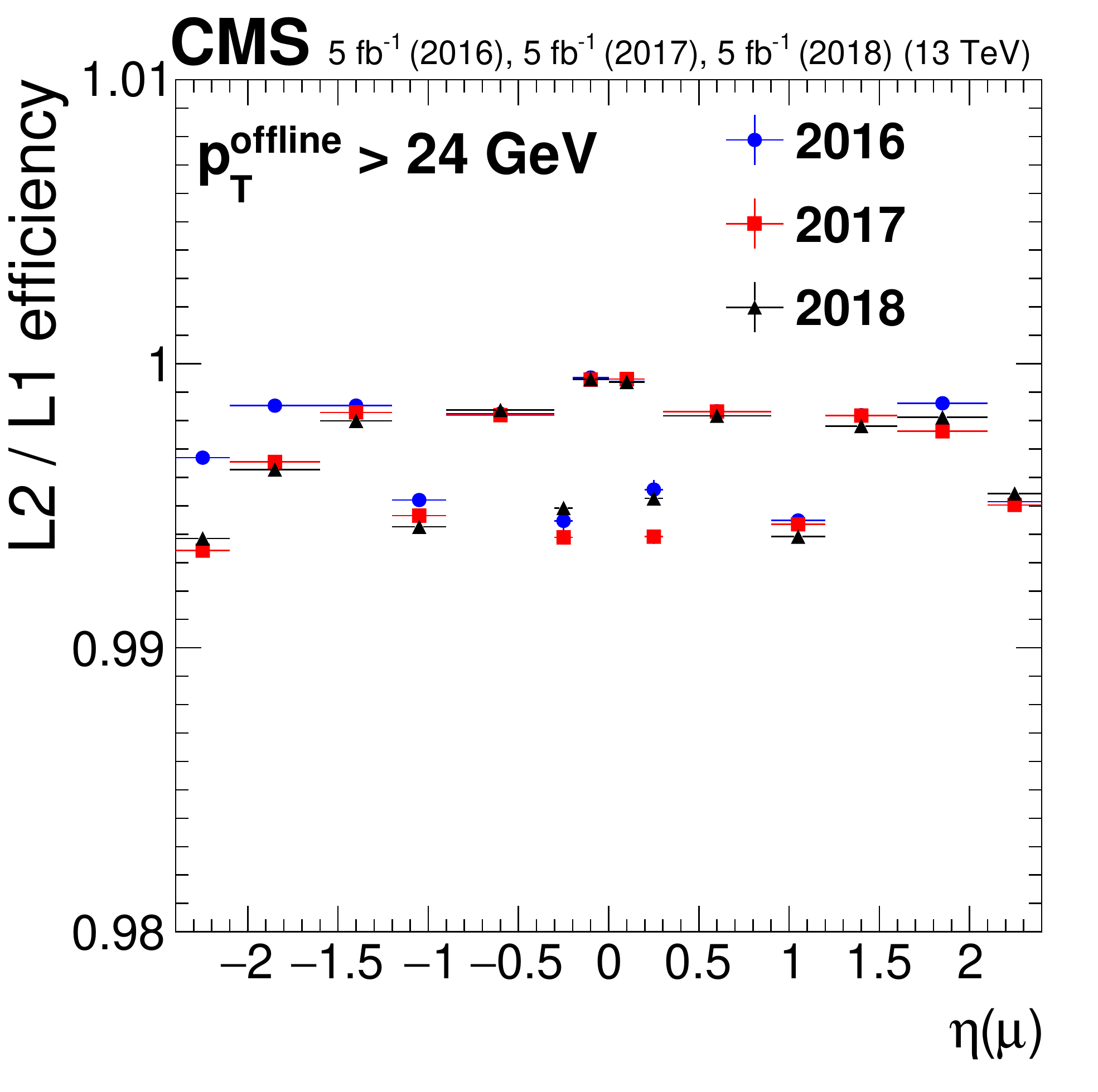}
  \includegraphics[width=0.45\textwidth]{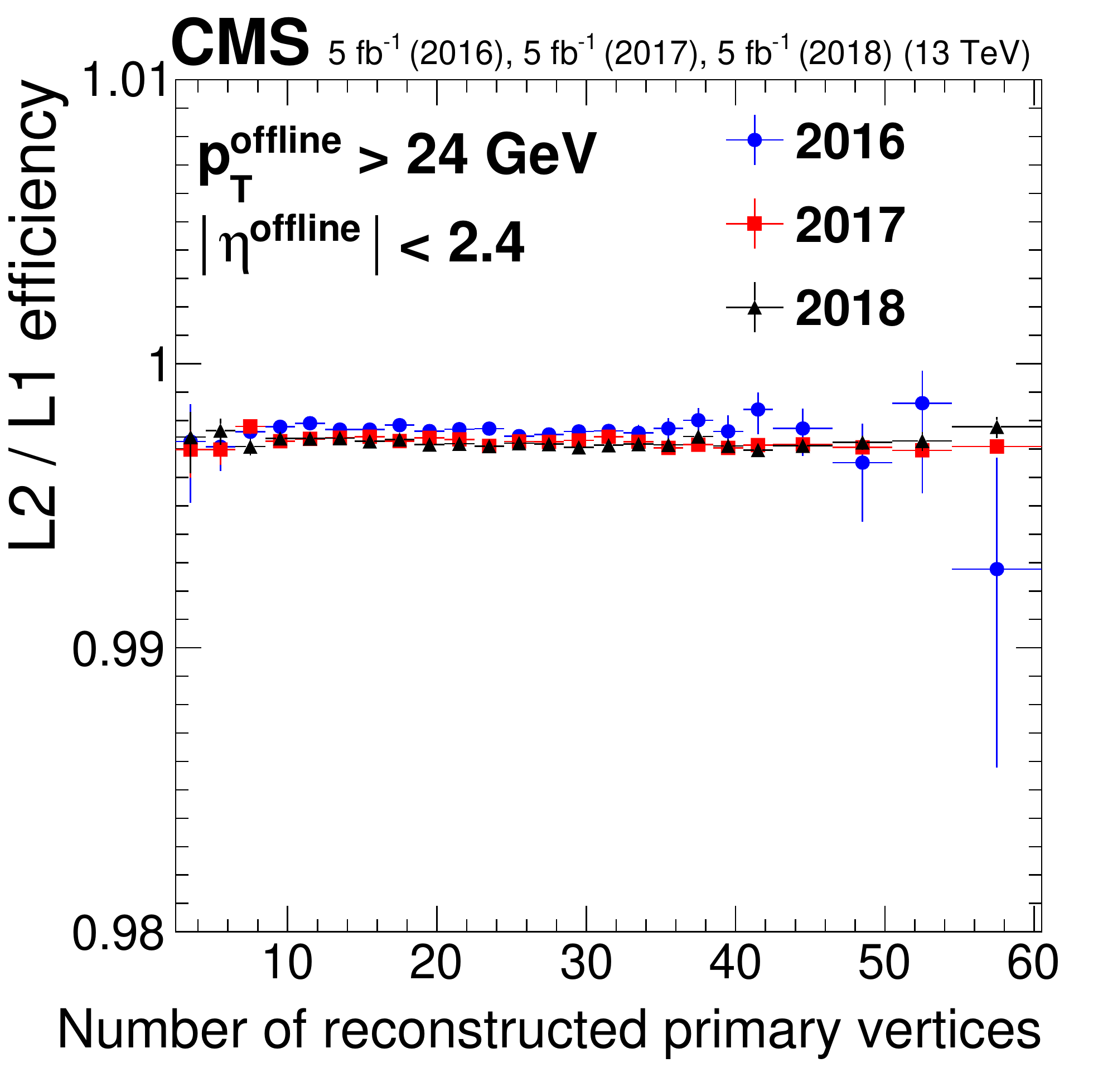}
  \caption{Efficiency of the L2 reconstruction with respect to L1 muons ($\pt>22\GeV$ and ``tight'' quality) as a
    function of \pt (upper left) and $\eta$ (upper right) of the probe muons. The dependence with the number of primary vertices is also shown (lower middle). The efficiency for 2016 data is shown in blue circles, for 2017 data in red squares, and for 2018 data in black triangles. Only statistical uncertainties are shown.}
  \label{fig:perf:L2Eff}
\end{figure}

The L2 reconstruction efficiency with respect to L1 muons is shown in Fig.~\ref{fig:perf:L2Eff} as
a function of \pt and $\eta$ of the probe muon, as well as the number of primary vertices for data collected in 2016, 2017 and 2018. The overall efficiency exceeds 99.5\% and is independent of \pt within uncertainties, as is expected since the standalone muon reconstruction that is part of the offline reconstruction, and therefore the probe object, is very close to the L2 reconstruction algorithm. Because the L1 and L2 reconstruction algorithms are sensitive to the same detector effects, requiring the probe muon to be matched to an L1 muon masks the impact of these effects in this measurement. However, small residual effects remain because of the differences in the reconstruction of trigger primitives in the L1 versus DT and CSC segments in L2.
The efficiency drops for some values of $\eta$ ($0.2<\abs{\eta}<0.3$ and $0.9<\abs{\eta}<1.2$) are caused by
the cracks in between the wheels of the CMS muon system. In these cases, muon segments may be missing in some stations, or are reconstructed with lower quality at the edge of the muon chambers. Since the L1 and L2 muon reconstruction algorithms are affected by this in different ways, the L2 efficiency is slightly reduced with respect to L1. Furthermore a slight loss of efficiency is observed in 2017 and 2018 for
negative values of $\eta$ because of two disabled muon CSC chambers in the first station, each covering
$10^{\circ}$ in azimuth.

Since the observed relative differences between L1 and L2 are typically $<$1\%, the stability of the L2 reconstruction performance during the LHC Run 2 is clear. Furthermore, the efficiency is independent of the number of reconstructed vertices in the event, because particles from the additional $\Pp\Pp$ interactions rarely reach the muon system.

\subsection{Level-3 muon reconstruction}
\label{sec:L3}

Level-3 muons are reconstructed using all available information
regarding the trajectory of the muon from both the muon spectrometer and the inner tracking detector. The muon is
reconstructed by either matching a track in the inner tracker with an L2 muon and  
performing a combined track fit using the information from both the muon system and tracking detectors, or by
identifying an inner-detector track as a muon candidate by matching it to an L1 muon without performing a combined fit. These two
types of reconstructed muons are conceptually equivalent to the ``global'' and ``tracker'' muon algorithms used in the
offline muon reconstruction~\cite{Sirunyan:2018fpa}.

The first step in this procedure is the reconstruction of tracks in the inner tracker. It uses the same techniques as
the general CMS track reconstruction~\cite{Chatrchyan:2014fea}. The tracking starts by building seeds that are used for the pattern recognition step.
The trajectory is sequentially propagated to the next detector layer and compatible hits are added using the Kalman
filter technique~\cite{Fruhwirth:1987fm}. This propagation happens either from the center of detector towards the muon
system (``inside-out'') or starting with the outer tracker layers and moving inwards towards the
interaction point (``outside-in''). 
The final track parameters are obtained from a refit that is 
performed once all hits associated with the trajectory are known.
Poor-quality tracks are rejected at the end of reconstruction.

During Run 2 data taking, different L3 reconstruction algorithms were used: during 2016, two distinct
approaches were used to build an L3 muon, the ``cascade'' and the ``tracker muon'' algorithms,
described in Sections~\ref{sec:cascade} and~\ref{sec:tkMu}, respectively. During 2017 and 2018, both algorithms
were replaced by a common approach called ``iterative'' algorithm, described in Section~\ref{sec:iterL3}.

\subsubsection{L3 muon seeded by L2: the cascade algorithm}
\label{sec:cascade}
The cascade algorithm reconstructs muon trajectories in the inner tracker based on L2 muons. 
Three different algorithms are used to create track seeds based on the L2 information: ``outside-in state'',
``outside-in hit'', and ``inside-out hit'', described in more detail below. These three algorithms are used in sequence, such
that later (more precise and computationally demanding) reconstruction algorithms are only run if the previous algorithms were not successful in reconstructing a track
for a given L2 muon, leading to the name ``cascade''. This exclusive approach reduces the required processing time. Muon tracks reconstructed during the early steps with a transverse impact parameter larger than 0.2\,cm are rejected to allow later steps to properly reconstruct them again and ensure that poorly reconstructed tracks are not used in the final selection. 

\begin{itemize}

\item The \textit{outside-in state} algorithm propagates the track state of the L2 muon onto the surface of the outermost tracker layer. A ``stepping helix'' propagator is used, which splits the propagation into steps of finite helix length and includes the magnetic field, as well as energy losses and multiple scattering in material. This propagated state, with its uncertainties enlarged by $\eta$- and \pt-dependent scale factors ranging from 3 to 10, is used to identify detector modules in a given tracker layer compatible with the state within $\chi^2 < 40$, independent of the presence of hits. The $\chi^2$ is calculated based on the position of the state on the tracker layer and the closest edge of the module, taking into account the uncertainties of the state. A seed is created from a state updated with the position of a compatible module and is used to start the pattern recognition. If no compatible module is found in a layer, the state is propagated further inwards until a compatible module is found or all possible layers are exhausted. The pattern recognition uses the Kalman filter, propagating layer-by-layer towards the interaction point and adding compatible hits to the trajectory, followed by the final track fit. 

\item The \textit{outside-in hit} algorithm propagates the track state of the L2 muon to the outermost tracker layer. However, instead of using the propagated state directly as the track seed, it first searches for compatible hits in that layer, using the same criteria to identify compatible detector modules to check for hits as in the previous case. Up to five of the most compatible hits are then used to update the track state and create one seed per hit. Again, the algorithm moves inwards layer-by-layer until a layer with compatible hits is found or all layers are exhausted. If the seed positions projected onto the beam axis fulfill $\abs{z} < 25$\,cm with respect to the center of the detector, they are used to initiate the same steps of pattern recognition and track fitting as the outside-in state algorithm.

\item The \textit{inside-out hit} algorithm uses the L2 muon information to define a region of interest (ROI) on the innermost pixel layer. The size of the region is given by the minimum of either a parameterization as a function of $\eta$ and \pt of the L2 muon or the uncertainty in the L2 parameters propagated to the pixel detector. However, a minimal size of $0.1{\times}0.1$ in $\eta$ and $\phi$ is always used. Within these regions, compatible triplets (doublets) of pixel hits on three (two) pixel layers are formed whose transverse (longitudinal) impact parameters are required to be smaller than 0.2 (15.9)\,cm with respect to the beam spot. These hit multiplets are used as the seeds to initiate the pattern recognition from the pixel detector outwards.
\end{itemize}

The collection of tracker tracks created from the three cascade algorithms are then matched to the L2 muons. 
For matching, the tracker tracks and L2 muons are propagated onto a common surface and compared using $\eta$- and \pt-dependent criteria based on geometrical distance and $\chi^2$ compatibility. A combined fit is performed with the Kalman filter using information from
both the selected tracker track and the matched L2 muon.
The fit with the best $\chi^2$ per degree of freedom is used as the final
L3 muon. Approximately 90\% of accepted muons are reconstructed using the outside-in
state approach, whereas about 5\% each come from the outside-in hit and inside-out
hit approaches.

\subsubsection{L3 muon seeded by L1: the tracker muon algorithm} \label{sec:tkMu}
The tracker muon reconstruction algorithm solely relies on the
pixel and strip tracking detectors to reconstruct the muon tracks. In this algorithm, information
from the muon detectors is only used in two places, first to define the ROI for the track reconstruction in the tracking detector and later to tag the reconstructed tracks as muons by matching them to segments in the muon stations. This approach does not rely on the successful reconstruction of an L2 muon and therefore leads to improvement in the muon trigger efficiency in cases where the L2 muon was poorly reconstructed. When combined with the L2-seeded approaches, this makes the overall muon trigger reconstruction more robust.

The starting point of the inside-out track reconstruction is an L1 muon. An ROI is generated based on the L1 parameters,
extrapolated back to the interaction point, and the track reconstruction algorithms are confined to this region, to
reduce the computing time demands. To further reduce this timing, only L1 muons with a $\pt > 15\GeV$ are used to select the region. The number of ROIs in the tracker muon algorithm is limited to two per event, and ROIs are created from L1 muons in descending order of \pt. Tracks are built following an iterative tracking approach~\cite{Chatrchyan:2014fea},
where tracks are built first using triplets of pixel hits as seeds, which is then relaxed to using doublets of hits in a second iteration. The collections of tracks produced in both iterations passing a loose track quality selection based on the track quality and impact parameters are merged into a single collection. The resulting tracks are matched to segments in the muon system to tag them as ``tracker muons'', following the same
procedure that is used for the offline reconstruction~\cite{Sirunyan:2018fpa}.

\subsubsection{Performance of the cascade and tracker muon algorithms}
\begin{figure}\centering
\includegraphics[width=0.45\textwidth]{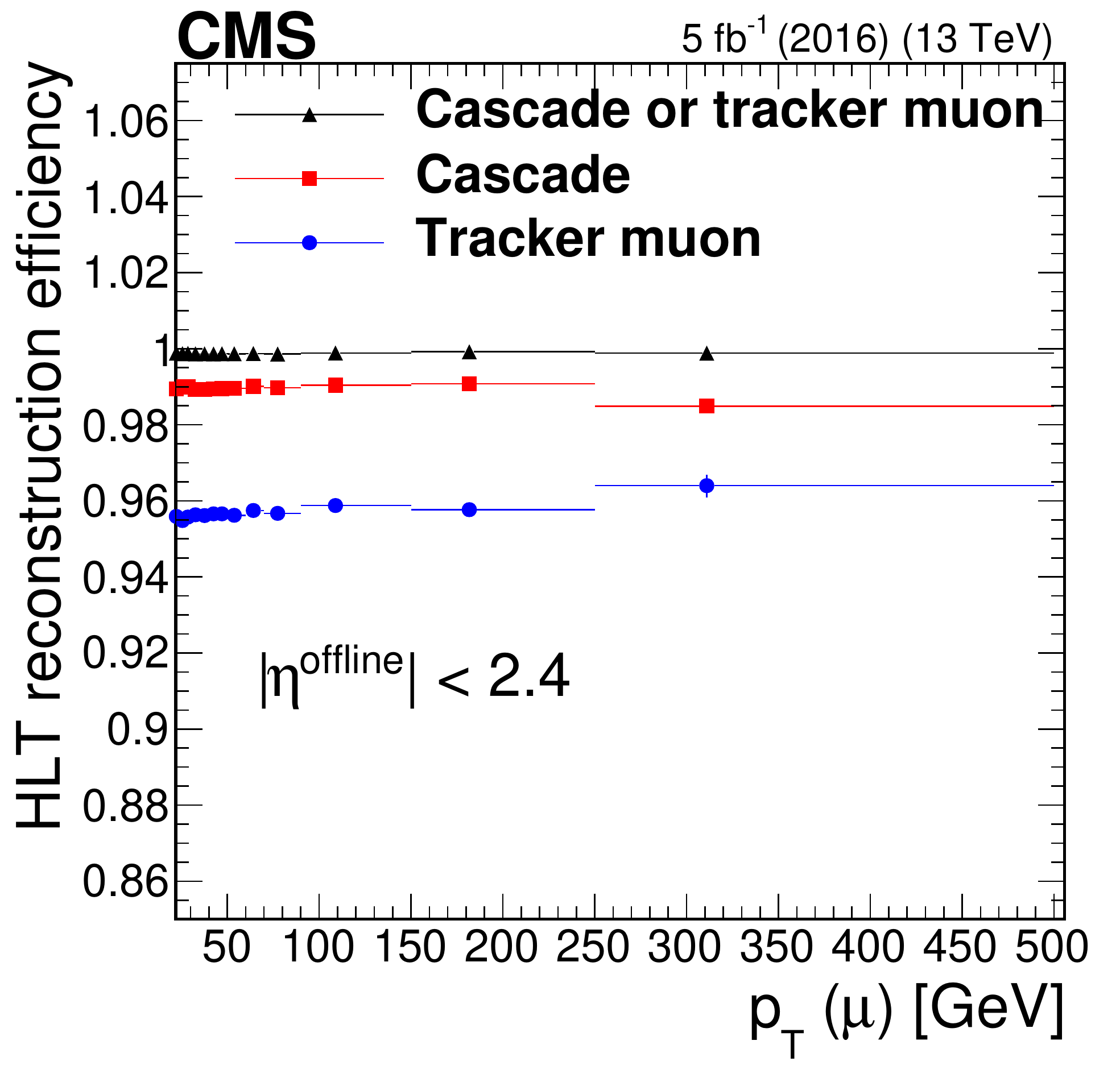}
\includegraphics[width=0.45\textwidth]{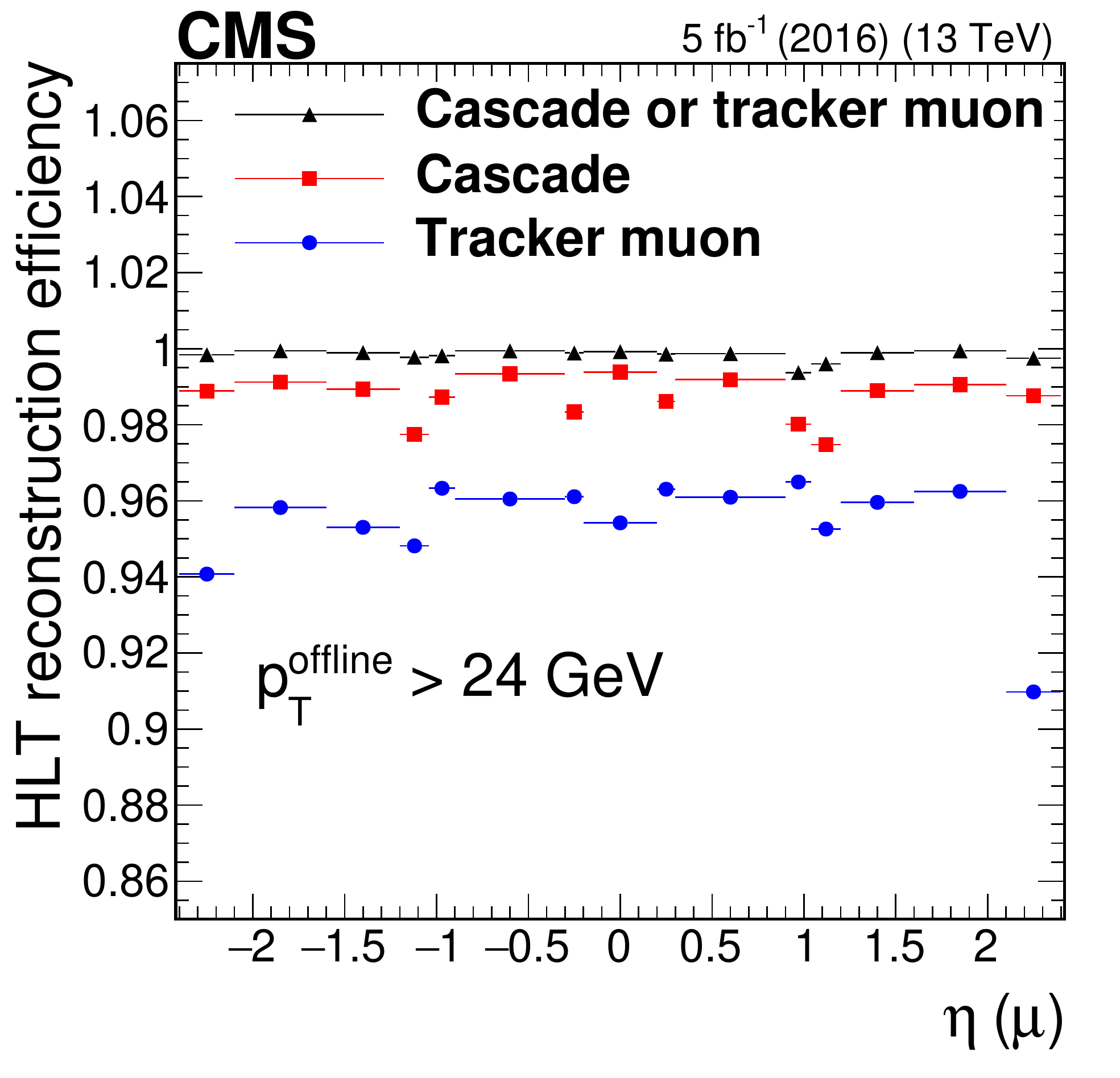}
\includegraphics[width=0.45\textwidth]{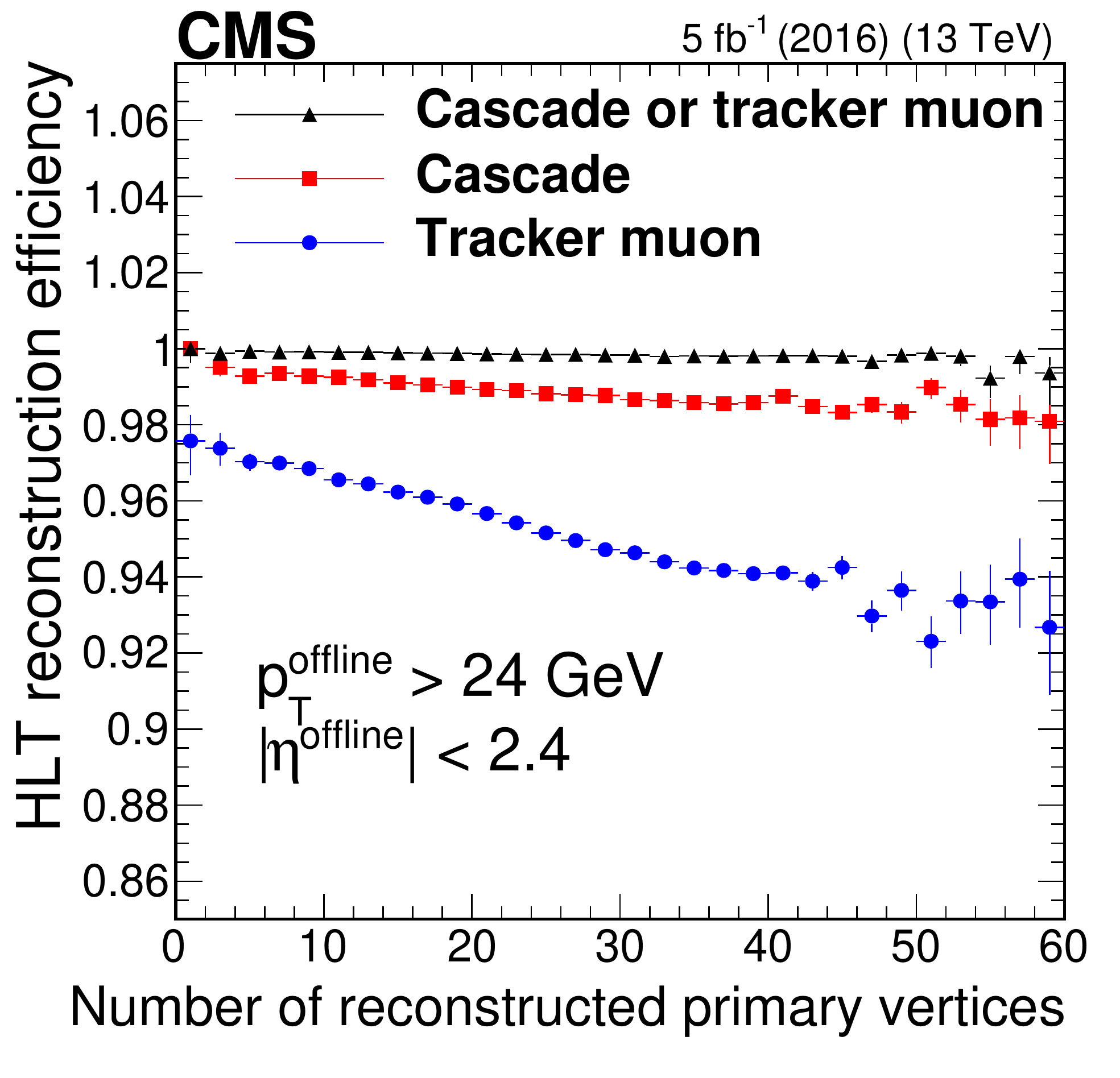}
\caption{Efficiency of the cascade and tracker muon L3 reconstruction algorithms, as well as the logical OR between the two, measured in data events as a function of \pt (upper left), muon $\eta$ (upper right), and the number of reconstructed primary vertices (lower middle). Only statistical uncertainties are shown.}
\label{fig:perf:L3TkMuCascade}
\end{figure}

The L3 muon reconstruction efficiency with respect to L1 muons for both the cascade and tracker muon algorithms is shown in Fig.~\ref{fig:perf:L3TkMuCascade}. Since each algorithm was used for independent triggers, the combined efficiency of the logical OR of both algorithms is also shown. The average efficiencies are about 96\% for the tracker muon algorithm and 99\% for the cascade algorithm. This difference is caused by two effects. First, quality requirements on the muon tracks are applied in the tracker muon algorithm, which are not present in the cascade algorithm. Secondly, as the track reconstruction in the cascade algorithm is seeded predominantly outside-in, it is able to reconstruct tracks with very few hits in the pixel detector, which are missed by the tracker muon algorithm. Combined, the two algorithms reach an efficiency of about 99.9\%, independent of $\eta$ and $\phi$ of the muons. Such high efficiencies are confirmed also in simulation for the combination of both algorithms. Although some loss of efficiency with increasing pileup is seen for each algorithm individually, the combination of the algorithms is not degraded at high pileup. The tracker muon algorithm is more strongly affected by higher pileup than the cascade algorithm since, whereas the occupancy of the pixel detector increases with pileup, it remains relatively low in the outer part of the tracker where most seeds for the cascade algorithms are built. In addition, the pixel detector in use in 2016 exhibited a loss of hit efficiency as the occupancy of the detector increased, resulting in a loss of efficiency for the creation of track seeds for the tracker muon algorithm. The upgraded pixel detector installed in 2017 does not exhibit this effect.

\subsubsection{L3 muon seeded by L2 and L1: the iterative algorithm}
\label{sec:iterL3}
The third muon reconstruction algorithm that was used during Run 2 is the so-called ``iterative'' algorithm.
With the installation of the new pixel detector before the beginning of the 2017 data taking, a redesign of the reconstruction
techniques was required to take advantage of the additional pixel layer in the upgraded detector~\cite{Dominguez:1481838}. The algorithm replaced the previously described cascade and tracker muon
approaches, combining the advantages of both into a single algorithm to make a unified trigger decision. This eliminates the need to consider two distinct triggers at analysis level. The iterative algorithm consists of three steps: one outside-in step seeded by L2 muons, one inside-out step seeded by
L2 muons, and a second inside-out step seeded by L1 muons. For the inside-out step seeded by an L2 muon, only muons that were not reconstructed as an L3 muon in
the outside-in step are used. For the L1-muon seeded step, in 2017 it was run only for L1 muons that could not be matched to an already reconstructed L3 muon. In 2018, all the L1 muons were used to maximize the efficiency
of the final algorithm. 

The outside-in reconstruction step uses L2 muons to start a search for track reconstruction seeds in the inner tracker.
Using the parameters from the L2 muon, the trajectory is propagated to the outermost tracker layers using the stepping helix propagator to find the seed,
similar to what is done with the outside-in steps of the cascade algorithm. To increase the efficiency of this step
with respect to the cascade algorithm, the iterative algorithm creates two trajectory states on the outer tracker surface for
seeding (one updated with the interaction point as an additional constraint and one not updated). In addition, up to five seeds are then generated by
finding a silicon strip detector hit compatible with the extrapolated state. After the pattern recognition step,
only tracks satisfying certain quality requirements~\cite{Chatrchyan:2014fea} are kept. 

Both inside-out steps make use of the iterative tracking techniques, similar to those used for the tracker muon
algorithm. The ROIs are generated using either L2 or L1 muons, depending on the step. Since the uncertainties in the parameters of an L2 muon are much smaller than those for the L1 muons, the size of the ROI is much smaller for the L2-seeded step. The first iteration is seeded using pixel
quadruplets built using the cellular automaton algorithm~\cite{Pantaleo:2293435}, targeting the reconstruction of
prompt tracks above 1.2\GeV. The second iteration is seeded by pixel triplets to recover efficiency from the previous
iteration. Finally, one last iteration, seeded by pixel doublets, was added for the data taking in 2018 to improve the resilience of the algorithm in situations where fewer than three hits might be available for seed building. In 2018, about 40\% of muon tracks created by this algorithm on average are reconstructed from quadruplet seeds, 50\% from triplet seeds, and 10\% from doublet seeds. 

Each reconstruction step produces a collection of tracker tracks that are then combined, removing any duplicates. For the L2-seeded steps, the tracker tracks are then matched to the L2 muon, and a combined fit of the
tracker and the muon track is performed to build an L3 muon. For the L1-seeded step, the tracker tracks are matched to hits in the muon system to create tracker muons. Duplicate tracks
already contained in the L3 muon collections from the previous steps are removed, and all muons are merged into one
single collection. This matching and refitting procedure is identical to that used in offline reconstruction to
produce the muon collection~\cite{Sirunyan:2018fpa}.

\begin{figure} \centering
\includegraphics[width=0.45\textwidth]{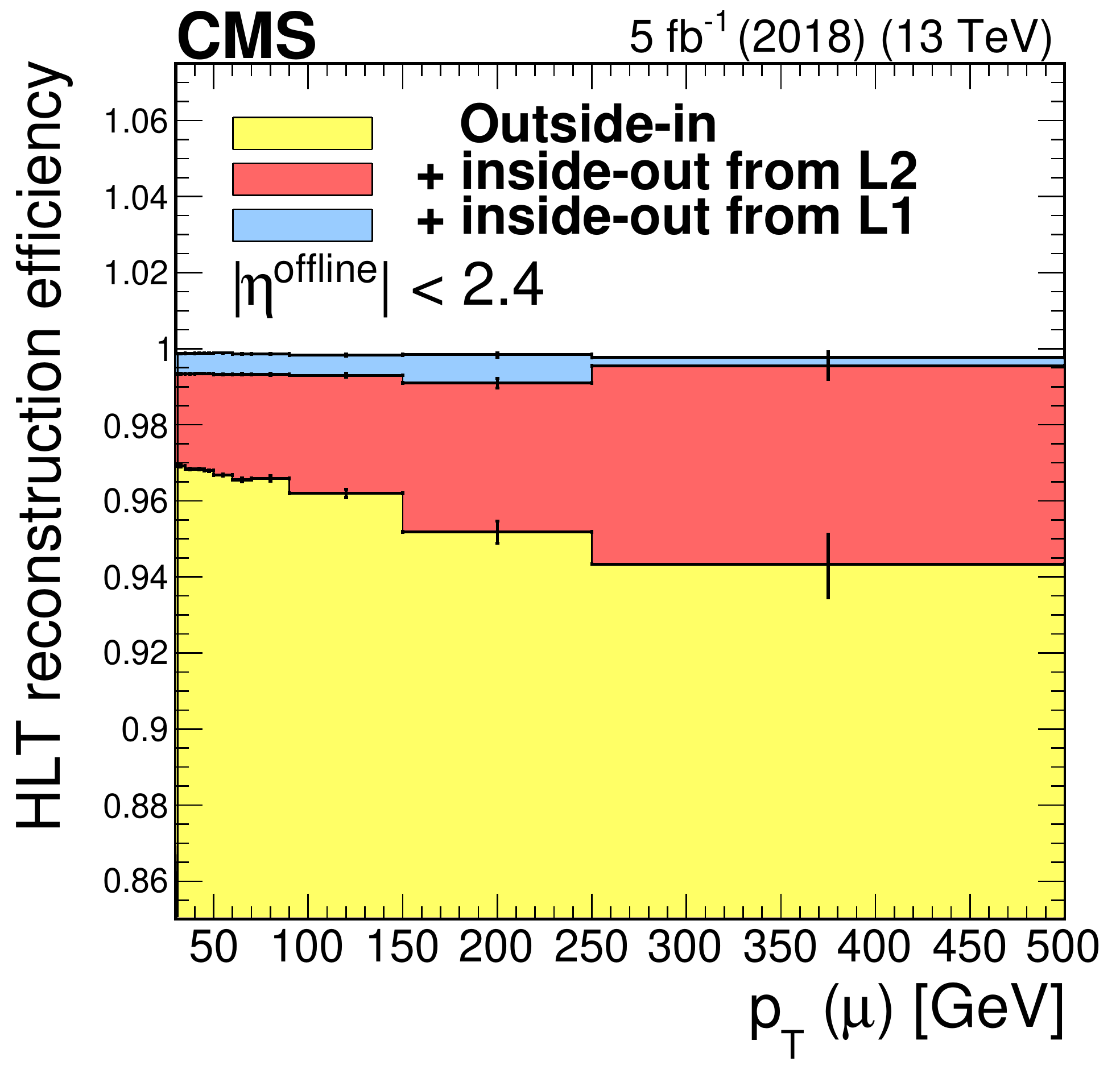}
\includegraphics[width=0.45\textwidth]{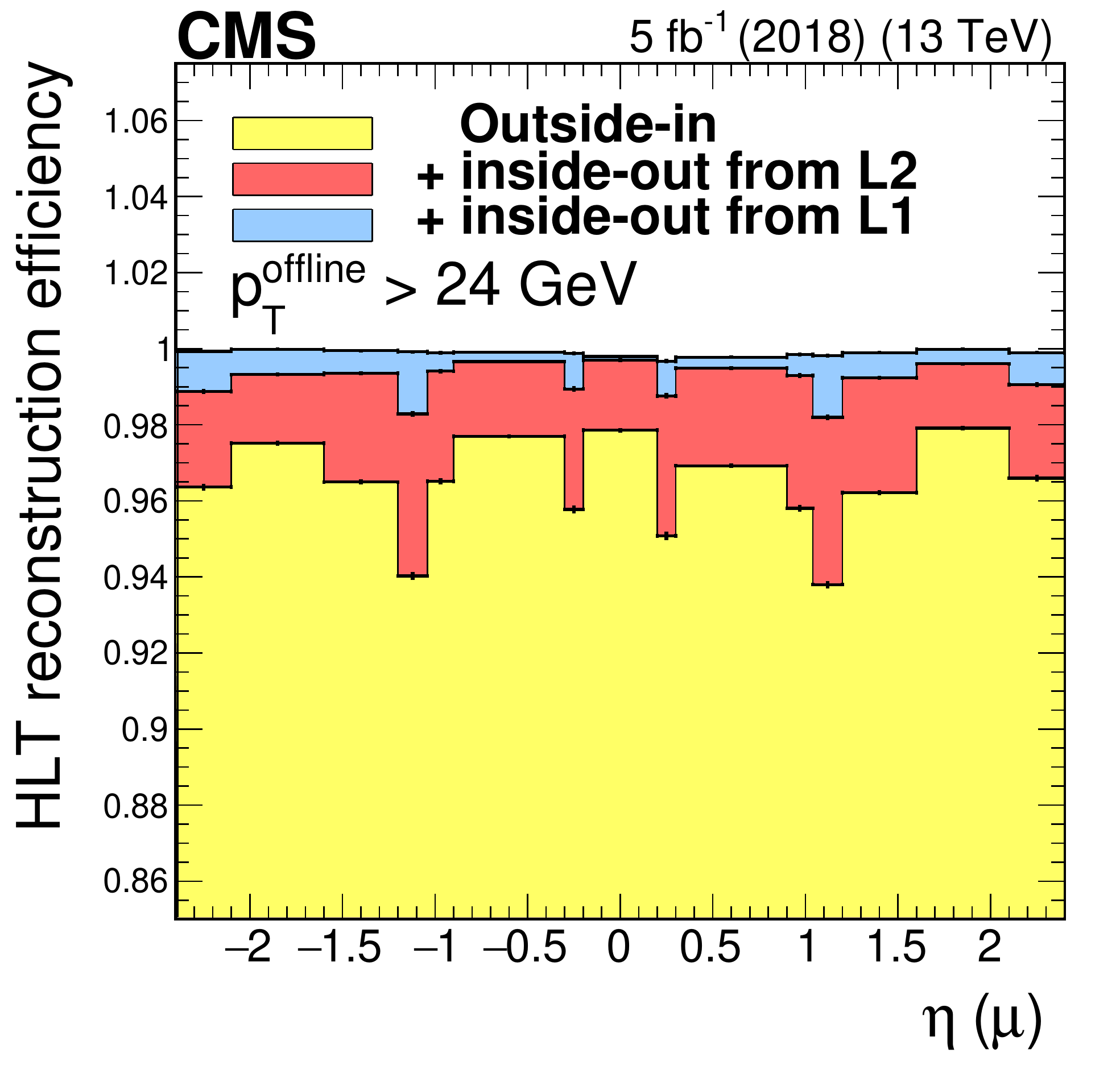}
\includegraphics[width=0.45\textwidth]{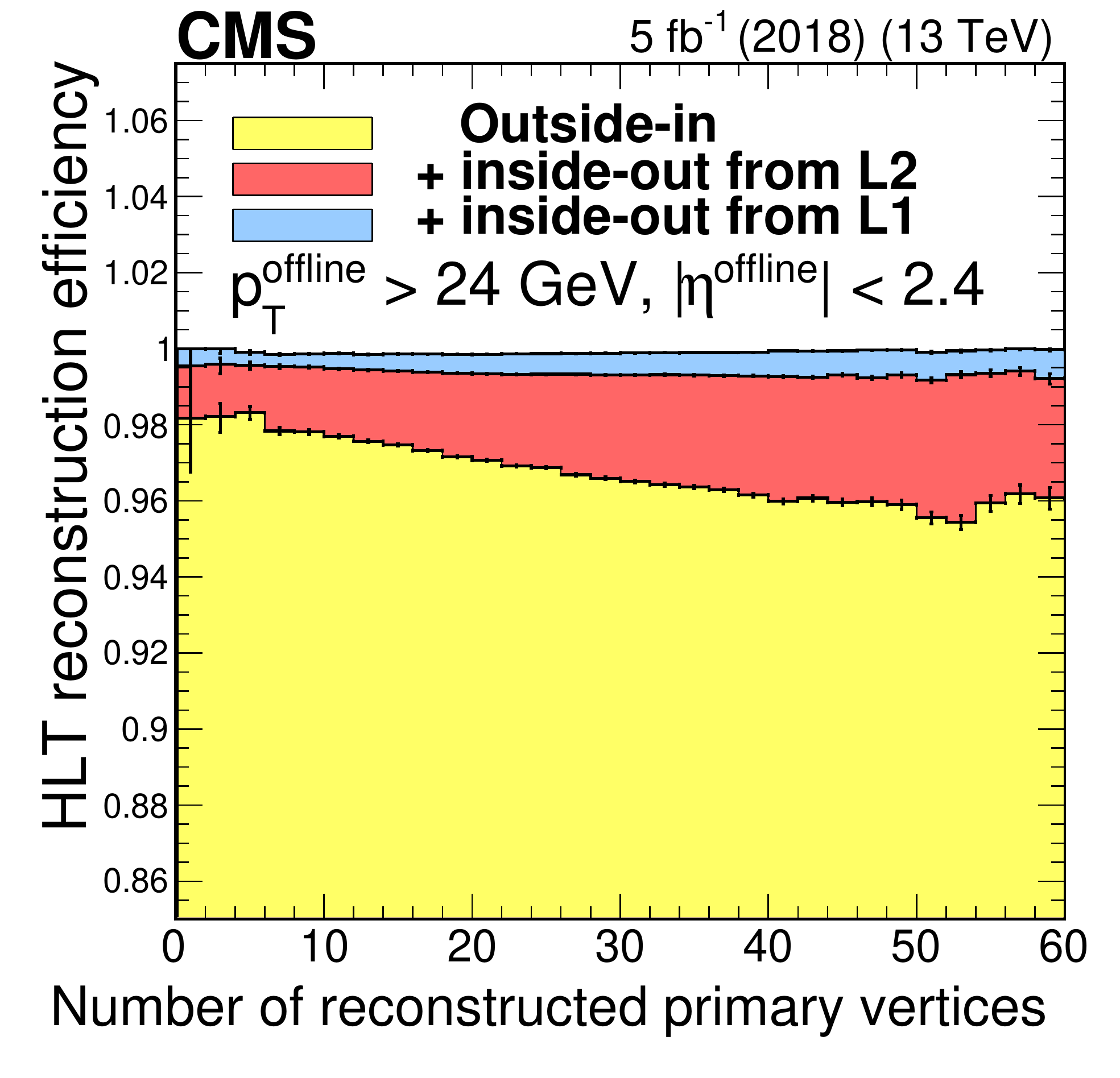}
\caption{Efficiency of the iterative algorithm with respect to L1 muons as a function of \pt (upper left), muon $\eta$ (upper right), and the number of reconstructed primary vertices (lower middle). The efficiency contributions from the different iterations are shown as stacked histograms. Only statistical uncertainties are shown.}
\label{fig:perf:IterL3}
\end{figure}

Figure~\ref{fig:perf:IterL3} shows the reconstruction efficiency with respect to L1 muons of the different steps
of the iterative algorithm, as a function of muon \pt and $\eta$ and of the number of
reconstructed primary vertices. This measurement is carried out on 2018 data using the best performing version of the algorithm deployed during data taking. The stacked histograms show the increase of tracking efficiency when the different
steps of the algorithm are successively added. The overall efficiency is almost 100\%, showing how the inclusion of the inside-out from L1 step
prevents inefficiencies from relying on the L2 reconstruction and the L2-seeded tracking. Around 97\% of the muons are reconstructed by the outside-in
step, and the rest of the efficiency is recovered by the two inside-out steps. The outside-in reconstruction
efficiency is most efficient at low \pt and with few reconstructed vertices. Since the scale factors used to enlarge the uncertainty of the L2 muons during the seeding stage were not optimized for high-\pt muons, the efficiency drops to about 94\% above 250\GeV. The inside-out steps recover the efficiency loss, ensuring excellent
performance at both high pileup and high \pt. 

Finally, to decrease the trigger rate and increase the purity of the L3 muon candidates, mild identification criteria were introduced
to the iterative L3 algorithm at the start of 2018. At least one muon segment
should match the extrapolated tracker track, just as for offline tracker muons~\cite{Sirunyan:2018fpa}. In addition, the tracker tracks of the muons are required to have 
more than five strip detector layers with measurements, and at least one hit in the pixel detector. Furthermore,
for muons with $\pt > 8\GeV$ the track is required to be matched to at least two muon stations, except
when fewer than two matches are expected from the detector geometry as occurs, e.g., in the cracks between the wheels of the muon system. 

\subsubsection{L3 muon reconstruction and identification performance}

\begin{figure}[htb] \centering
\includegraphics[width=0.45\textwidth]{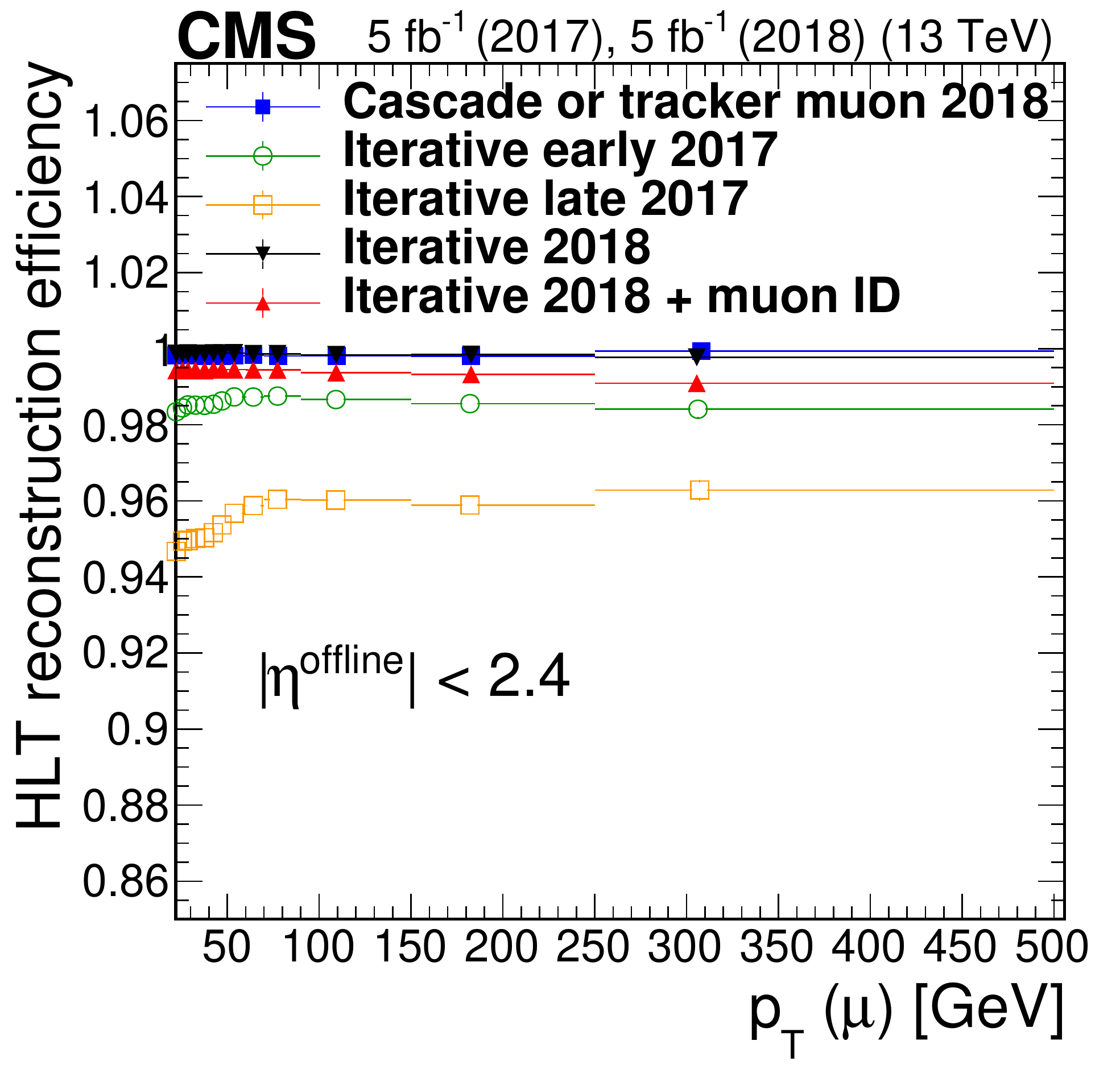}
\includegraphics[width=0.45\textwidth]{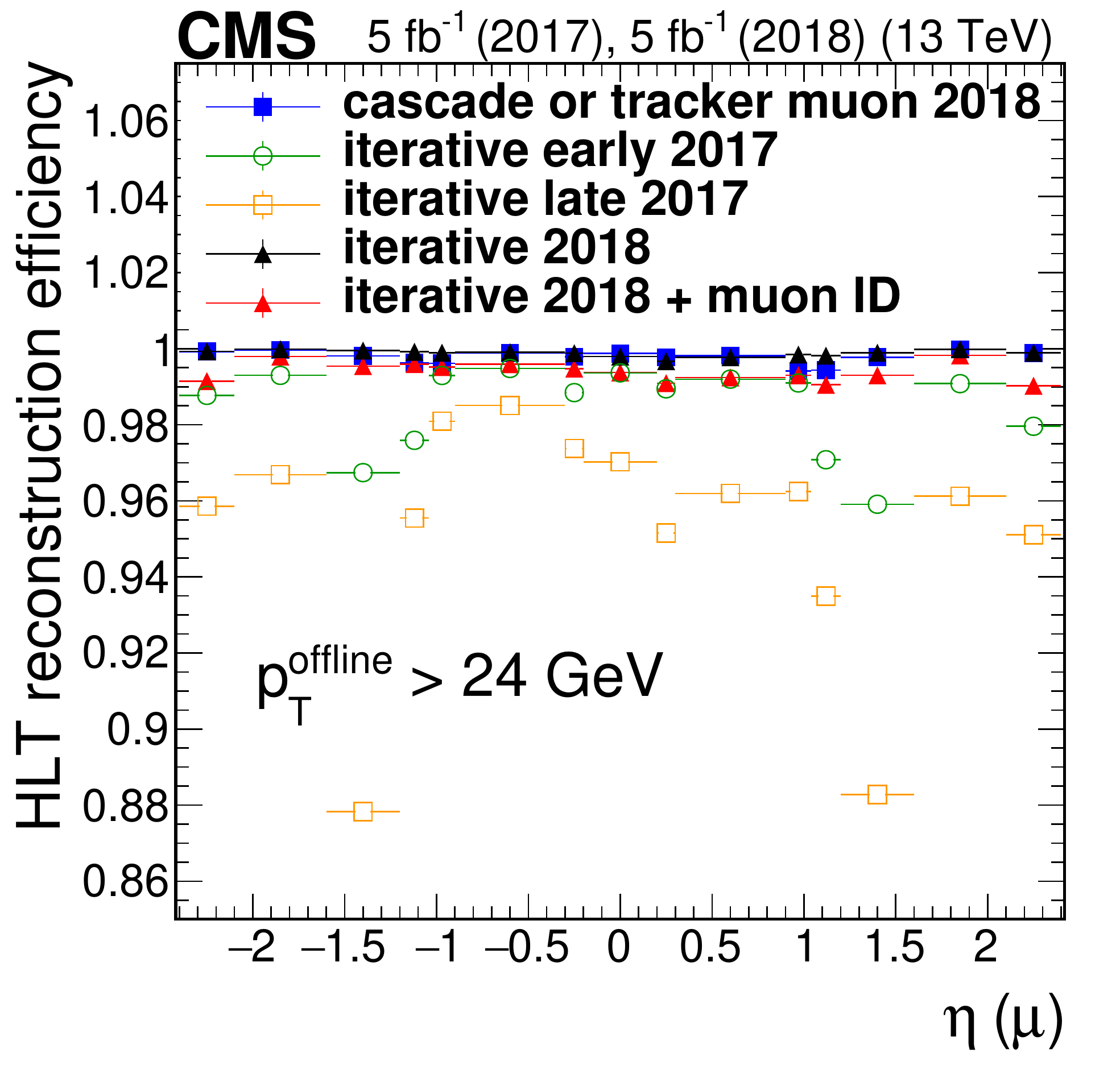}
\includegraphics[width=0.45\textwidth]{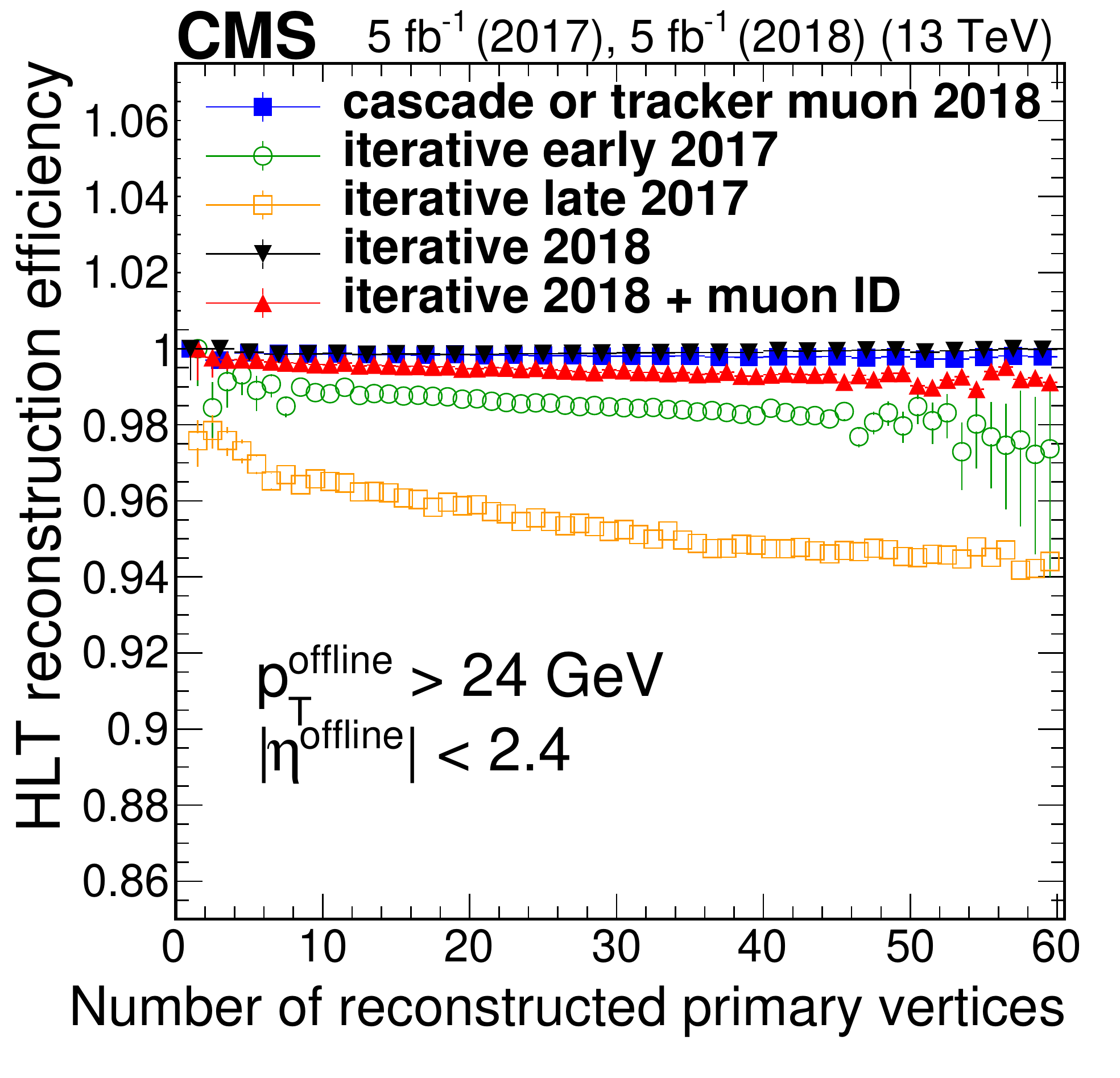}
\caption{Efficiency of the iterative algorithm with respect to L1 muons compared with the logical OR of the cascade and tracker muon reconstruction measured in data events as a function of \pt (upper left), muon $\eta$ (upper right), and the number of reconstructed vertices (lower middle). Only statistical uncertainties are shown.}
\label{fig:perf:L3Comp}
\end{figure}

To directly compare the performance of different algorithms,
Fig.~\ref{fig:perf:L3Comp} shows the efficiency of the iterative
and the logical OR of the cascade and tracker muon algorithms in data 
containing \PZ boson candidates, as a function of muon \pt and $\eta$, and the number of reconstructed primary vertices. 

For the iterative algorithm, we show the performance for 2018 data using the most efficient configuration of the algorithm, as well as two measurements for 2017 data using the version of the algorithm used during that year. The early 2017 data provide the best performance of the upgraded pixel detector achieved during that year, whereas the late 2017 data show the effect of inactive pixel detector modules on the reconstruction efficiency. The 2017 version of the iterative algorithm achieves an efficiency of up to 99\% on the early 2017 data. Efficiency losses are observed at low \pt, in the barrel-endcap overlap region of the detector and the very forward endcaps, and with a high number of reconstructed vertices. These trends are significantly more pronounced in the late 2017 data where the full efficiency drops to 96\% overall and as low as 88\% in the overlap region. The asymmetry between negative and positive $\eta$ values in the central part of the detector highlights the random distribution of the pixel module loss. The overlap region is especially affected since tracks in this region pass only three layers of the pixel detector so that any loss of modules has a particularly large effect. This motivated the significant changes to the iterative algorithm for the 2018 data taking.
 
For 2018, we show both the L3
reconstruction efficiency, as presented in Fig.~\ref{fig:perf:IterL3}, and
the full efficiency of the algorithm, including muon
identification described in the previous section. For a fair
comparison, the cascade and tracker muon algorithms have been run on 2018 data and the resulting efficiency has been measured. On these data, both algorithms achieve reconstruction efficiencies above 99.5\%, with
slightly higher efficiency in the case of the iterative
algorithm. However, after the application of the muon identification
selection, the efficiency of the iterative algorithm ends up slightly lower than the logical OR of the
cascade and tracker muon algorithms. This small loss of efficiency is expected
and acceptable considering the benefits to the purity and trigger rates in high pileup conditions that the muon identification selection provides, as discussed
in Section~\ref{sec:rate}. In all cases, the
efficiency is independent of $\eta$ and $\phi$ of the muon, and
the reconstruction is robust against pileup. A small pileup dependence
is observed for the muon identification, for which the efficiency
decreases by about 1\% as the number of reconstructed vertices increases up to 60.

Good \pt resolution for L3 muons with respect to the offline
reconstruction ensures a sharp turn-on of the trigger efficiency at
the nominal \pt threshold. This allows efficient rejection of low-\pt muons that might otherwise have their momentum overestimated and pass the trigger threshold. Moreover,
a sharp efficiency turn-on improves the offline analysis acceptance, since the turn-on
region is commonly rejected.

The \pt resolution has been measured and compared
for L2 muons and the different L3 reconstruction algorithms used during Run 2. It
is obtained from the residual with respect to the offline muon
reconstruction as
\begin{linenomath} 
\begin{equation}
    \text{residual} = \left( q^{\text{offline}}/p_{\mathrm{T}}^{\text{offline}} - q^{\text{trigger}}/p_{\mathrm{T}}^{\text{trigger}} \right) \frac{p_{\mathrm{T}}^{\text{offline}}}{q^{\text{offline}}}. 
\end{equation}
\end{linenomath}
where $q$ is the reconstructed electric charge of the muon. 
This quantity is measured in data using a T\&P-like selection by
selecting events with two offline muons passing the tight
identification and isolation criteria. The invariant mass of the
dimuon pairs must be in the range $81 < m_{\mu\mu} < 101$\GeV. If an
L3 muon track is matched to one of these offline muons within
$\Delta R < 0.1$, the residual is
computed. With these selection criteria, the charge misidentification probability for L2 muons compared with offline muons rises from 3\% to 10\% as \pt increases from 30\GeV to 200\GeV. For the different L3 reconstruction algorithms, the charge misidentification probability depends on the level of track quality requirements applied in the algorithm, ranging from $\mathcal{O}(10^{-3})$ for cascade to $\mathcal{O}(10^{-5})$ for the iterative algorithm after application of muon identification requirements.

\begin{figure} \centering
\includegraphics[width=0.45\textwidth]{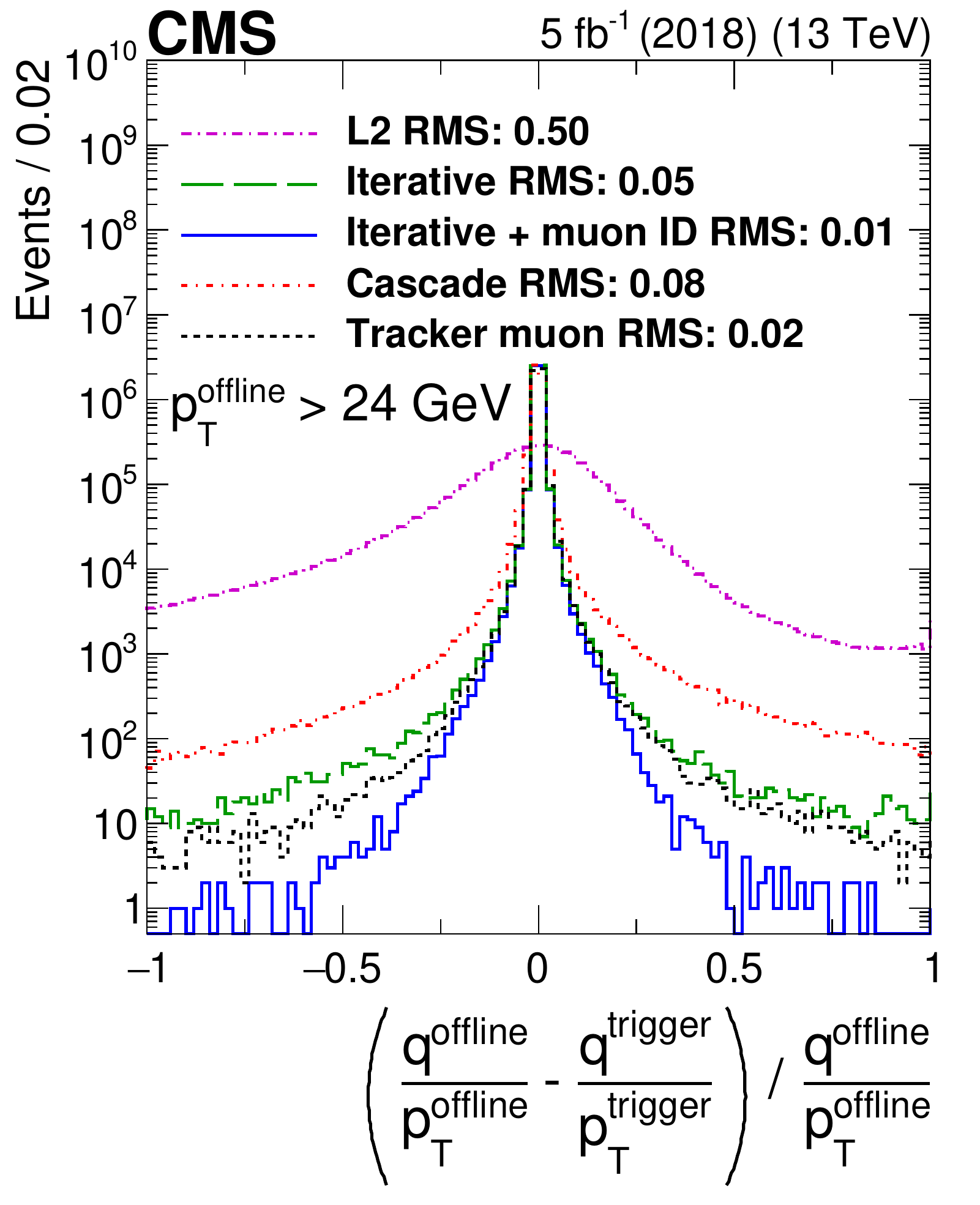}
\caption{The q/\pt residual distributions integrated over $\eta$ and \pt for all different HLT muon reconstruction algorithms used in 2018. Only statistical uncertainties are shown.}
\label{fig:perf:pTRes2018}
\end{figure}

The residual distributions, integrated over $\eta$ and \pt, are shown in Fig.~\ref{fig:perf:pTRes2018}. Since they use only the information from the muon system, the distribution for L2 muons is much wider than for the L3 reconstruction algorithms. The different L3 algorithms have comparable resolution in the core of the distributions, but differ significantly in the tails, as shown by the RMS values. The iterative L3 and the tracker muon
reconstruction have a better \pt resolution with fewer muons in the tails of the distribution than those for the cascade algorithm, since the former includes a track quality selection that is missing in the latter. The smallest tails are observed for the iterative L3
reconstruction after applying muon identification, because this removes a significant
number of poorly reconstructed muon tracks that survived the initial track selection at the reconstruction stage. From Fig.~\ref{fig:perf:pTRes2018}, no significant bias in the momentum scale is visible. For L2 muons, the scale bias is about 1\% toward higher \pt values compared with offline muons. For the cascade algorithm it is less than 0.1\% and even smaller biases are observed for the other L3 algorithms. 

\begin{figure} \centering
\includegraphics[width=0.45\textwidth]{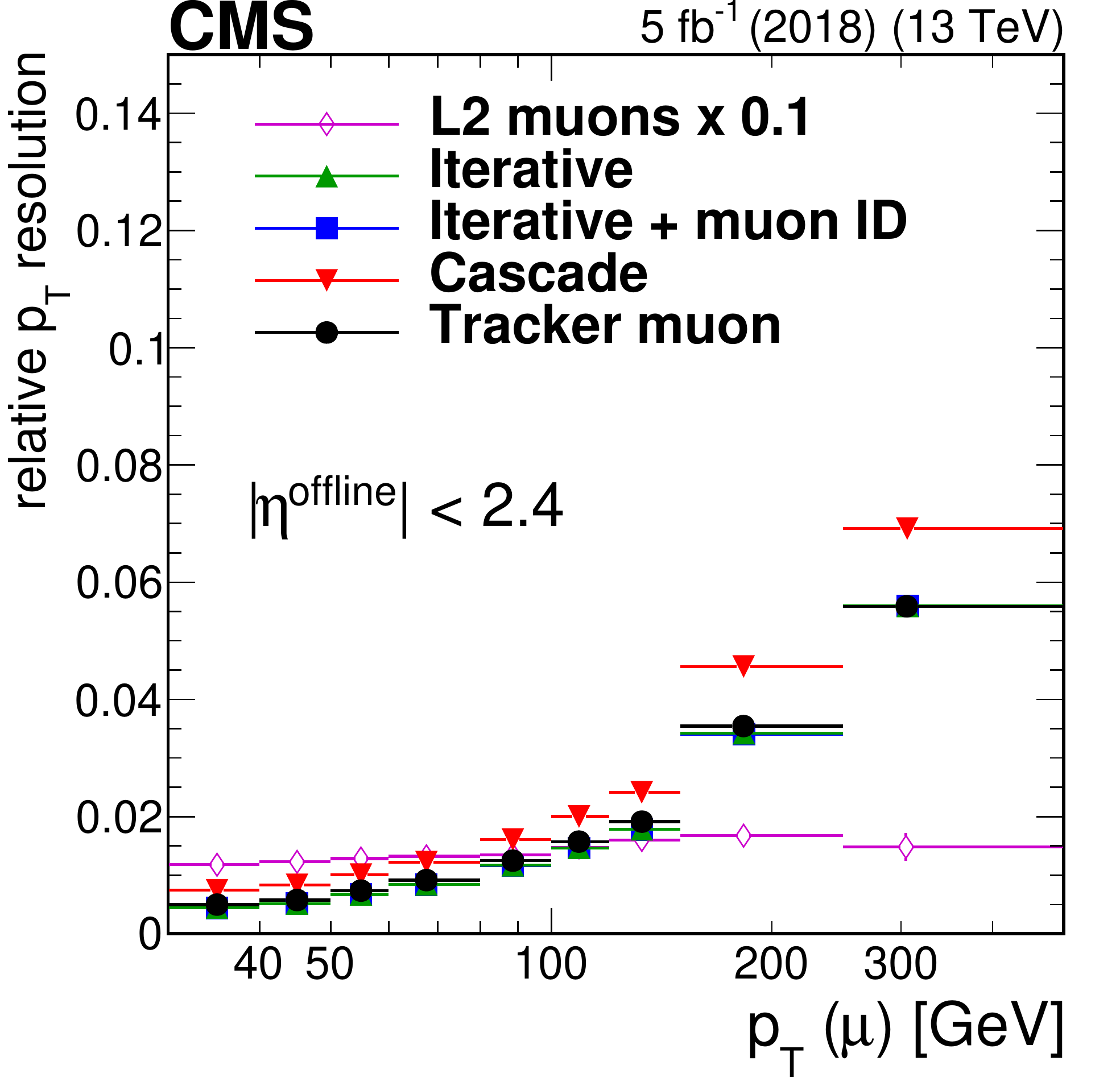}
\includegraphics[width=0.45\textwidth]{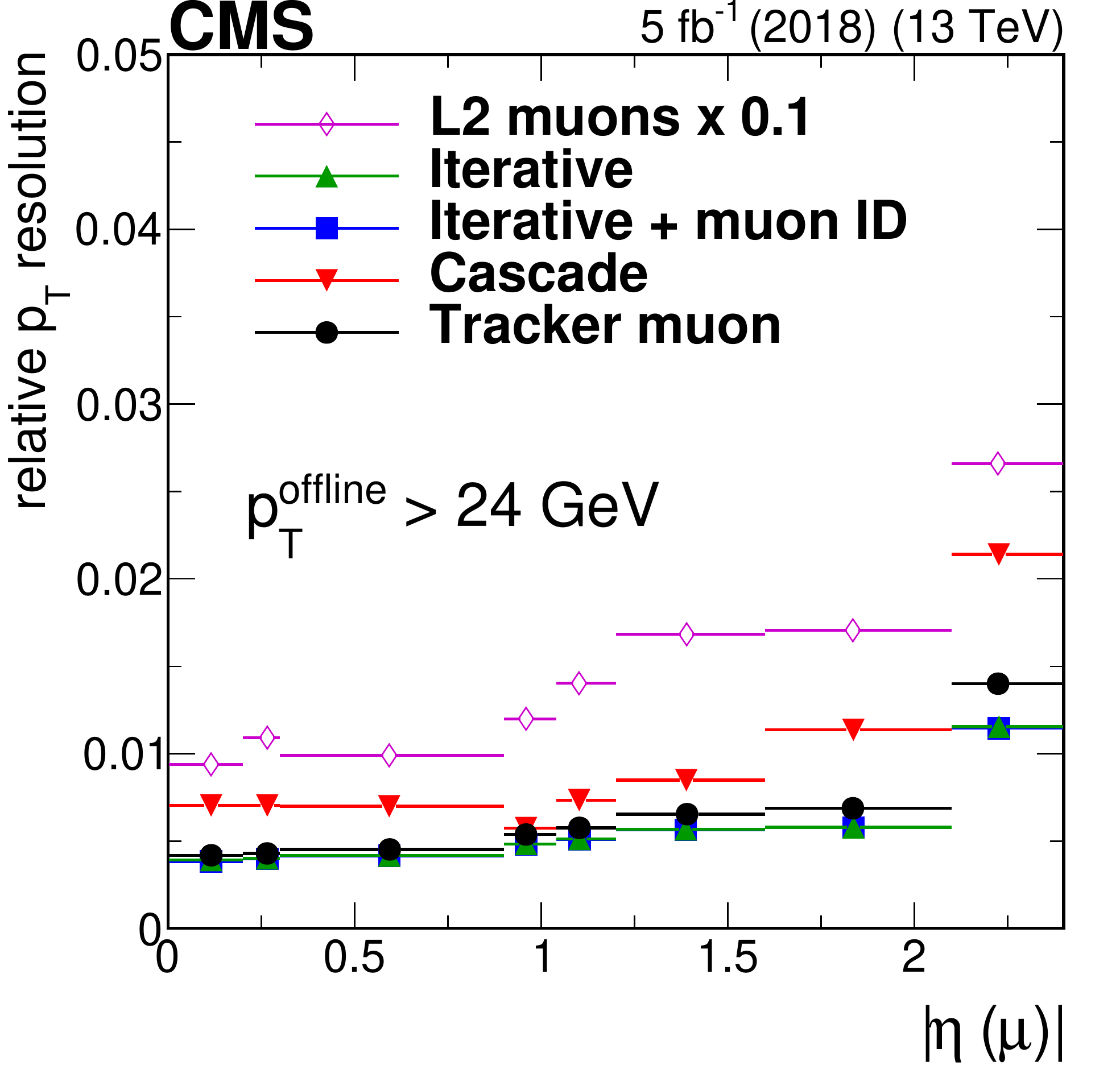}
\caption{Relative \pt resolution obtained from the standard deviations of the q/\pt residual distributions as a function of \pt (left) and $\eta$ (right) for all different HLT muon reconstruction algorithms used in 2018. Only statistical uncertainties are shown.}
\label{fig:perf:pTResVs2018}
\end{figure}

Figure~\ref{fig:perf:pTResVs2018} shows the resolution (standard deviation of the Gaussian core from a fit of a double-sided Crystal Ball function~\cite{Oreglia:1980cs}) of the residual distributions from
the L2 and L3 algorithms as a function of the
\pt and $\abs{\eta}$ of the offline muon. The resolution for L2 muons has been reduced by a factor of 10 for the purpose of these plots. Because the L3 reconstruction algorithms make use of the hits in the inner tracking detector, their \pt resolution is significantly better than that for the L2 reconstruction. Generally, the momentum resolution worsens
at high \pt, because the tracks are straighter with a smaller sagitta, and in the forward region, where the $\vec{v} \times \vec{B}$ effect causes the magnetic bending to be smaller when the track and magnetic field are nearly collinear. No significant difference is observed between the tracker muon and iterative L3 reconstruction algorithms, whereas the cascade algorithm has poorer \pt resolution because of its reliance on outside-in reconstruction, as discussed above. 

\subsubsection{Muon isolation}
Muons produced in the decays of vector or Higgs bosons, including the
\PW bosons in the decay chain of top quark pairs (\ttbar), as well as in many new physics
models, are mostly produced spatially isolated from other final-state
particles. Thus, the isolation of a muon from other particles in the event
is an effective procedure to suppress muons from decays in flight and meson decays, and to select a pure sample of
signal-like muons for the analysis of these kinds of final states,
allowing for reduced muon \pt thresholds for isolated muon triggers. 

The isolation of the muon is evaluated by considering the \pt of additional tracks reconstructed in the inner tracker and energy deposits in the
calorimeters, computed using a clustering algorithm based on the CMS particle-flow algorithm~\cite{CMS-PRF-14-001}, in a cone of
radius $\Delta R= 0.3$ around the muon itself. The muon track and the ionization energy it deposited in the calorimeters are excluded from the calculation. Furthermore, the contribution from tracks not originating at the same vertex as the muon is rejected. The estimated contribution from pileup to the energy deposits in the calorimeter is also subtracted. It is computed using the average energy density in the event, $\rho$, scaled to an effective area which relates the per-event $\rho$ to the fraction expected in the isolation cone, and subtracted from the particle-flow cluster sums.
The effective areas are computed separately for the barrel and endcap regions, and for electromagnetic and hadronic clusters.

The isolation is evaluated separately for tracks and ECAL and HCAL clusters, and sequential requirements
are applied on the resulting momentum or energy sums relative to the muon
\pt. Effective areas for the pileup subtraction and the relative isolation requirements have been adjusted for each year of data-taking, due to the different pileup conditions and minor changes in the clustering. The relative isolation requirements range from 0.08 (0.16) to 0.14 (0.22) for electromagnetic (hadronic) clusters, and from 0.07 to 0.09 for the track-based component. A looser version of the isolation based only on tracks, requiring the relative track isolation to be $<0.4$ has been used in some dimuon triggers. 

\begin{figure}\centering
\includegraphics[width=0.45\textwidth]{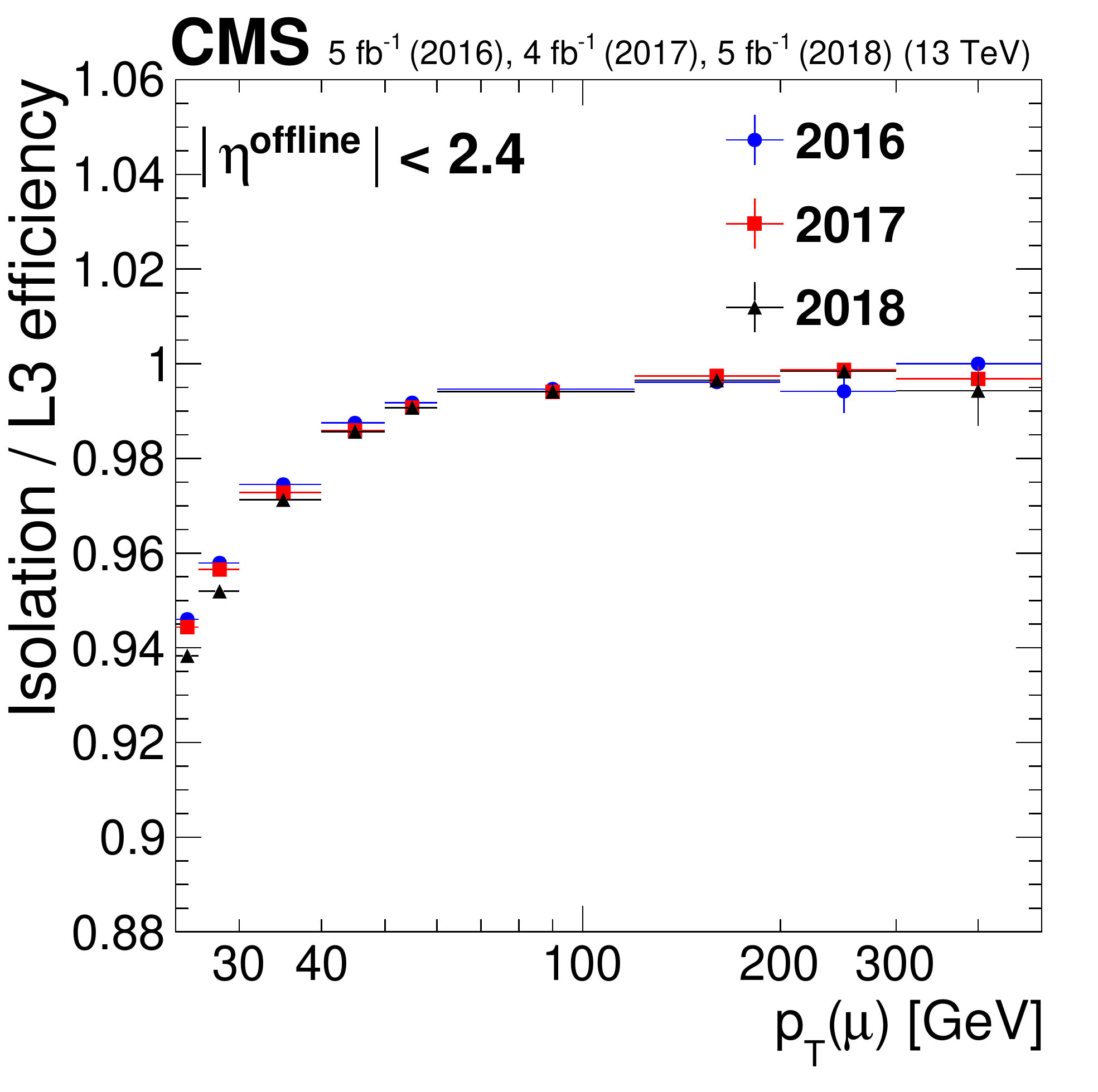}
\includegraphics[width=0.45\textwidth]{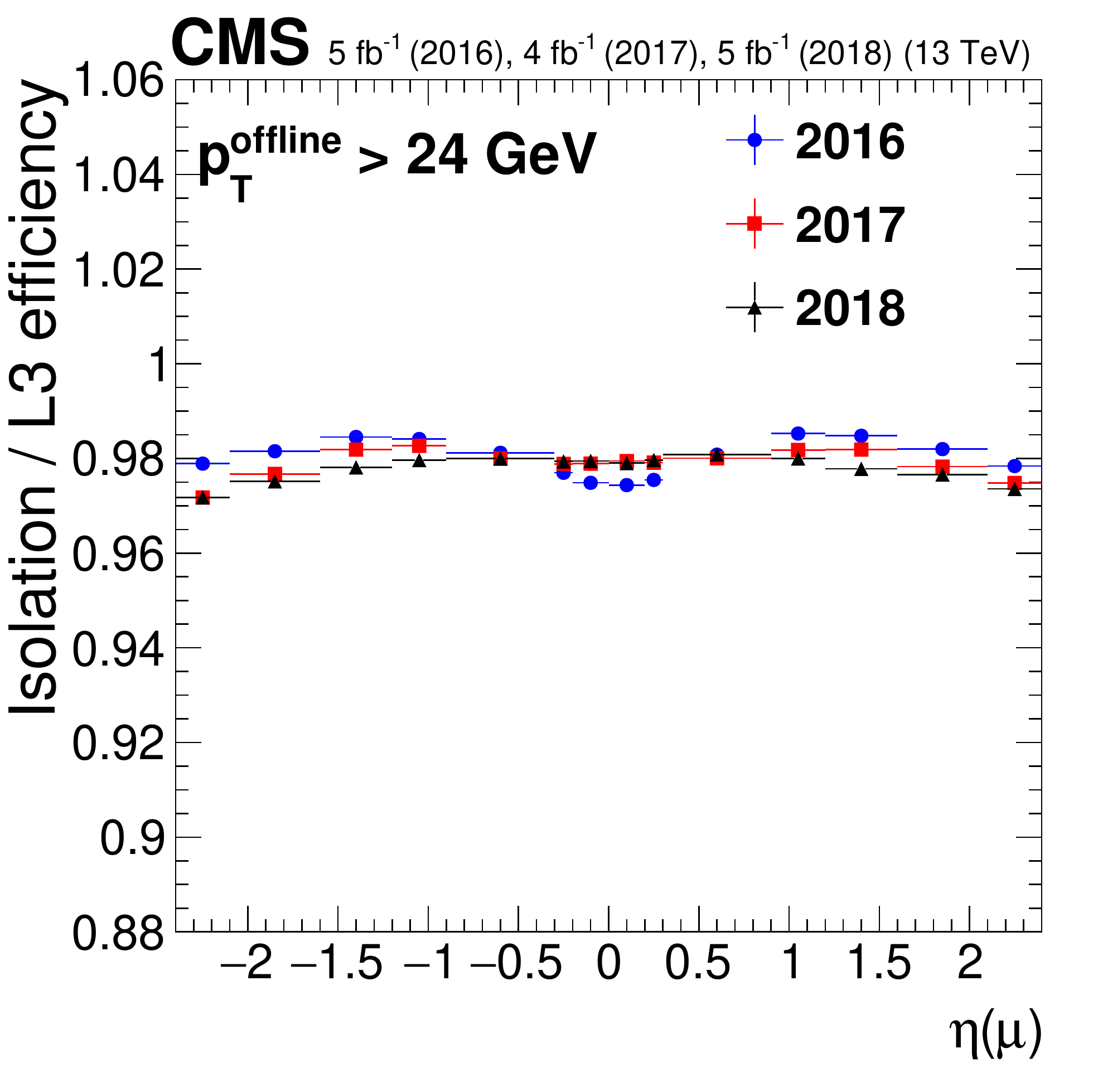}
\includegraphics[width=0.45\textwidth]{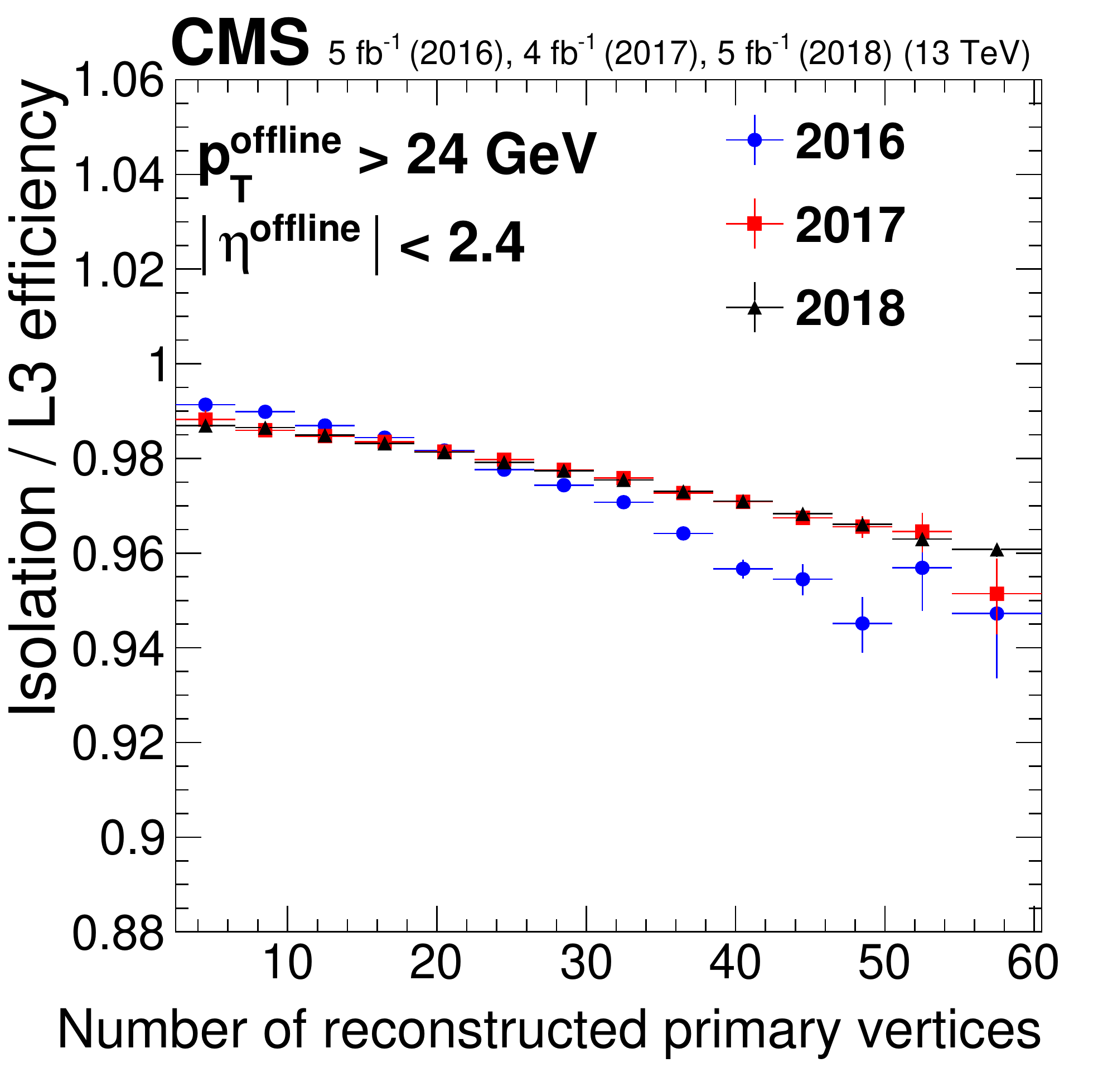}
\caption{Isolation efficiency as a function of muon \pt (upper left)), $\eta$ (upper right), and the number of reconstructed vertices (lower middle) during the Run 2 data taking. Only statistical uncertainties are shown.}
\label{fig:perf:Isolation}
\end{figure}

The efficiency of the isolation requirement with respect to L3 muons is measured in 2016, 2017, and 2018
data and shown in Fig.~\ref{fig:perf:Isolation}, as a function of muon
\pt, $\eta$, and the number of reconstructed vertices. The overall
efficiency is about 98\% on average for the three years. As a result of the adjustment of the isolation requirements for the different years, the 2016 data
show a slightly higher isolation efficiency in the endcaps, whereas the efficiency
is higher in 2017 and 2018 data in the central part of the detector. The
efficiency shows a slight drop at low \pt in both data sets and is
constant at high \pt. The isolation requirements were re-optimized in 2017 and 2018 to reach a more stable efficiency, as a function of pileup, as shown in Fig.~\ref{fig:perf:Isolation} (right).

\section{Performance of selected single- and double-muon reference triggers}

\label{sec:performance}

In this section, a selection of reference triggers
is described to illustrate the performance of the full trigger
sequence, including reconstruction, identification and, if
applicable, isolation. The triggers chosen are the two unprescaled
single-muon triggers, with and without an isolation requirement, and one
unprescaled double-muon trigger. The details of these triggers,
together with a shorthand name that will be used in the following are given here: 

\begin{itemize}
  \item \textit{Isolated single-muon trigger (IsoMu24)}: A single-muon trigger with a \pt threshold of 24\GeV that requires the muon to be isolated to achieve the low-\pt threshold. This is the main trigger for many analyses studying standard model processes, such as $\PW\to \PGm\Pgn$, $\PZ\to \PGm\PGm$, or semileptonic \ttbar decays.
  
  \item \textit{Nonisolated single-muon trigger (Mu50)}: A single-muon trigger with a \pt threshold of 50\GeV without an isolation requirement. The higher momentum threshold achieves acceptable trigger rates without applying an isolation requirement at the HLT, increasing the trigger efficiency slightly even for cases in which a high-\pt muon showered in the calorimeter. This trigger is therefore used in analyses targetingg events with particularly high-\pt leptons, such as searches for heavy new particles. Other uses include searches for Lorentz-boosted objects in which the muon may overlap with jets. To reach maximum efficiency for analyses using dedicated high-\pt muon identification criteria, this trigger is used in combination with two other triggers that use the cascade and the tracker muon algorithms with a \pt threshold of 100\GeV that were present during the entire data-taking period and kept as a backup solution when the iterative L3 reconstruction was introduced~\cite{MUO-17-001}. Therefore all performance measurements presented for Mu50 include the logical OR with these two backup triggers. 	 
    
  \item \textit{Double-muon trigger (Mu17Mu8)}: A double-muon trigger where
  each muon is required to pass the loose track-based isolation requirement. The \pt threshold
  applied to the muon with higher (lower) \pt is 17 (8)\GeV. The lower \pt
  requirement on the first muon compared with the single-muon triggers
  increases the acceptance for many dilepton processes, such as fully
  leptonic \ttbar decays, and many searches for new physics. To reduce the trigger rate due to pileup events, a requirement that the longitudinal distance between the two muons along the beam line must be less than 0.2 cm was added during the 2016 data taking, with negligible impact on the efficiency.
Similarly, during 2017 and 2018, an extra requirement on the invariant
mass of the dimuon system to be larger than 3.8\GeV was applied to reduce the rate from low-mass
dimuon resonances. 
\end{itemize}

In the following, the efficiencies for the two single-muon triggers and the low-\pt part of the double-muon trigger are
discussed in Section~\ref{sec:eff}. The HLT rate measurements and the composition of the selected muon sample are
described in Section~\ref{sec:rate}. Finally, the
processing time for the muon HLT reconstruction is discussed in
Section~\ref{sec:timing}. 

\subsection{Trigger efficiency}
\label{sec:eff}
\begin{figure}\centering
\includegraphics[width=0.45\textwidth]{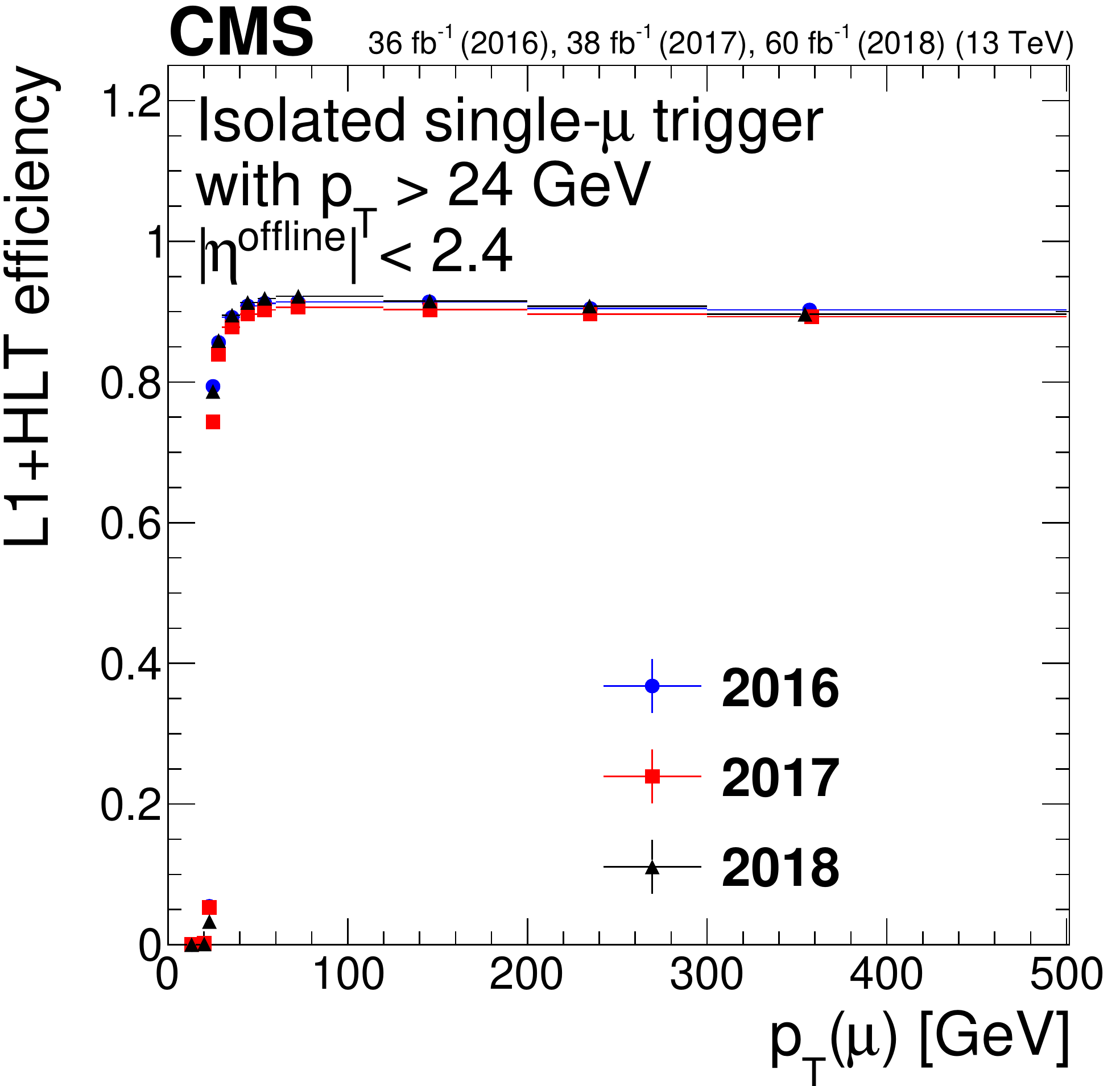}
\includegraphics[width=0.45\textwidth]{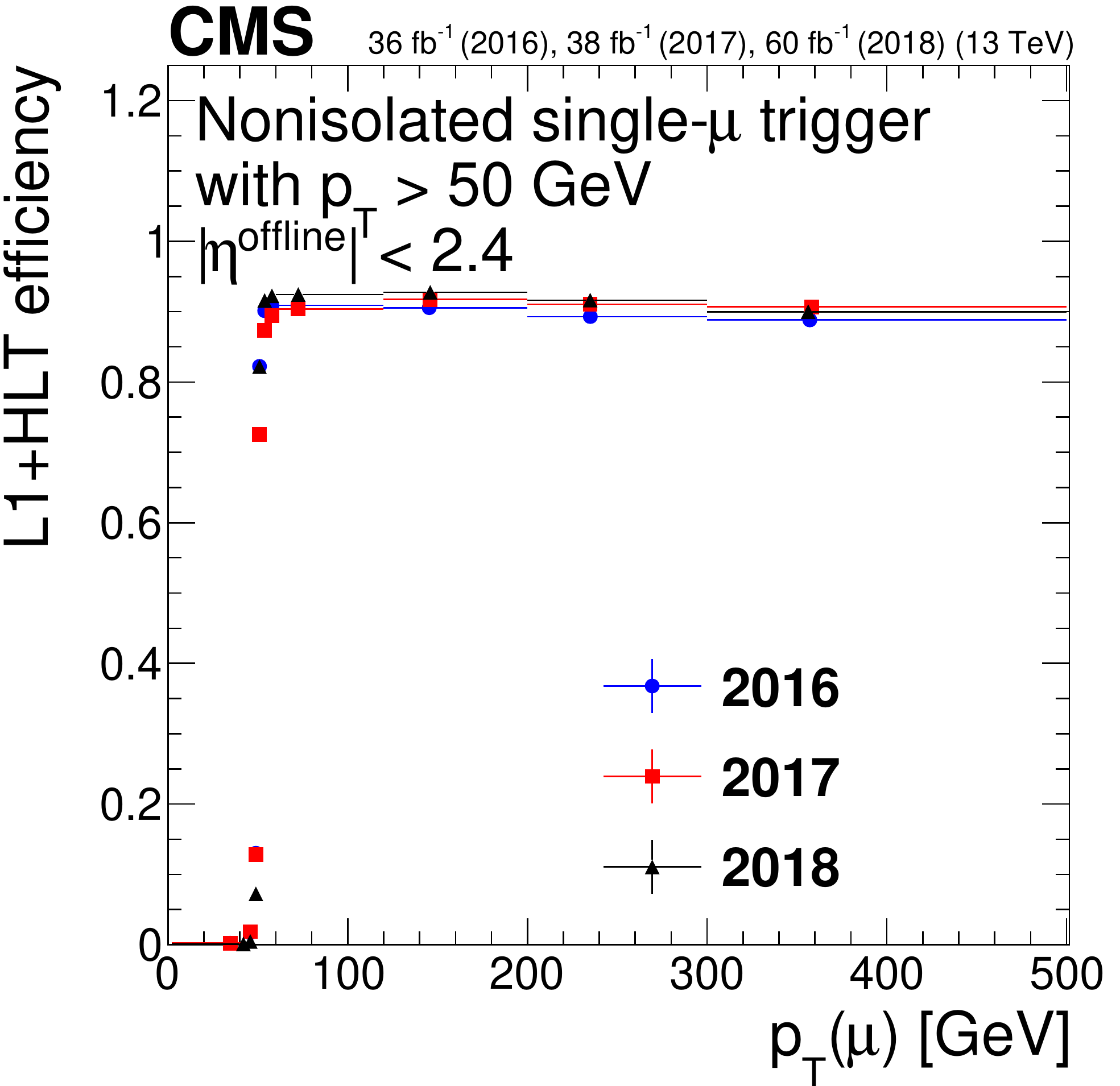}
\includegraphics[width=0.45\textwidth]{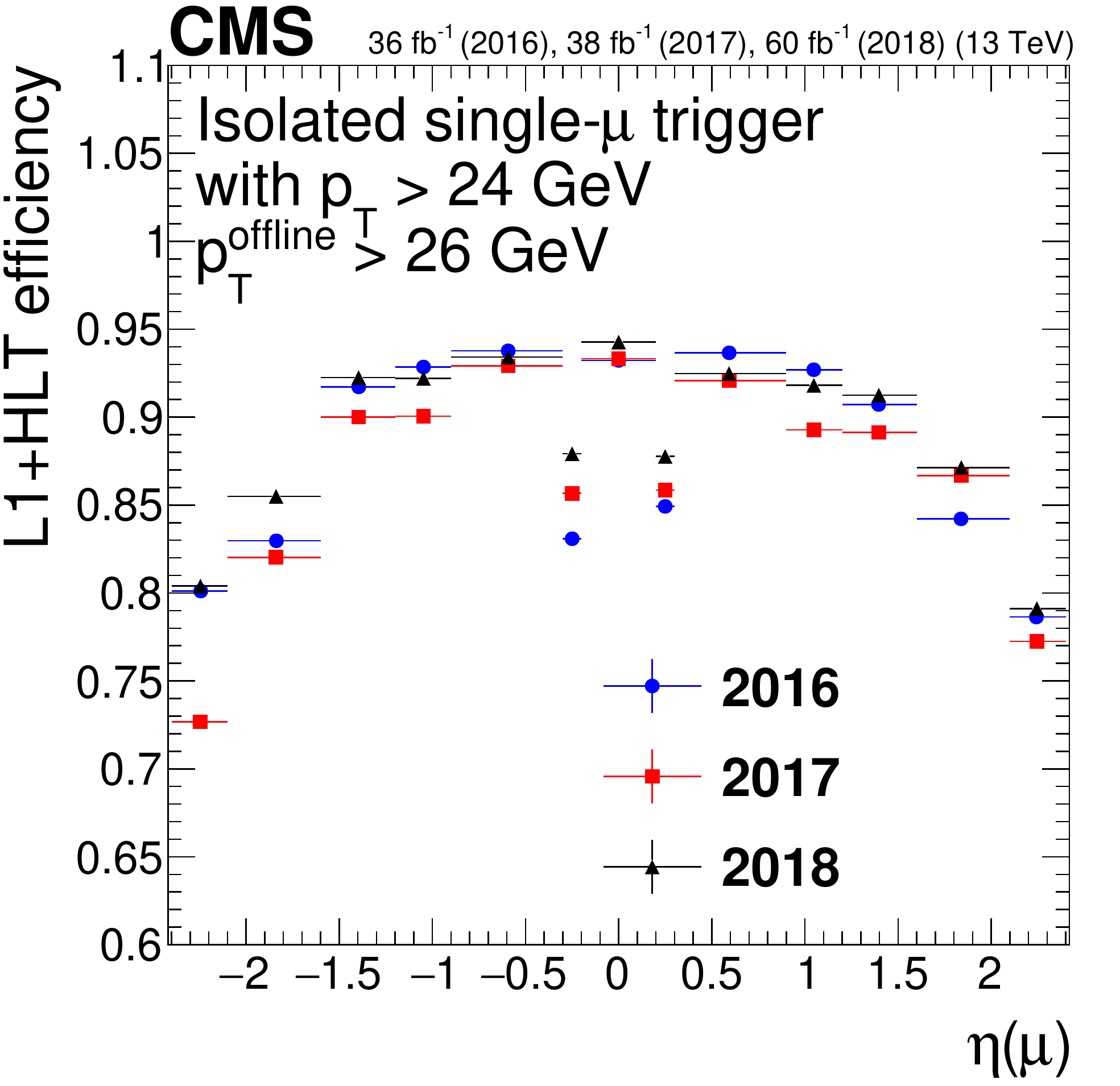}
\includegraphics[width=0.45\textwidth]{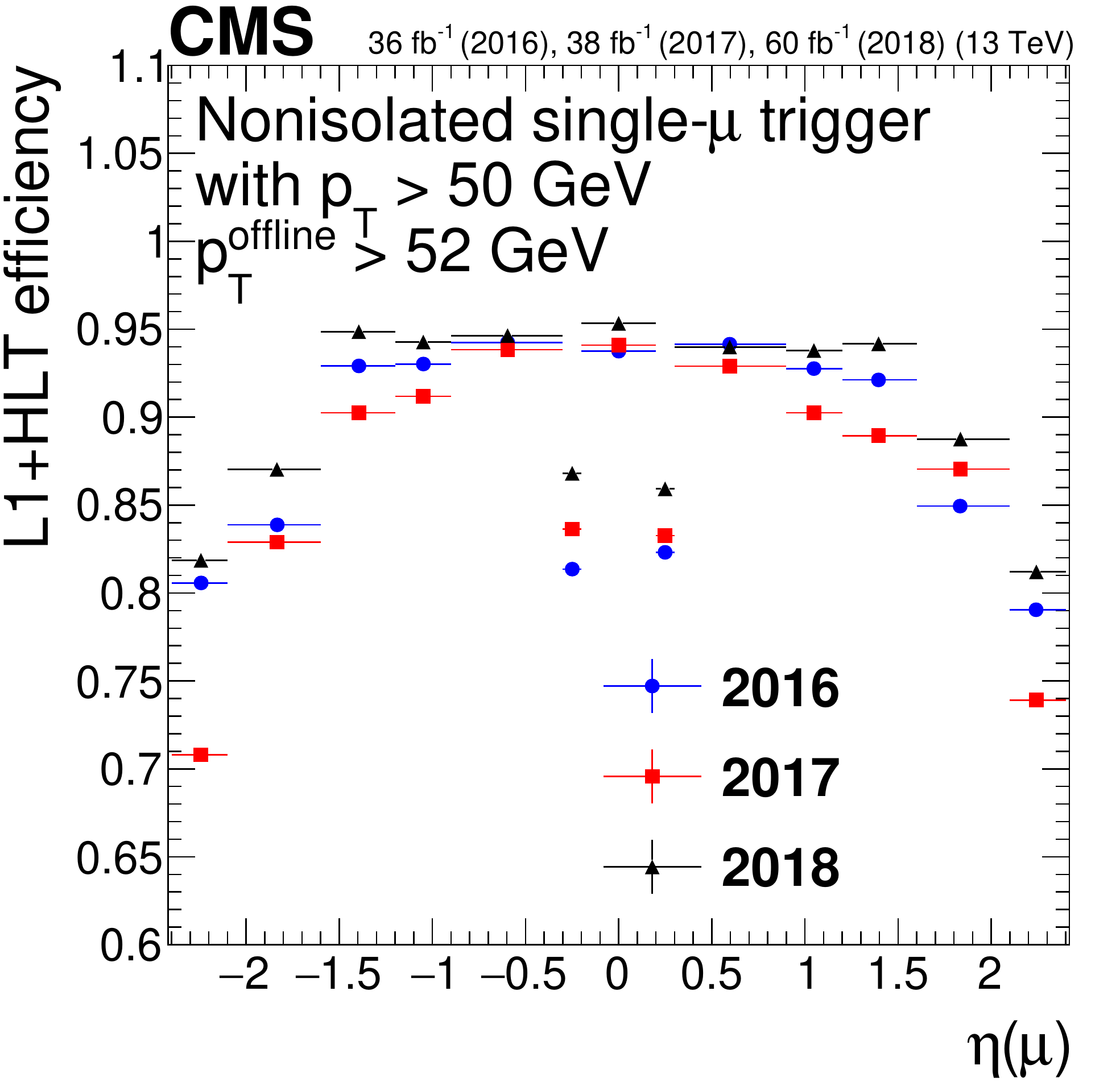}
\includegraphics[width=0.45\textwidth]{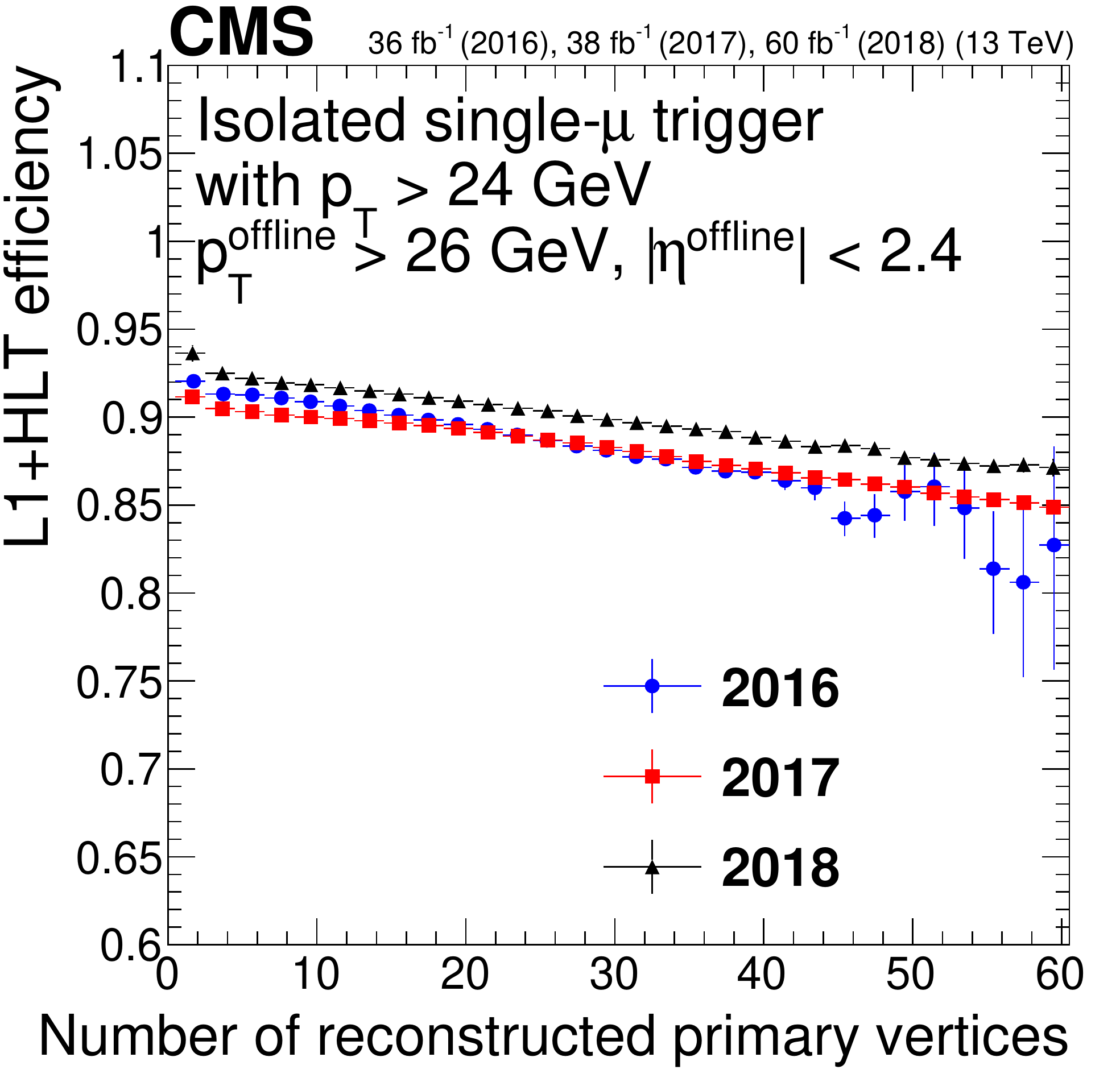}
\includegraphics[width=0.45\textwidth]{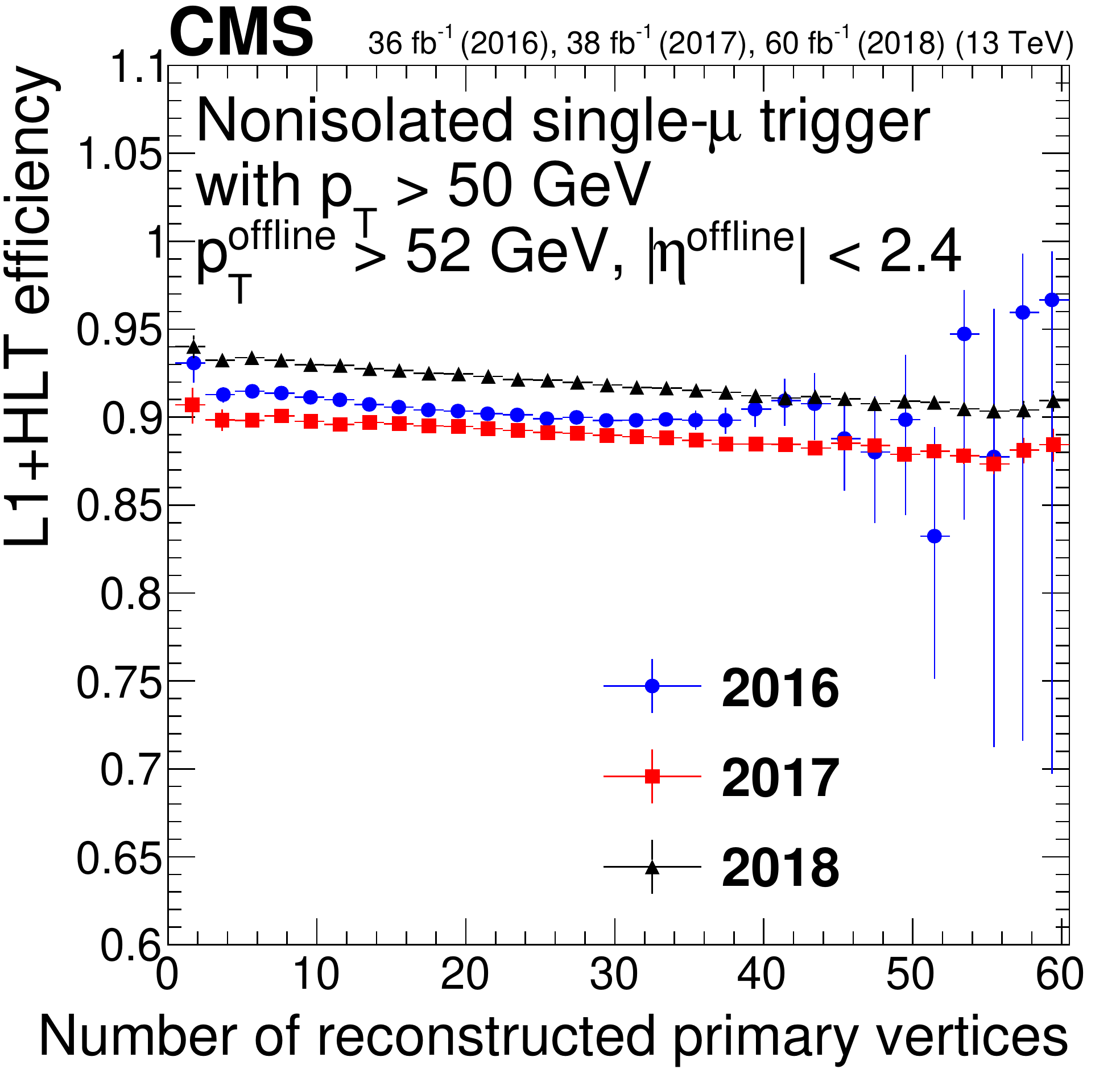}
\caption{Single-muon trigger efficiency for IsoMu24 (left) and Mu50 (right), as a function of muon \pt (upper row), $\eta$ (middle row), and the number of reconstructed vertices (lower row). Only statistical uncertainties are shown.}
\label{fig:perf:singleMuEff}
\end{figure}

The efficiencies of the isolated and nonisolated single-muon triggers are shown in
Fig.~\ref{fig:perf:singleMuEff} as a function of \pt and $\eta$ of the
probe muons and the number of reconstructed vertices. The full
data sets for 2016, 2017, and 2018 are used to illustrate the performance for the data set used in CMS analyses.
To count as passed in the efficiency calculation, the probe must be
matched to an HLT muon that satisfies the trigger under study. 

For the isolated single-muon triggers, the overall efficiencies for
muons with $\pt > 26 \GeV$ are close to 90\% in all years. The efficiency is about 1\% lower in 2017. This is attributed
to both the introduction of the first version of the iterative L3 reconstruction,
which was updated several times during the 2017 data-taking period to recover the performance, 
and the inactive pixel detector modules caused by the powering issue~\cite{DCDC}. These changes
offset the improvements in the L1 efficiency described above. In 2018, a higher
efficiency was achieved, even exceeding the 2016 value because of the more
robust iterative L3 algorithm and the repairs made to the pixel detector
between the 2017 and 2018 data-taking periods. There is a strong $\eta$
dependence of the efficiency. It is mostly constant in the barrel region
of the detector with the exception of the transition region between muon wheels at $\abs{\eta} \approx 0.2$, but the efficiency falls
significantly towards high values of $\abs{\eta}$ because of the difficulty to reconstruct L1 muons in that region, as shown in
Fig.~\ref{fig:perf:L1Eff}. The efficiency decreases as a function of
the number of reconstructed vertices mostly because of the isolation
requirement. 

For the nonisolated muon triggers, a
dedicated identification for high-\pt muons~\cite{Khachatryan:2014fba}
and an offline isolation requirement based only on inner tracker tracks are
used to select probe muons to reflect the
offline selection typically used together with this trigger. The overall efficiencies for muons with $\pt > 52 \GeV$ are 91\%, 89\%,
and 92\% for 2016, 2017, and 2018 data, respectively. The dependencies of
the efficiency on muon \pt, $\eta$, and the number of reconstructed
vertices are shown in Fig.~\ref{fig:perf:singleMuEff}. The observed trends are very similar to the case of the isolated triggers, with
the dependence on the number of reconstructed vertices reduced
because of the lack of isolation requirements.  

\begin{figure*}[!htb]
    \centering
    \includegraphics[width=0.45\textwidth]{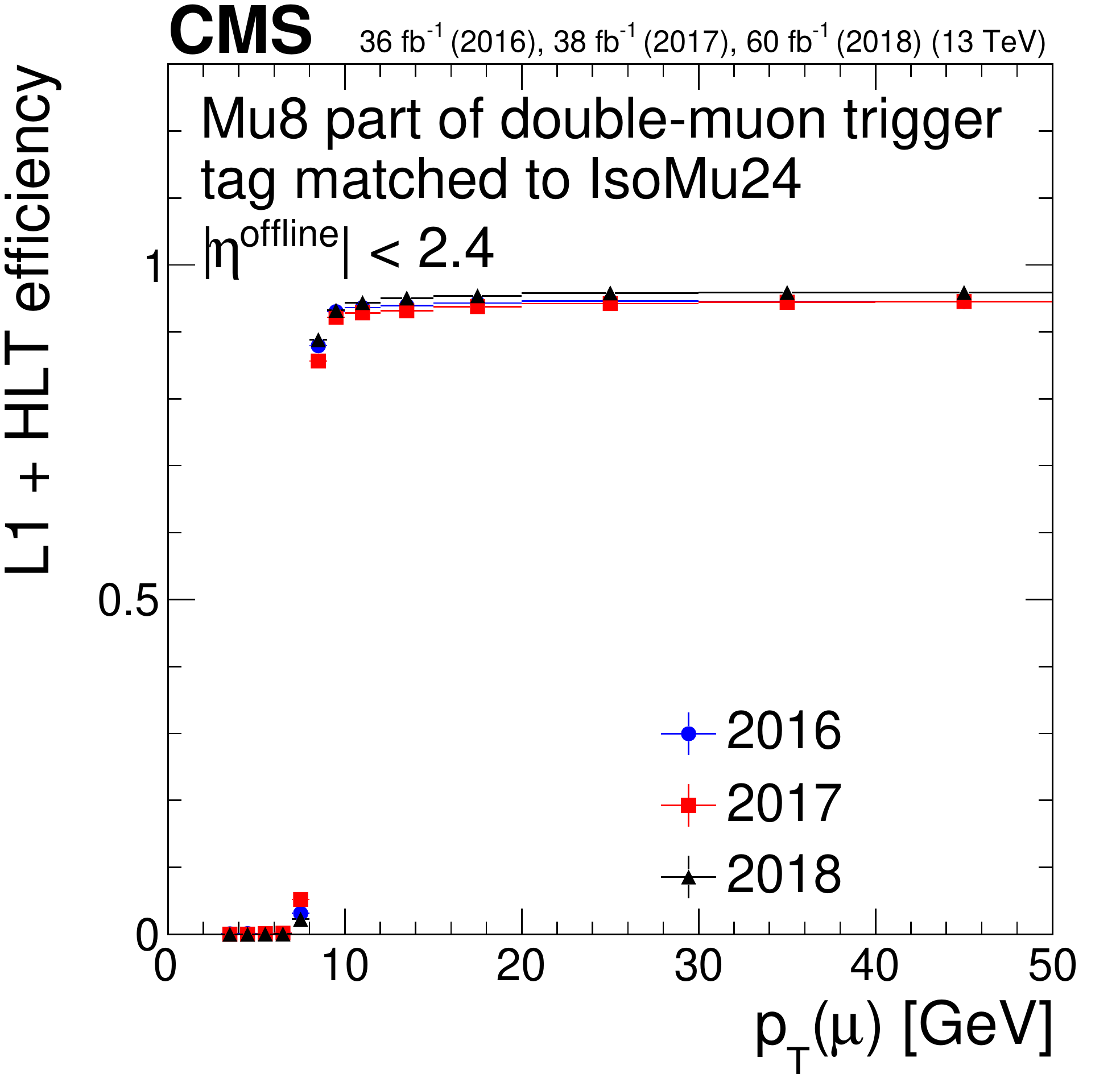}
    \includegraphics[width=0.45\textwidth]{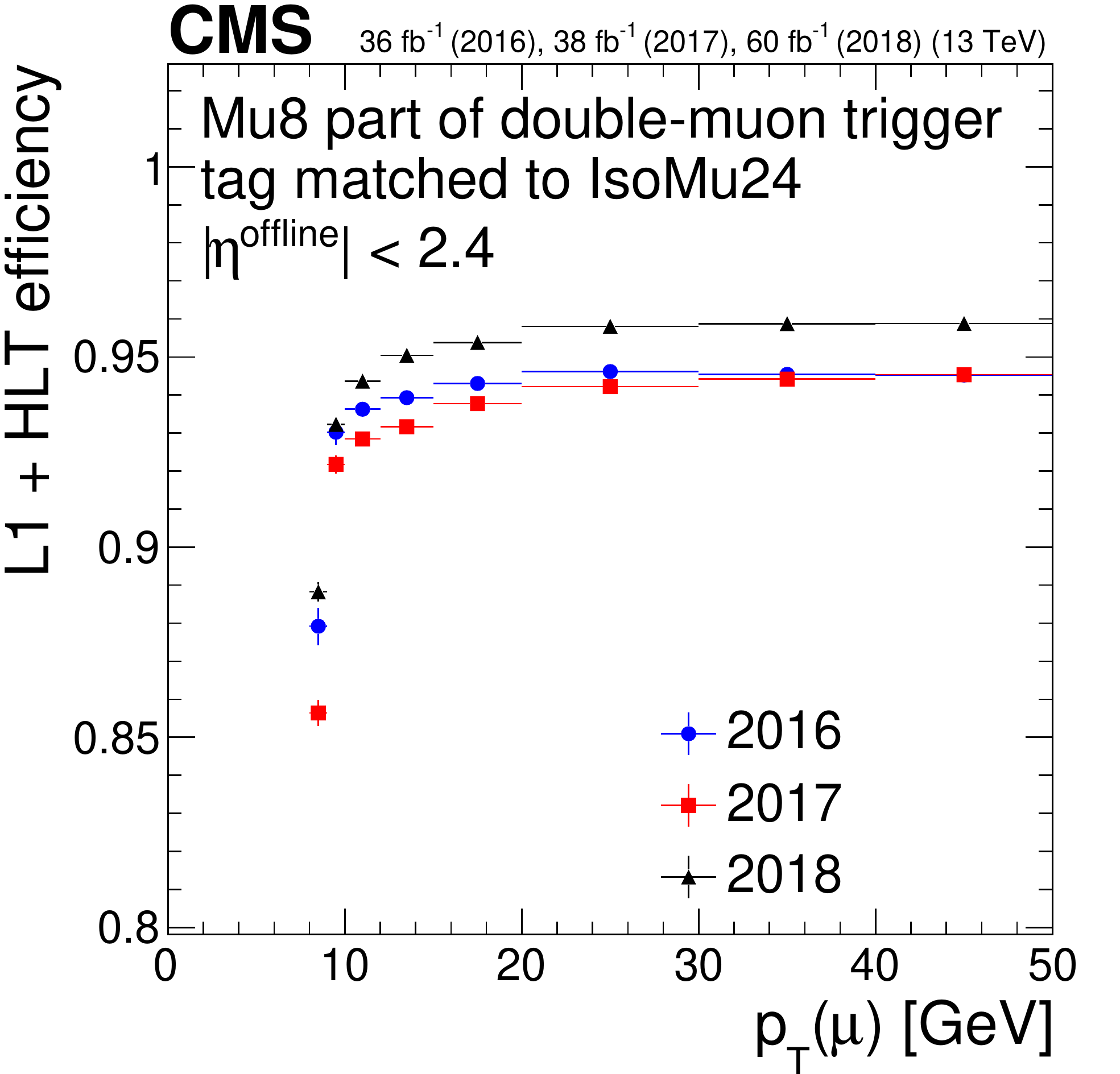}
    \caption{Trigger efficiency for the lower-\pt muon in double-muon triggers with respect to offline muons for 2016, 2017, and 2018. The right plot is the zoomed version of the left plot. Only statistical uncertainties are shown.
    }
    \label{fig:Mu8Eff}
\end{figure*}

The single-muon triggers used during Run 2 have a \pt threshold of 24\GeV. To probe the performance of the trigger at low \pt, the efficiency to trigger on the muon with lower \pt in a dimuon event has been measured. For this purpose, the same T\&P technique as described above is used, with the exception that the $\pt$ requirement on the probe muon is relaxed. As the tag muon is required to satisfy the IsoMu24 trigger, the efficiency of the probe muon to satisfy the part of the double-muon triggers that requires one muon with $\pt > 8\GeV$ is measured independently of the requirements on the higher-\pt muon. The measured efficiencies versus the probe muon $\pt$ for the three years are given in Fig.~\ref{fig:Mu8Eff}. The same dependence of the efficiency on the different years as for the single-muon triggers is observed.

\subsection{Trigger rate}
\label{sec:rate}
The previous sections show that the iterative algorithm is as
efficient as the trigger strategy used at the beginning of Run 2,
despite the slight loss of efficiency from the introduction of muon
identification criteria in 2018. This small loss comes with a significant
reduction in rate, thus the same \pt thresholds could be maintained
during the entire data taking despite the higher luminosity and more challenging pileup conditions in
2018. In the following, HLT trigger rates for the trigger configuration that was used at the beginning
(cascade OR tracker muon in 2016) and at the end (iterative in 2018) of Run 2 will be compared.  

The trigger rate is measured using dedicated data sets for trigger performance studies that are collected without HLT requirements, and therefore contain a representative mix of event topologies similar to those the HLT receives from the L1 trigger during data taking. The rate is defined as the frequency with which a trigger condition has been satisfied in a given data-taking period. Because the rate of muons produced is given by the overall production cross section times the instantaneous luminosity, a linear luminosity dependence of the trigger rate is expected, as long as reconstruction or selection effects do not introduce additional dependencies.
In the following, to facilitate the comparison between the triggers in 2016 and 2018, the rates for both years are scaled linearly to an
instantaneous luminosity of $10^{34}\percms$. Only statistical uncertainties are considered for these measurements. 

Each step of the reconstruction described in the previous sections (L1, L2, L3) reduces the rate and thus the total number of events that are kept for further analysis. The L1 trigger with $\pt > 22\GeV$ and tight quality requirements delivers a rate of $4506 \pm 8$ ($4147 \pm 10$)\unit{Hz} in 2016 (2018) at an instantaneous luminosity of $10^{34}\percms$. The difference stems from the improvements to the L1 trigger described earlier. In 2016, L2 muons with $\pt < 10\GeV$ were rejected. In 2017 and 2018, this requirement was removed. The L2 step introduces a rate reduction of about 50\% in 2016 and 15\% in 2018. For the IsoMu24 trigger, before applying isolation, only 8\% of the muons that were originally accepted by the L1 trigger are kept after the L3 step, and the isolation requirement reduces the L3 rate by a further factor of 3. The rate for this trigger is about 250\unit{Hz} at the highest instantaneous luminosity achieved in Run 2. The rates for the individual steps of the trigger are summarized in Table~\ref{tbl:RateSteps}.

\begin{table}[htbp]
\centering
\topcaption{Rates for the individual steps of the IsoMu24 trigger. The rates are
measured in the 2016 and 2018 data sets, respectively, and then scaled to
an instantaneous luminosity of 
$10^{34}\percms$. Only statistical uncertainties are
considered for each measurement. } 
\begin{tabular}{ccc}
                                      & Rate (2016) [Hz] & Rate (2018) [Hz]  \\ \hline
 L1 rate, $\pt > 22\GeV$    & $4506 \pm 8$ & $4147 \pm 10$    \\
 L2 rate, ($\pt > 10\GeV$ in 2016)   & $2238  \pm 6$ & $3559 \pm 10$  \\
 L3 rate, $\pt > 24\GeV$  &  $377  \pm 2$  & $363 \pm 3$ \\
 Isolated L3 rate, $\pt > 24\GeV$  &  $125 \pm 1$ & $126 \pm 2$    \\
\end{tabular}
\label{tbl:RateSteps}
\end{table}

The rates for the standard isolated single-muon trigger, nonisolated
single-muon trigger, and double-muon triggers for 2016 and 2018 are shown in
Table~\ref{tbl:Rate}. The isolated triggers already select a very pure muon sample in 2016, so no further rate reduction from the
improvement of the L3 reconstruction algorithm in 2018 is observed. However for the nonisolated triggers and the double-muon triggers, a
respective rate reduction of around 10\% and 30\% from 2016 to 2018 is observed. 
For the single-muon trigger this rate reduction is achieved mainly by
the identification requirements introduced for the 2018 data
taking. For the double-muon trigger the rate has also been reduced by
the introduction of a lower bound on the invariant mass of the two
muons of 3.8\GeV, which rejects the $\JPsi$ meson peak.

\begin{table}[htbp]
\centering
\topcaption{Rates for the 2016 and 2018 trigger strategies. The rates are
measured in the 2016 and 2018 data sets, respectively, and then scaled to
an instantaneous luminosity of 
$10^{34}\percms$. Only statistical uncertainties are
considered for each measurement. } 
\begin{tabular}{ccc}
                                      & Rate (2016) [Hz] & Rate (2018) [Hz]  \\ \hline
 IsoMu24   & $124.9 \pm 1.3$ & $126.0 \pm 1.8$    \\
 Mu50   & $28.7  \pm 0.6$ & $24.2 \pm 0.8$  \\
 Mu17Mu8 &  $21.2  \pm 0.5$  & $14.9 \pm 0.6$
\end{tabular}
\label{tbl:Rate}
\end{table}

\begin{figure}[htb]
\centering
\includegraphics[width=0.45\textwidth]{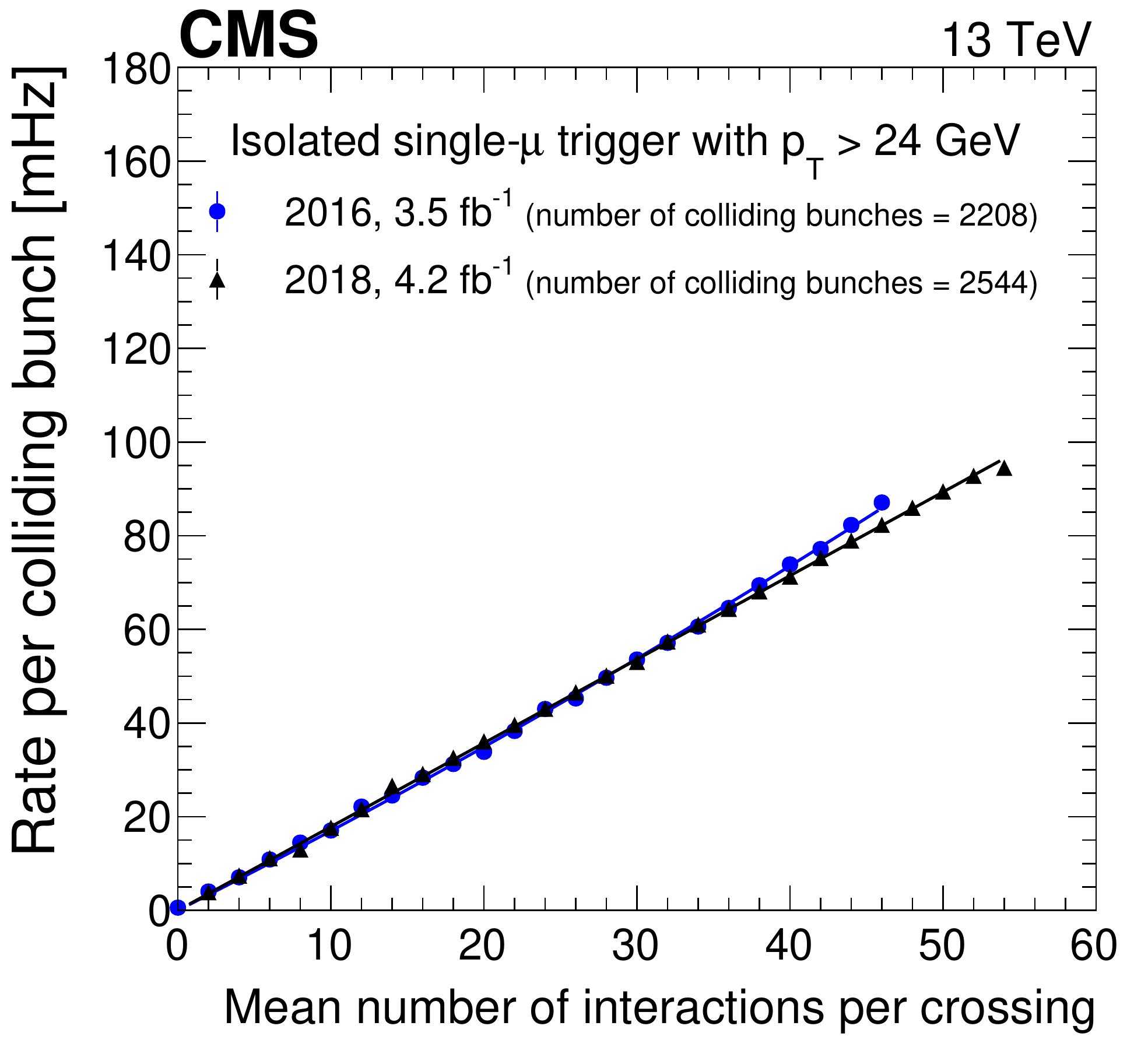}
\includegraphics[width=0.45\textwidth]{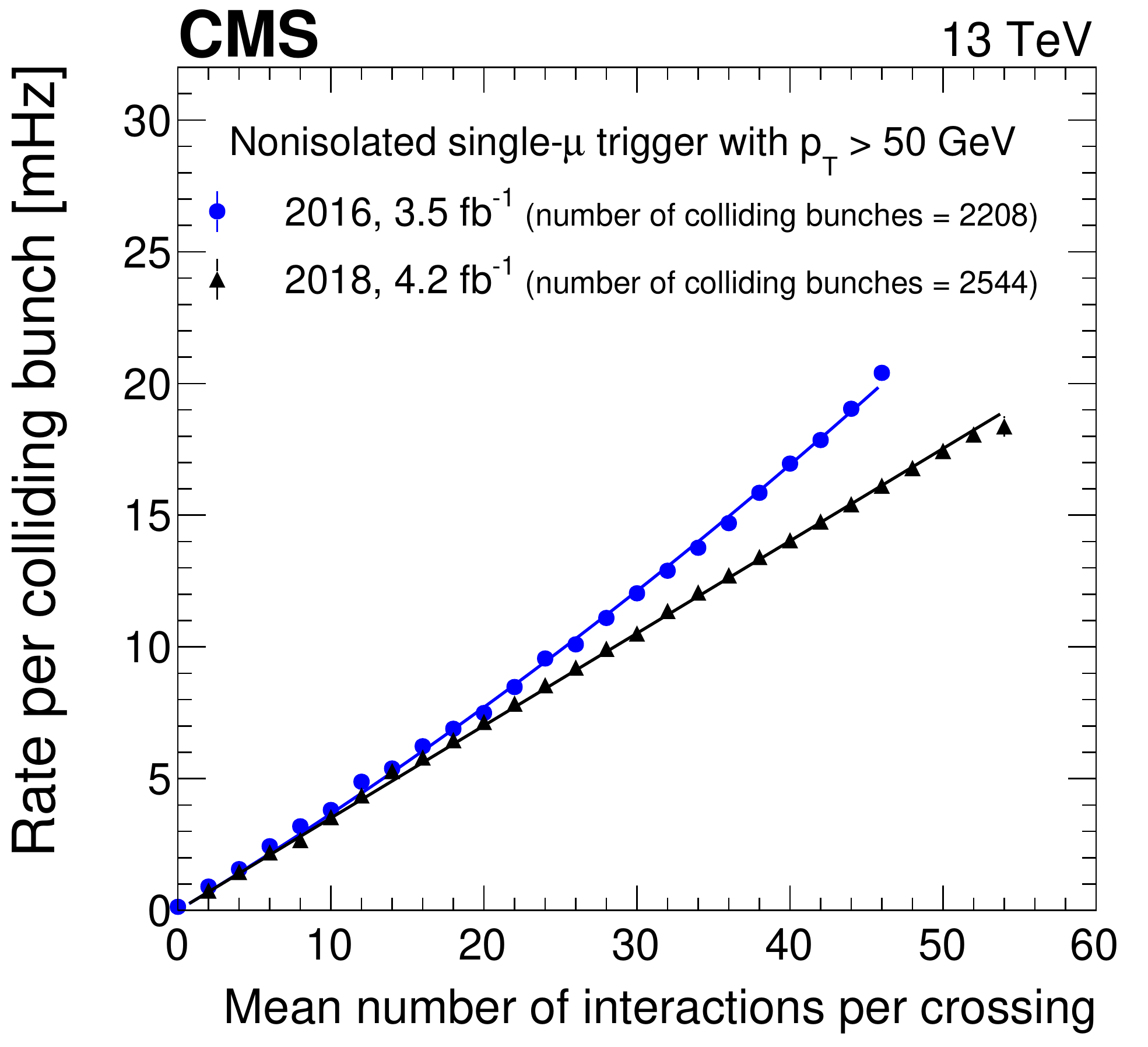}
\includegraphics[width=0.45\textwidth]{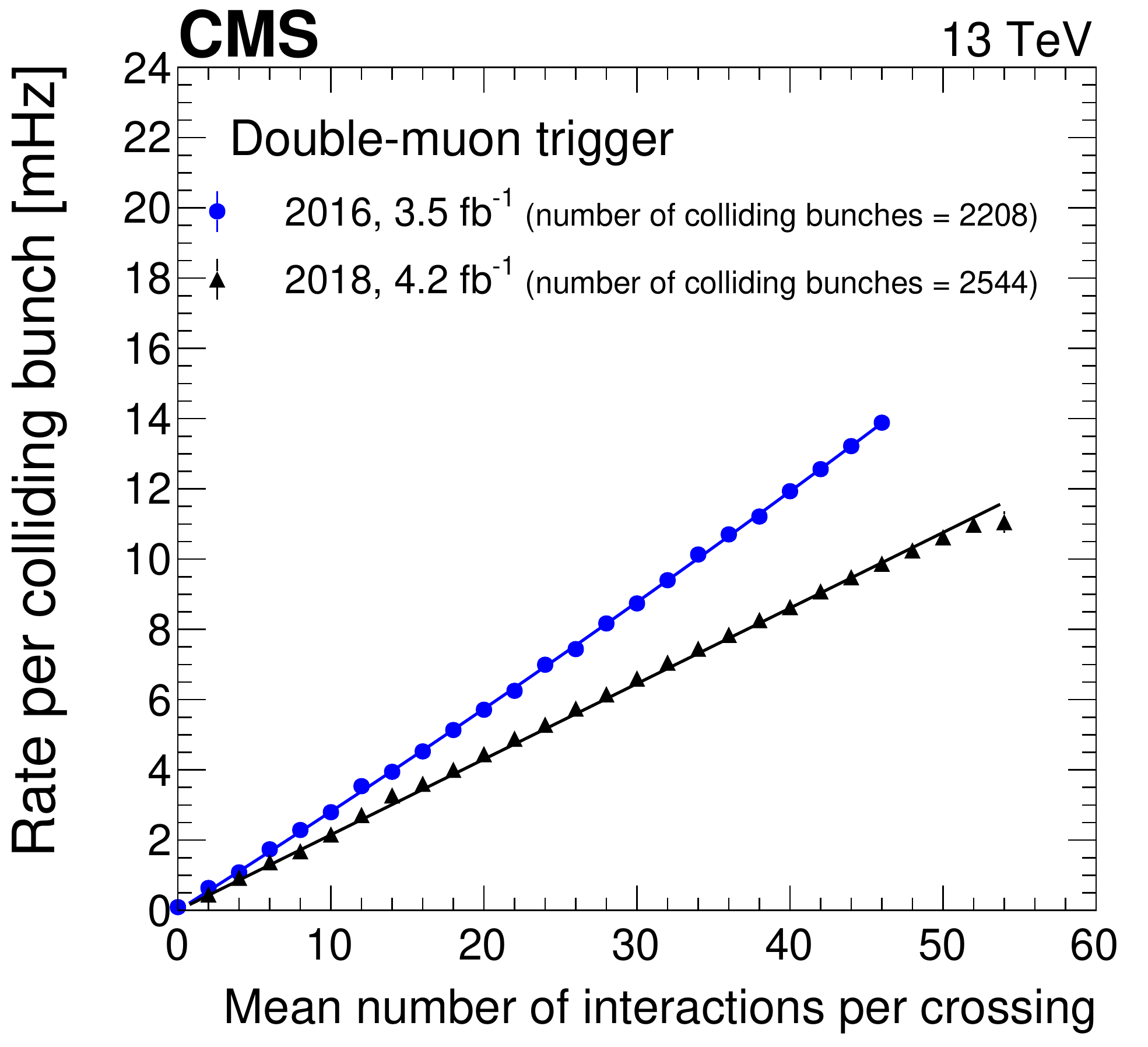}
\caption{Dependence of the trigger rate per colliding bunch on the mean number of interactions per bunch crossing for the isolated (upper left) and nonisolated (upper right) single-muon triggers as well as the double-muon trigger (lower middle), comparing the 2016 and 2018 trigger configurations. Quadratic (linear) fits to the 2016 (2018) data points are shown as colored lines. Only statistical uncertainties are shown.}
\label{fig:perf:rate}
\end{figure}

Figure~\ref{fig:perf:rate} shows the trigger rates, normalized to the number of colliding bunches to account for the different beam conditions between the years, as a function of the mean number of interactions per bunch crossing $n_{\mathrm{int}}$ for the isolated and nonisolated single-muon triggers as well as the double-muon triggers. 

The mean number of interactions per bunch crossing is calculated from the instantaneous luminosity and the inelastic $\Pp\Pp$ cross section. The data used for this analysis was taken only in an LHC configuration with 2208 colliding bunches in 2016 and 2544 bunches in 2018, so it is a subset of the total data set. The expected linear dependence is observed for all triggers except the nonisolated single-muon trigger in 2016, where a faster growth occurs at the highest number of interactions observed in that year. This effect is also slightly visible for the isolated single-muon trigger. This results from a slightly higher rate of nonmuon objects accepted by the trigger in high-pileup events in 2016, which was remedied in 2018. As discussed above, there is no significant difference between the rates observed for the isolated single-muon trigger in 2016 and 2018. In the case of the nonisolated single-muon trigger, the rate is similar between the years at low luminosity, but the increase of the rate with luminosity is significantly reduced in 2018. For the double-muon trigger there is a significant difference already at low luminosity, originating from the introduction of the invariant mass requirement discussed above. The dependency of the rate on the number of interactions has been obtained from parabolic fits for 2016 and linear fits for 2018 and the resulting parameters are summarized in Table~\ref{tab:rateSlope}.

\begin{table}
\centering
\topcaption{Increase of the trigger rate as a function of the number of $\Pp\Pp$ interactions in an event ($n_{\mathrm{int}}$) in units of mHz/interaction, parametrized with 2nd and 1st order polynomials, respectively, for 2016 and 2018.} 
\begin{tabular}{ccc}
 & 2016 & 2018 \\ \hline
 IsoMu24 & ($1.65\pm0.02$) $n_{\mathrm{int}}$ + ($0.0047\pm0.0007$) $n_{\mathrm{int}}^2$   & ($1.787\pm0.004$) $n_{\mathrm{int}}$\\
 Mu50 & ($0.348\pm0.007$) $n_{\mathrm{int}}$ + ($0.0019\pm0.0002$) $n_{\mathrm{int}}^2$ &   ($0.351\pm0.001$) $n_{\mathrm{int}}$ \\
 Mu17Mu8 & ( $0.276\pm0.002$) $n_{\mathrm{int}}$ + ($0.0006\pm0.0001$) $n_{\mathrm{int}}^2$ &  ($0.215\pm0.001$) $n_{\mathrm{int}}$\\
\end{tabular}
\label{tab:rateSlope}
\end{table}

The limit of an average of about 1\unit{kHz} on the total rate of events accepted by the HLT, imposed by limited storage capacity and computing resources for offline reconstruction, make the
trigger bandwidth a coveted resource. Therefore it is important to
ensure that the muon sample accepted by the HLT is as pure in genuine muons as possible; otherwise the trigger rate would be wasted on other
objects misreconstructed as muons. To illustrate the improvements made
in this regard during Run 2, the composition of the selected sample has
been measured in an unbiased sample of events accepted by the L1 trigger by geometrically matching the triggered muons to offline reconstructed muons within $\Delta R < 0.1 (0.3)$ for L3 (L1 and L2) muons. The triggered muons are then sorted into three categories:
\begin{itemize}
\item \textit{Unmatched}: the triggered muon could not be matched to an offline muon passing identification requirements.
\item \textit{Nonisolated muon}: the triggered muon is matched to an offline muon passing identification but not isolation requirements.
\item \textit{Isolated muon}: the triggered muon is matched to an offline muon passing both identification and isolation requirements. 
\end{itemize}
For this purpose we use loose identification requirements designed to accept both isolated and nonisolated muons~\cite{Sirunyan:2018fpa}, while a tight isolation requirement is used to ensure that only truly prompt muons are counted. 

\begin{figure}
\centering
{\includegraphics[width=0.45\textwidth]{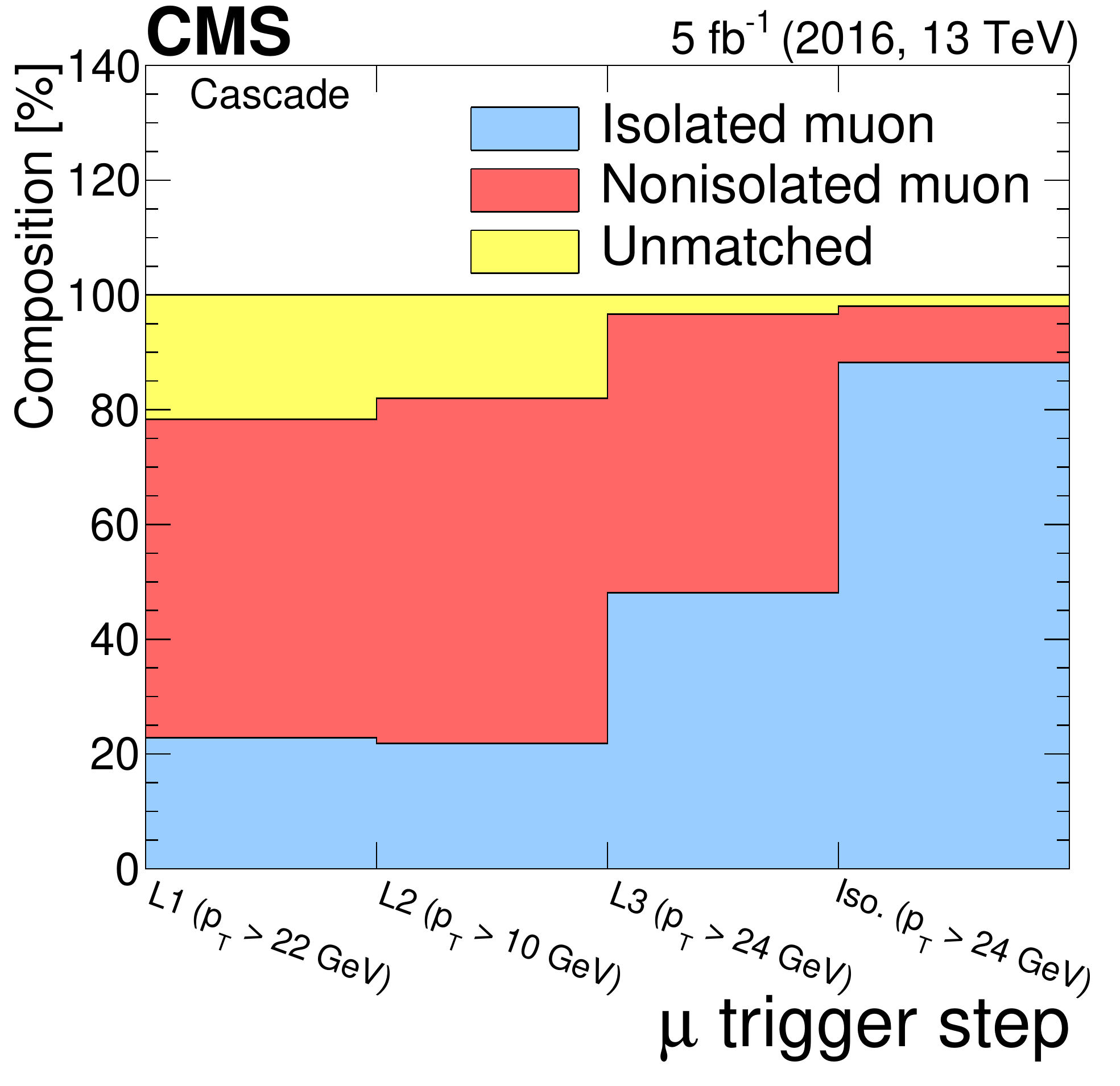}}
{\includegraphics[width=0.45\textwidth]{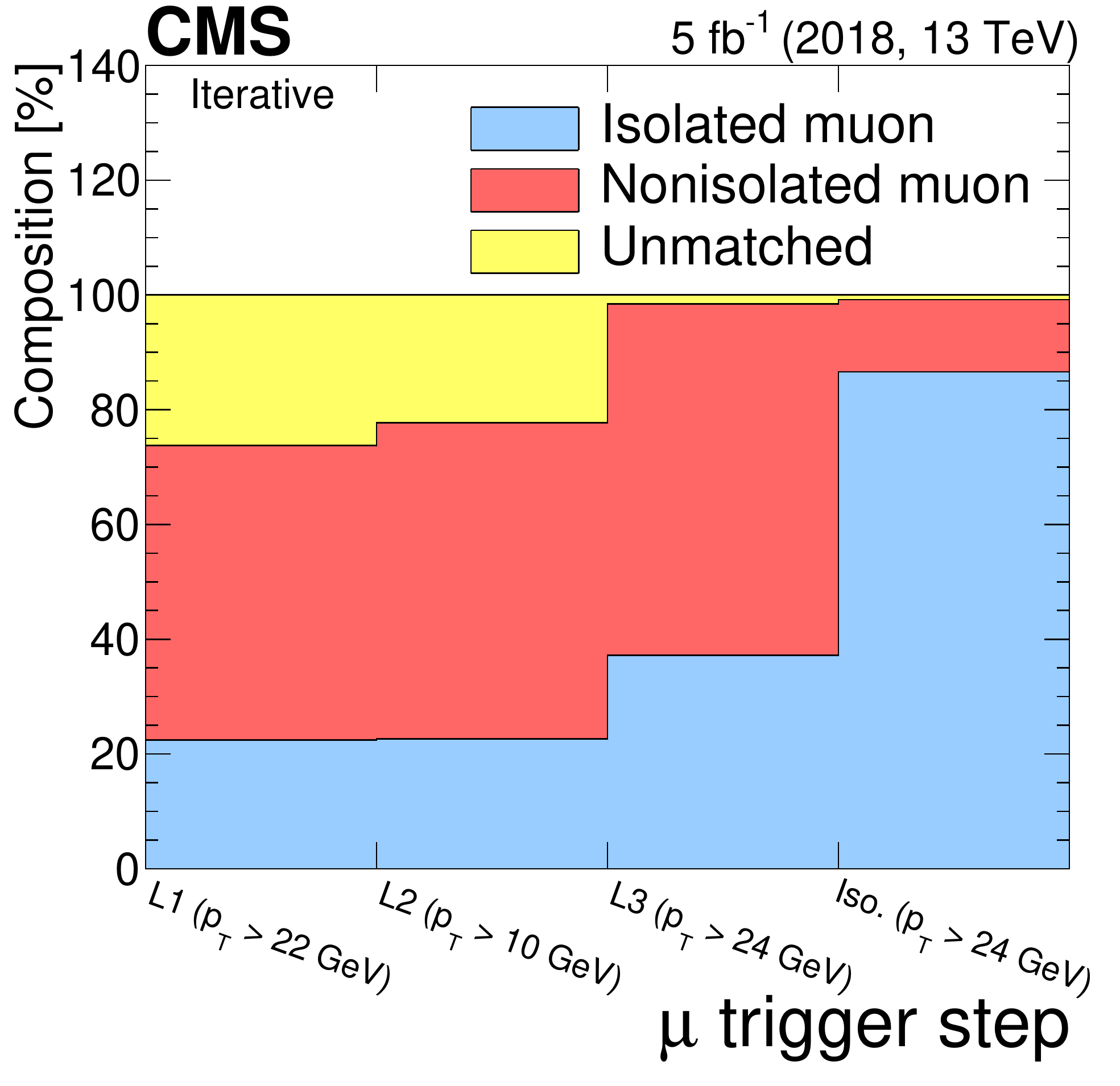}}
\caption{The composition of the selected muon sample for the 2016 (left) and 2018 (right) data.}
\label{fig:perf:contamination}
\end{figure}

The composition of the selected sample for the four steps of the trigger sequence, L1, L2, L3, and isolation, is shown in Fig.~\ref{fig:perf:contamination} for the cascade algorithm measured on 2016 data and the iterative algorithm measured on 2018 data. For each step there is a minimum \pt requirement on the muons that is given by the values used in the trigger sequence. The thresholds are 22\GeV for the L1 muons used to seed the single-muon trigger, 10\GeV for the L2 muons used to seed L3 reconstruction (a very loose cut that was only applied in 2016), and 24\GeV for L3 muons since this is the final trigger threshold.
The isolated-muon fraction is sequentially increased with every step of the reconstruction, reaching up to values
of about 80\% at the end of the trigger sequence. Comparing 2016 and 2018, one can see a significant decrease in
the unmatched fraction by factors of 2--3 because of the improved reconstruction algorithm and the application of
identification criteria at L3. A slight increase in the fraction of unmatched L1 and L2 muons in 2018 compared to 2016 is observed. This effect is most pronounced in the barrel-endcap overlap region; it results from the use of lower quality DT segments, which improve the momentum assignment and, hence, the total rate, but also increase the fraction of nonmuon candidates that fire the trigger.

\subsection{Processing time}
\label{sec:timing}
The high input rate from L1 severely limits the computing resources available to make a decision to keep or reject an event at the HLT. Fast
algorithms are therefore required to keep the average time to process an event within the budget, which is given by the available computing resources for the HLT and increased from about 200\unit{ms} to about 300\unit{ms} from 2016 to 2018. The processing time for
the muon triggers has been measured using data from 2016 and
2018 on a CMS reference machine equipped with a Intel\textregistered ~Xeon\textregistered ~E5-2650 v2 CPU running at 2.60\unit{GHz}, which is slower than the latest processing nodes installed in the HLT. Events were processed concurrently on four CPU cores in four threads. To ensure comparability, events from runs with similar pileup conditions have been chosen. For 2016 the number of pileup interactions ranges from 45 to 50 with a mean value of 46.3, for 2017 from 47 to 49 with a mean of 47.4, and for 2018 from 46 to 48 with a mean of 47.3. The overall processing	
time for the full suite of triggers used by CMS in those data sets is 181\unit{ms/event} for 2016, 180\unit{ms/event} for 2017, and
296\unit{ms/event} for 2018. The large increase in 2018 is mostly driven by algorithms implemented to mitigate the inoperative modules in the pixel detector. The muon triggers discussed here account for 4.1\% of this processing time in 2016, which increases to 5.2\% in 2018. 

\begin{figure}
\centering
{\includegraphics[width=0.55\textwidth]{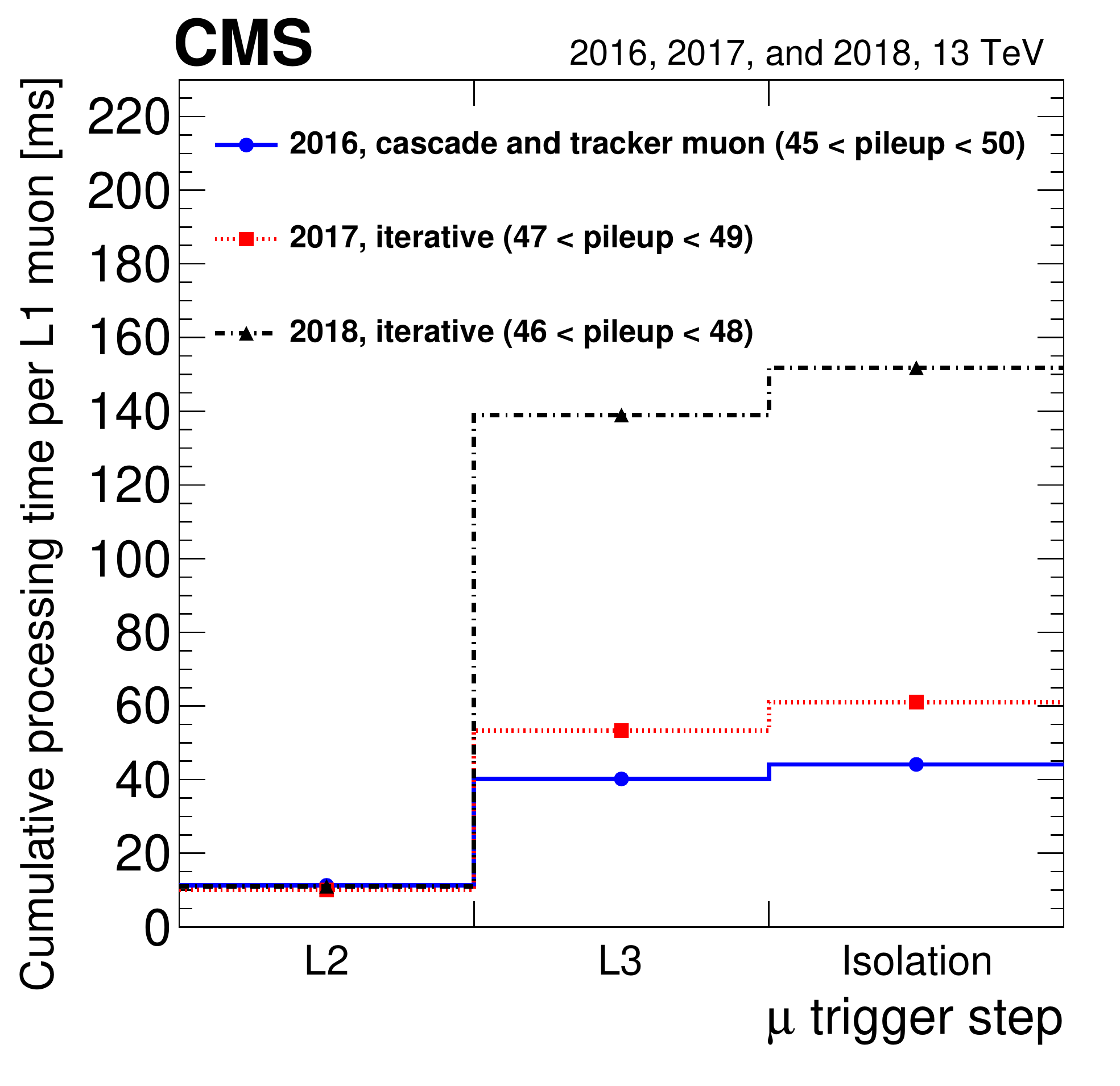}}
\caption{Cumulative average processing time of each of the trigger reconstruction steps for the IsoMu24 trigger.}
\label{fig:perf:timing}
\end{figure}

The cumulative average processing time of every step of the HLT reconstruction (L2 and L3) of the IsoMu24
trigger is shown in Fig.~\ref{fig:perf:timing}. The distributions are normalized to the number of L1 muons that are used to seed the reconstruction. No difference between the years is observed in the reconstruction time for L2 muons. The 2016 L3 reconstruction algorithms (cascade and tracker muon) require around 29\unit{ms/muon}, while the average processing 
time of the 2017 reconstruction is 43\unit{ms/muon} and that of the 2018 reconstruction algorithm (iterative) is 128\unit{ms/muon}. 

The increase in 2017, and especially 2018, is driven by several effects. In 2016, the inside-out reconstruction using multiple tracking iterations was run once per muon candidate in the tracker muon reconstruction. In the iterative approach, this inside-out reconstruction was initially only run for muons that failed in the outside-in step in 2017. In 2018, however, it was run for all L1 muons, and also seeded by those L2 muons that did not result in a track in the outside-in step. The iterative tracking in the inside-out reconstruction is not only run more often in 2018, it is also slower in 2017 and 2018 compared to 2016 because of the increased number of pixel detector layer combinations in the upgraded pixel detector, resulting in a larger number of track seeds. In 2018, it was further slowed down by the inclusion of a step that was seeded by a pair of pixel hits. This step accounts for an increase of about 40\unit{ms/muon} in the processing time. Finally, the 2018 version of the outside-in reconstruction creates significantly more track seeds, resulting in an increase in the time spent on track building.

The isolation step is also slower by about 7\unit{ms}, driven both by the slower track reconstruction with the upgraded pixel detector, as well as by an increase in the time to reconstruct isolation information in the calorimeters because of a change in HCAL reconstruction.

\begin{figure}
\centering
{\includegraphics[width=0.45\textwidth]{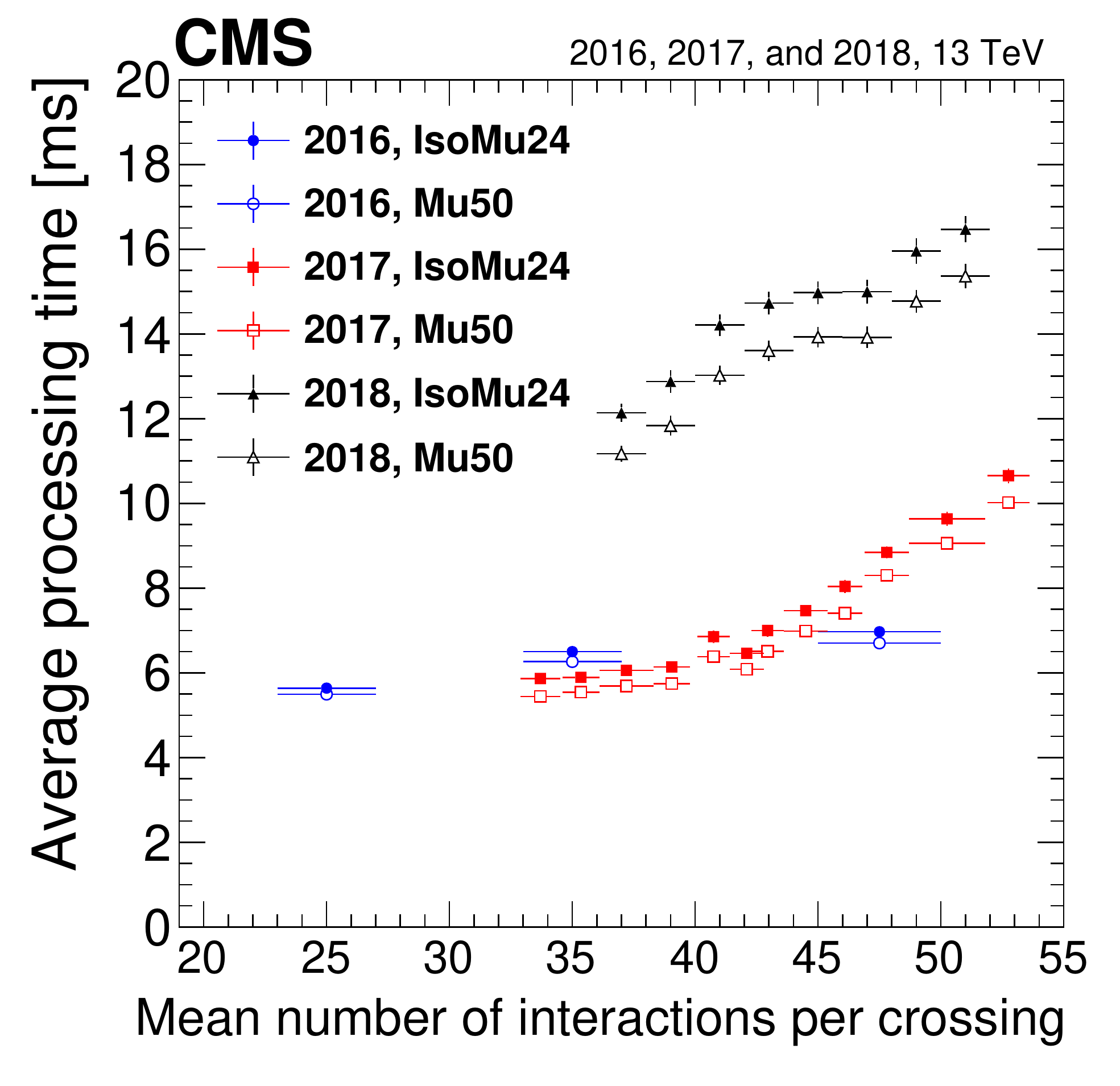}}
\caption{Average processing time of the full trigger for the IsoMu24 (filled symbols) and the Mu50 trigger (open symbols) as a function of pileup. The processing time for the combination of cascade and tracker muon measured in 2016 data is shown in blue circles, while they are shown in red squares and black triangles for the iterative algorithm measured in 2017 and 2018 data. Only statistical uncertainties are shown.}
\label{fig:perf:timingPU}
\end{figure}

The pileup dependence of the average processing time is shown in Fig.~\ref{fig:perf:timingPU} for both the IsoMu24 and Mu50 triggers. The processing times in this case are lower than in Fig.~\ref{fig:perf:timing} because an event sample representative of the average composition encountered during data taking is used and events with no L1 muons are included; these lower the average time. These results are comparable to the processing time for the full HLT menu discussed above. The average number of L1 muons per event in these data sets does not vary between the years by more than a few percent. This makes these values representative of the processing time during nominal operation of the HLT. The measurement is performed for the combination of the cascade and tracker muon algorithms in 2016 data and for the iterative algorithm in 2018 data. Because of the different running conditions in 2016 and 2018, the measurements span a different pileup range. There is a moderate increase in the processing time between Mu50 and IsoMu24 caused by the need to compute the isolation in case of the IsoMu24. The time needed to compute the isolation does not significantly depend on the pileup. The iterative algorithm does not only have a higher average processing time, but also a stronger pileup dependence compared to the cascade and tracker muon algorithms. This is due to the greater reliance of the iterative algorithm on track seeding in the pixel detector, especially with the doublet-seeded iterations, where the increased hit occupancy at higher pileup increases the number of tracks seeds found inside the ROIs. 

\section{Summary}
\label{sec:summary}
To maintain the excellent performance of the CMS muon trigger system
in the much harsher running conditions of LHC 
Run 2, especially the high-pileup environment in 2017 and 2018, the
reconstruction algorithms and identification 
criteria for muons in the high-level trigger system were continually improved during
the Run 2 data-taking period. In particular, the algorithm used to
reconstruct muon tracks in the inner tracker was improved in several
steps during the 2017 and 2018 data-taking periods. Upgrades to the
hardware-based Level-1 trigger further improved the trigger efficiency for
muons. This facilitated an increase in the purity of the selected
muon sample by introducing muon identification criteria in the high-level trigger system,
while maintaining, or even slightly improving, the overall
efficiency. The plateau efficiency for both isolated and nonisolated single muons
above the transverse momentum threshold is around 90\%.
Compared with previous versions of the algorithm, a
significant improvement in the momentum resolution is achieved. For
triggers without an isolation requirement, a reduction in trigger rate
of 10\% for the same trigger threshold and instantaneous luminosity is observed compared with the
beginning of Run 2.  
Because these improvements come at a cost of increased processing time needed to run the algorithms, this leaves room for further optimization for the upcoming LHC Run 3 with the goal to maintain the excellent physics performance while improving algorithmic efficiency.

\begin{acknowledgments}
  We congratulate our colleagues in the CERN accelerator departments for the excellent performance of the LHC and thank the technical and administrative staffs at CERN and at other CMS institutes for their contributions to the success of the CMS effort. In addition, we gratefully acknowledge the computing centers and personnel of the Worldwide LHC Computing Grid and other centers for delivering so effectively the computing infrastructure essential to our analyses. Finally, we acknowledge the enduring support for the construction and operation of the LHC, the CMS detector, and the supporting computing infrastructure provided by the following funding agencies: BMBWF and FWF (Austria); FNRS and FWO (Belgium); CNPq, CAPES, FAPERJ, FAPERGS, and FAPESP (Brazil); MES (Bulgaria); CERN; CAS, MoST, and NSFC (China); COLCIENCIAS (Colombia); MSES and CSF (Croatia); RIF (Cyprus); SENESCYT (Ecuador); MoER, ERC PUT and ERDF (Estonia); Academy of Finland, MEC, and HIP (Finland); CEA and CNRS/IN2P3 (France); BMBF, DFG, and HGF (Germany); GSRT (Greece); NKFIA (Hungary); DAE and DST (India); IPM (Iran); SFI (Ireland); INFN (Italy); MSIP and NRF (Republic of Korea); MES (Latvia); LAS (Lithuania); MOE and UM (Malaysia); BUAP, CINVESTAV, CONACYT, LNS, SEP, and UASLP-FAI (Mexico); MOS (Montenegro); MBIE (New Zealand); PAEC (Pakistan); MSHE and NSC (Poland); FCT (Portugal); JINR (Dubna); MON, RosAtom, RAS, RFBR, and NRC KI (Russia); MESTD (Serbia); SEIDI, CPAN, PCTI, and FEDER (Spain); MOSTR (Sri Lanka); Swiss Funding Agencies (Switzerland); MST (Taipei); ThEPCenter, IPST, STAR, and NSTDA (Thailand); TUBITAK and TAEK (Turkey); NASU (Ukraine); STFC (United Kingdom); DOE and NSF (USA).
  
  \hyphenation{Rachada-pisek} Individuals have received support from the Marie-Curie program and the European Research Council and Horizon 2020 Grant, contract Nos.\ 675440, 724704, 752730, and 765710 (European Union); the Leventis Foundation; the Alfred P.\ Sloan Foundation; the Alexander von Humboldt Foundation; the Belgian Federal Science Policy Office; the Fonds pour la Formation \`a la Recherche dans l'Industrie et dans l'Agriculture (FRIA-Belgium); the Agentschap voor Innovatie door Wetenschap en Technologie (IWT-Belgium); the F.R.S.-FNRS and FWO (Belgium) under the ``Excellence of Science -- EOS" -- be.h project n.\ 30820817; the Beijing Municipal Science \& Technology Commission, No. Z191100007219010; the Ministry of Education, Youth and Sports (MEYS) of the Czech Republic; the Deutsche Forschungsgemeinschaft (DFG), under Germany's Excellence Strategy -- EXC 2121 ``Quantum Universe" -- 390833306, and under project number 400140256 - GRK2497; the Lend\"ulet (``Momentum") Program and the J\'anos Bolyai Research Scholarship of the Hungarian Academy of Sciences, the New National Excellence Program \'UNKP, the NKFIA research grants 123842, 123959, 124845, 124850, 125105, 128713, 128786, and 129058 (Hungary); the Council of Science and Industrial Research, India; the HOMING PLUS program of the Foundation for Polish Science, cofinanced from European Union, Regional Development Fund, the Mobility Plus program of the Ministry of Science and Higher Education, the National Science Center (Poland), contracts Harmonia 2014/14/M/ST2/00428, Opus 2014/13/B/ST2/02543, 2014/15/B/ST2/03998, and 2015/19/B/ST2/02861, Sonata-bis 2012/07/E/ST2/01406; the National Priorities Research Program by Qatar National Research Fund; the Ministry of Science and Higher Education, project no. 0723-2020-0041 (Russia); the Programa Estatal de Fomento de la Investigaci{\'o}n Cient{\'i}fica y T{\'e}cnica de Excelencia Mar\'{\i}a de Maeztu, grant MDM-2015-0509 and the Programa Severo Ochoa del Principado de Asturias; the Thalis and Aristeia programs cofinanced by EU-ESF and the Greek NSRF; the Rachadapisek Sompot Fund for Postdoctoral Fellowship, Chulalongkorn University and the Chulalongkorn Academic into Its 2nd Century Project Advancement Project (Thailand); the Kavli Foundation; the Nvidia Corporation; the SuperMicro Corporation; the Welch Foundation, contract C-1845; and the Weston Havens Foundation (USA).
\end{acknowledgments}

\bibliography{auto_generated}  
\cleardoublepage \appendix\section{The CMS Collaboration \label{app:collab}}\begin{sloppypar}\hyphenpenalty=5000\widowpenalty=500\clubpenalty=5000\input{MUO-19-001-authorlist.tex}\end{sloppypar}
%%% END EDITABLE REGION %%%
% skeleton_end
\end{document}

%% file: MUO-19-001-authorlist.tex
\vskip\cmsinstskip
\textbf{Yerevan Physics Institute, Yerevan, Armenia}\\*[0pt]
A.M.~Sirunyan$^{\textrm{\dag}}$, A.~Tumasyan
\vskip\cmsinstskip
\textbf{Institut f\"{u}r Hochenergiephysik, Wien, Austria}\\*[0pt]
W.~Adam, T.~Bergauer, M.~Dragicevic, A.~Escalante~Del~Valle, R.~Fr\"{u}hwirth\cmsAuthorMark{1}, M.~Jeitler\cmsAuthorMark{1}, N.~Krammer, L.~Lechner, D.~Liko, I.~Mikulec, F.M.~Pitters, J.~Schieck\cmsAuthorMark{1}, R.~Sch\"{o}fbeck, M.~Spanring, S.~Templ, W.~Waltenberger, C.-E.~Wulz\cmsAuthorMark{1}, M.~Zarucki
% ...
\vskip\cmsinstskip
\dag: Deceased\\
1:  Also at Vienna University of Technology, Vienna, Austria\\
% ...


%% file: MUO-19-001_temp.bbl
\providecommand{\href}[2]{#2}\begingroup\raggedright\begin{thebibliography}{10}%
\makeatletter
\providecommand{\hrefCMSnoop }[0]{\@secondoftwo}%
\makeatother
\providecommand{\doi}{\texttt{doi:}\begingroup \urlstyle{tt}\Url}

\bibitem{Chatrchyan:2012xi}
\hrefCMSnoop {}{{CMS Collaboration}, ``{Performance of CMS muon reconstruction
  in pp collision events at $\sqrt{s}=7$ TeV}'',} \textit{ JINST} \textbf{ 7}
  (2012) P10002,
  \href{http://dx.doi.org/10.1088/1748-0221/7/10/P10002}{\doi{10.1088/1748-0221/7/10/P10002}},
\href{http://www.arXiv.org/abs/1206.4071}{\texttt{arXiv:1206.4071}}.
%%CITATION = ARXIV:1206.4071;%%.

\bibitem{Khachatryan:2016bia}
\hrefCMSnoop {}{{CMS Collaboration}, ``{The CMS trigger system}'',} \textit{
  JINST} \textbf{ 12} (2017) P01020,
  \href{http://dx.doi.org/10.1088/1748-0221/12/01/P01020}{\doi{10.1088/1748-0221/12/01/P01020}},
\href{http://www.arXiv.org/abs/1609.02366}{\texttt{arXiv:1609.02366}}.
%%CITATION = ARXIV:1609.02366;%%.

\bibitem{Dominguez:1481838}
\href {https://cds.cern.ch/record/1481838}{{CMS Collaboration}, ``{CMS}
  technical design report for the pixel detector upgrade'',} Technical Report
  CERN-LHCC-2012-016. CMS-TDR-11, 2012.
\newblock \url{https://cds.cern.ch/record/1481838}.

\bibitem{Sirunyan:2018fpa}
\hrefCMSnoop {}{{CMS Collaboration}, ``{Performance of the CMS muon detector
  and muon reconstruction with proton-proton collisions at $\sqrt{s}=$ 13
  TeV}'',} \textit{ JINST} \textbf{ 13} (2018) P06015,
  \href{http://dx.doi.org/10.1088/1748-0221/13/06/P06015}{\doi{10.1088/1748-0221/13/06/P06015}},
\href{http://www.arXiv.org/abs/1804.04528}{\texttt{arXiv:1804.04528}}.
%%CITATION = ARXIV:1804.04528;%%.

\bibitem{TRG-17-001}
\hrefCMSnoop {}{{CMS Collaboration}, ``Performance of the {CMS} {Level}-1
  trigger in proton-proton collisions at $\sqrt{s} =$ 13 {TeV}'',} \textit{
  JINST} \textbf{ 15} (2020) P10017,
  \href{http://dx.doi.org/10.1088/1748-0221/15/10/P10017}{\doi{10.1088/1748-0221/15/10/P10017}},
  \href{http://www.arXiv.org/abs/2006.10165}{\texttt{arXiv:2006.10165}}.

\bibitem{MUO-17-001}
\hrefCMSnoop {}{{CMS Collaboration}, ``{Performance of the reconstruction and
  identification of high-momentum muons in proton-proton collisions at
  $\sqrt{s} =$ 13 TeV}'',} \textit{ JINST} \textbf{ 15} (2020) P02027,
  \href{http://dx.doi.org/10.1088/1748-0221/15/02/P02027}{\doi{10.1088/1748-0221/15/02/P02027}},
\href{http://www.arXiv.org/abs/1912.03516}{\texttt{arXiv:1912.03516}}.
%%CITATION = ARXIV:1912.03516;%%.

\bibitem{Chatrchyan:2008zzk}
\hrefCMSnoop {}{{CMS Collaboration}, ``The {CMS} experiment at the {CERN}
  {LHC}'',} \textit{ JINST} \textbf{ 3} (2008) S08004,
\href{http://dx.doi.org/10.1088/1748-0221/3/08/S08004}{\doi{10.1088/1748-0221/3/08/S08004}}.
%%CITATION = JINST,3,S08004;%%.

\bibitem{Chatrchyan:2014fea}
\hrefCMSnoop {}{{CMS Collaboration}, ``{Description and performance of track
  and primary-vertex reconstruction with the CMS tracker}'',} \textit{ JINST}
  \textbf{ 9} (2014) P10009,
  \href{http://dx.doi.org/10.1088/1748-0221/9/10/P10009}{\doi{10.1088/1748-0221/9/10/P10009}},
\href{http://www.arXiv.org/abs/1405.6569}{\texttt{arXiv:1405.6569}}.
%%CITATION = ARXIV:1405.6569;%%.

\bibitem{CMS-DP-2020-032}
\href {https://cds.cern.ch/record/2723305}{{CMS Collaboration}, ``Track impact
  parameter resolution in the 2017 dataset with the {CMS} phase-1 pixel
  detector'',} CMS Detector Performance Note CMS-DP-2020-032, 2020.
\newblock \url{https://cds.cern.ch/record/2723305}.

\bibitem{DPGPerformance}
\hrefCMSnoop {}{{CMS Collaboration}, ``The performance of the {CMS} muon
  detector in proton-proton collisions at $\sqrt{s} = 7$ {TeV} at the {LHC}'',}
  \textit{ JINST} \textbf{ 8} (2013) P11002,
  \href{http://dx.doi.org/10.1088/1748-0221/8/11/P11002}{\doi{10.1088/1748-0221/8/11/P11002}},
  \href{http://www.arXiv.org/abs/1306.6905}{\texttt{arXiv:1306.6905}}.

\bibitem{Bayatian:2006nff}
\href {http://cds.cern.ch/record/922757}{{CMS Collaboration}, ``{CMS} physics:
  Technical design report volume 1: Detector performance and software'',}
  {Technical Report} CERN-LHCC-2006-001, CMS-TDR-8-1, 2006.
\newblock \url{http://cds.cern.ch/record/922757}.

\bibitem{DCDC}
\hrefCMSnoop {}{F.~Faccio, S.~Michelis, and G.~Ripamonti, ``{Failure of
  FEAST2.1-based modules in the CMS pixel detector system (physics run
  2017)}''.}
  \url{https://espace.cern.ch/project-DCDC-new/_layouts/15/start.aspx#/Reports},
  2018.

\bibitem{CMS_WZ}
\hrefCMSnoop {}{{CMS Collaboration}, ``Measurement of the inclusive {W} and {Z}
  production cross sections in pp collisions at $\sqrt{s}=7${\TeV} with the
  {CMS} experiment'',} \textit{ JHEP} \textbf{ 10} (2011) 132,
  \href{http://dx.doi.org/10.1007/JHEP10(2011)132}{\doi{10.1007/JHEP10(2011)132}},
  \href{http://www.arXiv.org/abs/1107.4789}{\texttt{arXiv:1107.4789}}.

\bibitem{L1Phase1UpgradeTDR}
\href {https://cds.cern.ch/record/1556311}{{CMS Collaboration}, ``{CMS}
  technical design report for the {Level}-1 trigger upgrade'',} Technical
  Report CERN-LHCC-2013-011, CMS-TDR-012, 2013.
\newblock \url{https://cds.cern.ch/record/1556311}.

\bibitem{Chatrchyan:2009ae}
\hrefCMSnoop {}{{CMS Collaboration}, ``{Performance of CMS muon reconstruction
  in cosmic-ray events}'',} \textit{ JINST} \textbf{ 5} (2010) T03022,
  \href{http://dx.doi.org/10.1088/1748-0221/5/03/T03022}{\doi{10.1088/1748-0221/5/03/T03022}},
\href{http://www.arXiv.org/abs/0911.4994}{\texttt{arXiv:0911.4994}}.
%%CITATION = ARXIV:0911.4994;%%.

\bibitem{Fruhwirth:1987fm}
\hrefCMSnoop {}{R.~{Fr\"uhwirth}, ``{Application of Kalman filtering to track
  and vertex fitting}'',} \textit{ Nucl. Instrum. Meth. A} \textbf{ 262} (1987)
  444,
\href{http://dx.doi.org/10.1016/0168-9002(87)90887-4}{\doi{10.1016/0168-9002(87)90887-4}}.
%%CITATION = NUIMA,A262,444;%%.

\bibitem{Pantaleo:2293435}
\href {https://cds.cern.ch/record/2293435}{F.~Pantaleo, ``{New track seeding
  techniques for the CMS experiment}''}.
\newblock PhD thesis, Hamburg University, 2017.
\newblock CERN-THESIS-2017-028.

\bibitem{Oreglia:1980cs}
\href {http://www.slac.stanford.edu/cgi-wrap/getdoc/slac-r-236.pdf}{M.~J.
  Oreglia, ``A study of the reactions $\psi^\prime \to \gamma \gamma \psi$''}.
\newblock PhD thesis, Stanford University, 1980.
\newblock {SLAC} Report {SLAC-R-236}.

\bibitem{CMS-PRF-14-001}
\hrefCMSnoop {}{{CMS Collaboration}, ``Particle-flow reconstruction and global
  event description with the {CMS} detector'',} \textit{ JINST} \textbf{ 12}
  (2017) P10003,
  \href{http://dx.doi.org/10.1088/1748-0221/12/10/P10003}{\doi{10.1088/1748-0221/12/10/P10003}},
\href{http://www.arXiv.org/abs/1706.04965}{\texttt{arXiv:1706.04965}}.
%%CITATION = ARXIV:1706.04965;%%.

\bibitem{Khachatryan:2014fba}
\hrefCMSnoop {}{{CMS Collaboration}, ``{Search for physics beyond the standard
  model in dilepton mass spectra in proton-proton collisions at $ \sqrt{s}=8 $
  TeV}'',} \textit{ JHEP} \textbf{ 04} (2015) 025,
  \href{http://dx.doi.org/10.1007/JHEP04(2015)025}{\doi{10.1007/JHEP04(2015)025}},
\href{http://www.arXiv.org/abs/1412.6302}{\texttt{arXiv:1412.6302}}.
%%CITATION = ARXIV:1412.6302;%%.

\end{thebibliography}\endgroup
